# Multi-objective Design of Uniform Sparse MIMO Arrays

Suleyman Gokhun Tanyer[1,2], Paul Dent[1], Murtaza Ali[1], Curtis Davis[1], SenthinelKumar Rajagopal[1], Peter Driessen[2]

([1]) Uhnder, Inc., 3409 Executive Center Dr., Austin, TX 78731, USA

([2]) Dept. of Electr. and Computer Eng., University of Victoria, Victoria BC, V8P 5C2, Canada

**ABSTRACT** The problem of multi-objective design of sparse MIMO arrays for better multi-target detection capabilities is considered. A novel approach for efficient utilization of the antenna design resources; namely, the number of available array elements, and array aperture are studied for the angular beamforming performance metrics such as, beam width, the peak to sidelobe ratio (PSLR), and grating lobe limited field of view. The limiting constraints, the physical size of elements, and mutual coupling are also considered. Thinning of fully populated uniform MIMO antenna arrays to form effective uniform sparse arrays (USA), as the term proposed here, are examined as for their capability of improving usable field of view (uFOV), beamwidth (BW), and peak-to-side lobe-ratio (PSLR) using fewer physical antenna elements, simultaneously.

Sparse arrays require much less array elements to outperform uniform linear (ULA) and rectangular arrays (URA) for their beamwidths. Further, a rigorous design procedure considering the physical size limitations is currently unavailable. Here, we present a practical design architecture of such a uniform sparse array under multiple contradicting objectives. Angular resolution performances of the novel uniform sparse arrays are compared with the standard ULA and URAs. It is shown that design of an array with small inter-element spacings avoiding any grating lobes is possible even when the physical size of the elements are very large. Expanding the available elements to a much larger apertures provides much better angular resolution. However, it is also shown that these advantages come with the cost of increased side lobes. Both the simulated and the measured results demonstrate the superiority of the proposed uniform sparse array design compared to classical fully populated uniform arrays.

**INDEX TERMS** uniform sparse arrays, sparse MIMO array processing, radar signal processing, mutual coupling, angular resolution, side lobe reduction.

I. INTRODUCTION

Sensor arrays play an important role in spatio-temporal signal processing with applications spanning across multiple fields such as electromagnetic, acoustics, ultrasonic and seismic processing systems. Fundamental technology used in some of those applications includes radar, sonar, navigation, wireless communications, electronic surveillance and radio astronomy [1–3]. By exploiting the diverse waveform from its multiple antennas, multiple-input multiple-output (MIMO) radar transmits a probing waveform that can be chosen at will to maximize the power around areas of interest. Antenna pattern design for MIMO radar has become a popular research topic in recent years [1–8]. There is a vast amount of applications in signal processing and communication systems where the discrete signal samples are sparse in some domain such as time, frequency, or space. However, novel methods seem to outperform conventional techniques, regarding both complexity and optimality [9].

Sparse antenna array has several advantages in the high resolution thinned configurations for phased array radar and MIMO radar. However, because of array thinning, the side lobe level increases and subsequently leads to the grating lobe [4]. In this case, maintaining the approximation performance and preventing the grating lobe present two major challenges for a MIMO radar with sparse antenna arrays [1, 4, and 10].

Each field has developed similar tools, algorithms, and reconstruction methods for sparse signal processing which are; sampling: random sampling of bandlimited signals, compressed sensing (CS) [11], channel Estimation in Orthogonal Frequency Division Multiplexing (OFDM) [11–13], and finally the calibration for the mutual couping between antenna elements [14–18].

In this work, without any loss of generality our focus is on automotive radars, for which in the last few years it have been transformed from being a niche sensor to becoming standard even in middle-class cars. With Euro-NCAP ratings now requiring automated braking and pedestrian safety functionality, radar is often identified as the best suited sensor for this purpose. A selection of technology trends addressing some common requirements can be found in [19, and 20]

The remainder of this paper is organized as follows: In Section II, we derive the MIMO radar system model and extend the radiation and receiving mechanism of a radar formulated for sparse arrays. In Section III, classical Fourier based steering matrix formulation is derived for sparse arrays. In Section IV, multi-objective design and optimization of a sparse MIMO array is studied. Performance parameter definitions, smart initialization for fast convergence and minimization of mutual coupling effects are proposed. Finally, results and conclusions are drawn in Section V. Performance of the method is illustrated on four uniform sparse arrays, namely, AnkaraA/B and CoruhA/B arrays, and interesting directions for future work are pointed out.

## II. SPARSE MIMO RADAR SYSTEMS

For a multiple-input multiple-output (MIMO) radar system each transmitter (TX) signal is rendered distinguishable from every other TX by using appropriate differences in the modulation, such as, different digital code sequences. Each receiver (RX) correlates with each TX signal, producing a number of correlated outputs equal to the product of the number of RXs with the number of TXs. The outputs are deemed to have been produced by a number of virtual receivers (VRXs), which can exceed the number of physical RXs.

A transmit signal reflected from a reflecting object, and received and correlated by a receiver results in a digitized skin return signal that is the same as would have been produced by a transmit and receive antenna colocated at coordinates which are the sum of the actual transmit and receive antenna coordinates; such signals, called VRX signals, are further combined according to delay and doppler shift to resolve targets in the four dimensions of azimuth, elevation, range and doppler.

Antenna aperture and the desired field of view defined in the spherical coordinate system are illustrated in Fig. 1.

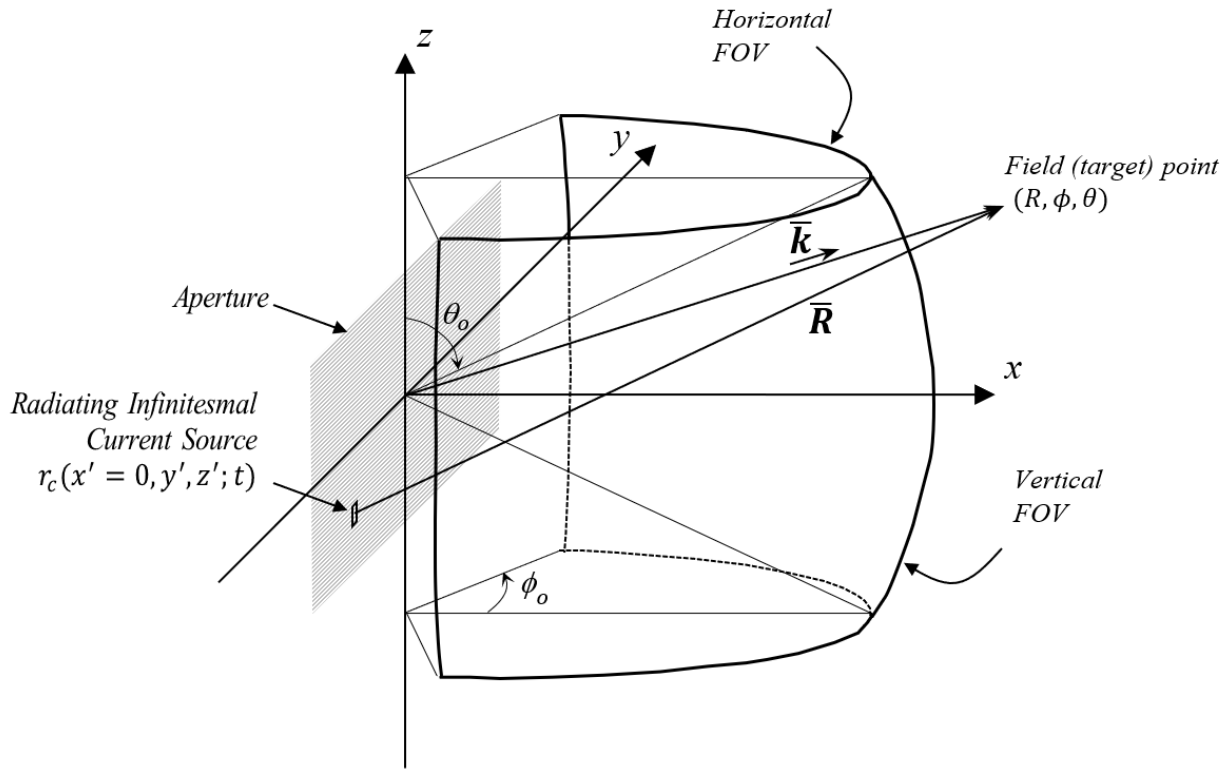

**Fig. 1.** Problem geometry for a planar antenna aperture.

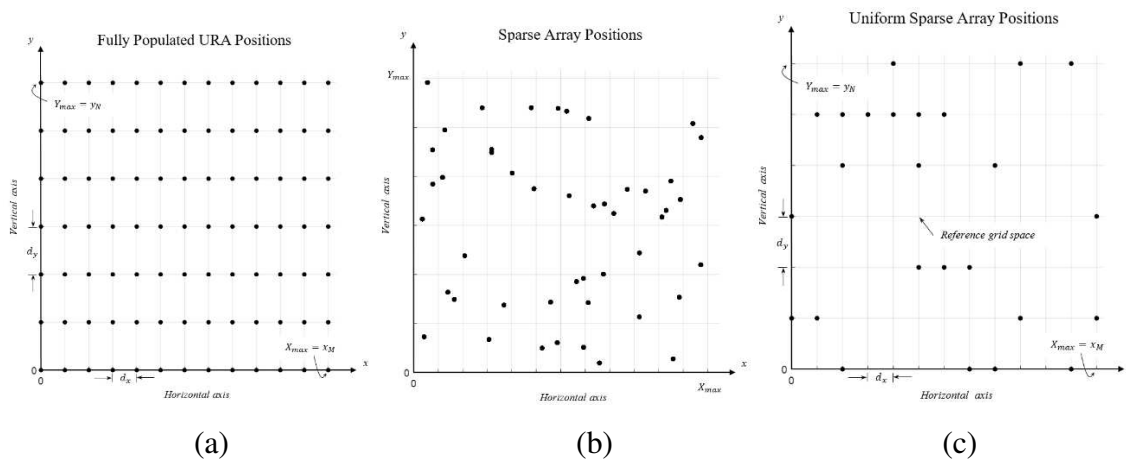

(a) (b) (c)

**Fig. 2**. Uniform rectangular array (URA) and sparse array positions. (a) Fully-populated reference URA, (b) a general sparse array, (c) uniform sparse array (USA).

Antenna aperture is generally shared by the individual transmitting and receiving elements. Their positions for the uniform linear and rectangular arrays are well-known. Sparse array (SA) elements in general can be located anywhere inside a given physical aperture. However, uniform sparse array (USA) elements, as the term proposed here, can only be located on the grid points defined by a reference fully populated uniform array with largest possible inter-element spacing values as shown in Fig. 2.

### A. A Radiating Antenna Aperture

Let us examine a radiating planar aperture where the radiated signal strength is proportional to the surface integral across the aperture of the current at each point weighted by a phase factor (with reference to the origin) that is a function of the position in the aperture and the field point for which the response is to be calculated. If the current distribution is $r(y,z)$ then the relative gain in the field point $(R, \phi, \theta)$ is given by [21]

$$s(R, \phi, \theta; t) = \int_S A\, r_c(y', z'; t) e^{j\beta_r(y', z'; t)} e^{-j\bar{k} \bullet \bar{R}}\, dy'\, dz' \qquad (1)$$

where $S$ is the planar antenna aperture, $A$ is the scaling constant, $r_c$ is the excitation current where $r_c(y', z'; t) = |r(y', z'; t)| e^{j\beta_r(y', z'; t)}$ with a phase $\beta_r$ with respect to the origin, $(x' = 0, y', z')$ is the primed cartesian (source) coordinates, $(R, \phi, \theta)$ is the unprimed spherical (field) coordinates, $\bar{k}$ is the wave vector where $k = |\bar{k}| = |k\hat{a}_R| = 2\pi/\lambda_0 = (2\pi f_0)/c$, $\lambda_o$ and $f_0$ are the wavelength and the center frequency for the narrow band transmission respectively, $\hat{a}_R$ is the radial unit vector in the direction of $(\phi, \theta)$, $(\bullet)$ is the scalar (dot) product operator, $\bar{R}$ is the exact one way source to the field poin displacement vector, and dependence to the field point brings the variables $R, \phi, \theta$ to the left side.

For field points in the far field the vectors $\bar{k}$ and $\bar{R}$ are approximately parallel (Fig. 1)

$$\bar{k} \bullet \bar{R} \cong R + y' \sin\phi \sin\theta + z' \cos\theta, \qquad (2)$$

and the radiated signal with respect to the reference phase $\beta(t)$ at the origin (1) simplifies to

$$s(\phi, \theta) = \int_S r_c(x', y') e^{-jk(y' \sin\phi \sin\theta + z' \cos\theta)} dy'\, dz' \qquad (3)$$

where the scaling factor is ignored. The above integral becomes

$$s(u, v) = \int_S r_c(x, z) e^{-jk(y'u + z'v)} dy'\, dz' \qquad (5)$$

where $(u, v) = (\sin\phi \sin\theta, \cos\theta)$. Note that $s$ is the inverse Fourier transform of the aperture distribution $r$, and $r$ is the Fourier transform of the broadside pattern $s$ where this pattern is defined only for the real angles for which $u^2 + v^2 < 1$.

For a single infinitesimal current element at an arbitrary location $(x_s, y_s)$ on the aperture

$$r_c(y', z') = r_0 \delta(y' - y_s, z' - z_s), \qquad (6)$$

the radiated field simplifies to

$$s(u,v) = r_o e^{-jk(y_s u + z_s v)} \tag{7}$$

where the current source amplitude directly contributes to the field with its propagation delay.

Let us assume a rectangular aperture defined by $0 < y < Y_{max}$, and $0 < z < Z_{max}$, and uniform spatial samples are at the source points $(md_y, nd_z)$ for $m = 0, 1, 2, \ldots (M-1)$, and $n = 0, 1, 2 \ldots (N-1)$, and $Y_{max} = (M-1)d_y$ and $Z_{max} = (N-1)d_z$, and $d_y$ and $d_z$ are the horizontal and vertical sampling intervals (Fig. 2). The infinitesimal currents are located on the rectangular $[M \times N]$ grid points

$$r_c(y', z') = r_c(m,n)\delta(y' - md_y, z' - nd_z), \quad \text{for } 0 < m < M-1, 0 < n < N-1. \tag{8}$$

The superposition of individual contributions is accurate if mutual coupling between the source elements are ignored the radiated field can be calculated as follows,

$$s(u,v) = \sum_{m=0}^{M-1}\sum_{n=0}^{N-1} r_c(m,n) e^{-jk(md_y u + nd_z v)} = \sum_{m=0}^{M-1}\sum_{n=0}^{N-1} r_c(m,n) e^{-j2\pi(md_{\lambda,y} u + nd_{\lambda,z} v)} \tag{9}$$

where $d_{\lambda,y} = d_y/\lambda$, and $d_{\lambda,z} = d_z/\lambda$. For a uniform sparse array using the same physical antenna aperture

$$s_s(u,v) = \sum_{m=0}^{M-1}\sum_{n=0}^{N-1} r_s(m,n) e^{-j2\pi(md_{\lambda,y} u + nd_{\lambda,z} v)} \tag{10}$$

where $r_s$ are the sparse samples of $r_c$, and are non zero only at the physically available positions. For a non uniform sparse array those exponential coefficients are not the regular Fourier coefficients.

### B. A Receiving Sparse MIMO Antenna Array

For a single-input single-output (SISO) radar the induced current at the receiver, $r$, due to multiple targets can be assumed to be in the form

$$r(y_{tx}, z_{tx}, y_{rx}, z_{rx}; t) = \sum_{t=1}^{T} A\sigma_{c,t} f_{tx}(\phi_t, \theta_t; t) f_{rx}(\phi_t, \theta_t; t) e^{j\bar{k}\bullet\bar{R}_{tx} + j\bar{k}\bullet\bar{R}_{rx}} \tag{11}$$

where $A$ is the two-way spread factor, $\sigma_{c,t}$ is the complex-valued skin return with real-valued radar cross section $\sigma_t$ and the phase return $\gamma_t$ for the $t$'th target at the field coordinates $(\phi_t, \theta_t)$, $f_{tx}$ and $f_{rx}$ are the antenna patterns for the transmitting and receiving elements, $\bar{R}_{tx}$ and $\bar{R}_{rx}$ are the exact transmitter to target and target to receiver displacement vectors forming the two-way propagation path, respectively and other parameters are defined in (1).

Assuming uniform antenna elements and no tapering over the elements, ignoring common terms, the received signal in the far field with respect to the reference phase $\beta(t)$ at the origin (13) can be simplified similar to (4)

$$\begin{aligned} r_{j,i}(p) &= r(y^p_{vrx}, z^p_{vrx}) \quad \text{for } p = 1, 2, \ldots P, \\ &= r(y^i_{tx}, z^i_{tx}; y^j_{rx}, z^j_{rx}) \\ &= \sum_{t=1}^{T} \sigma_{c,t} \, e^{jk(y^p_{vrx} \sin \phi_t \sin \theta_t + z^p_{vrx} \cos \theta_t)} \end{aligned} \quad (12)$$

where $r_{j,i}$ is the received signal at the $j$'th RX due to the $i$'th transmitted signal, $r$ is the received signal at the $p$'th VRX where $y^p_{vrx} = y^i_{tx} + y^j_{rx}$, $z^p_{vrx} = z^i_{tx} + z^j_{rx}$ are the coordinates of the $p$'th VRX created by the $i$'th RX and $j$'th TX couple located at the points $(y^i_{tx}, z^i_{tx})$ and $(y^j_{rx}, z^j_{rx})$ for $1 \leq i \leq M$ and $1 \leq j \leq N$ as illustrated in Fig. 2. In this fully populated uniform rectangular array (URA) all possible TX-RX couples create $MN$ virtual elements located over a virtual aperture of $2Y_{max} \times 2Z_{max}$ where a sorted list for those couples $(m, n)$ can be created using an order number $p$ where $1 \leq p \leq P = MN$, and each $p$ corresponds to a single $(m, n)$ pair assuming no overlapping occurs. Note that for a sparse MIMO array the number of physical elements, $P_s$, could be much less than $P$ (Fig. 2).

As an illustrative example let us set both inter-element spacings $d_x$ and $d_z$ to $\lambda/2$, and this column-wise vectorization of the received signal at $P$ virtual elements yields the weighted sum of $T$ vectors which are commonly referred to as the steering vectors

$$r(p) = \sum_{t=1}^{T} \sigma_{c,t} \, r_t(p; u_t, v_t), \quad \text{for} \quad p = 1, 2, \ldots P \quad (13)$$

where we can define the steering vector, $r_t$, as in general the phase distribution induced on virtual array elements for a single target in the far-field at a given target angle its elements are in the form

$$r_t(p; u, v) = e^{j2\pi(mu + nv)}, \quad \text{for} \quad p = 1, 2, \ldots P. \quad (14)$$

Once again for a uniform sparse MIMO array sharing the same grid space with a fully-populated array only a few number of VRXs could be available and (15) reduces to

$$r_s(p_s) = \sum_{t=1}^{T} \sigma_{c,t} \, r_t(p_s; u_t, v_t) \quad \text{for} \quad p_s = 1, 2, \ldots P_s < P \quad (15)$$

## III. ANGLE BEAMFORMING FOR A SPARSE MIMO RADAR

Recalling the properties of Fourier transform and its inverse discussed in Section II. B and using (8), (10), and (11) the received signal pattern for a full-populated uniform array for some $(\phi, \theta)$ can be calculated simply by

$$g(\phi, \theta) = \boldsymbol{br} \tag{16}$$

where the received signal at the virtual array elements form the received signal (column) vector $\boldsymbol{r}$ of length $P$, and $\boldsymbol{b}$ is the steering (column) vector for the same array evaluated at $(\phi, \theta)$. One can oversample $(u, v)$ uniformly for the received signal pattern

$$u_m = \frac{2m'}{Mq_\phi} - 1, \qquad \text{for} \quad 0 \leq m' < Mq_\phi, \tag{17}$$

$$v_n = \frac{2n'}{Nq_\theta} - 1, \qquad \text{for} \quad 0 \leq n' < Nq_\theta. \tag{18}$$

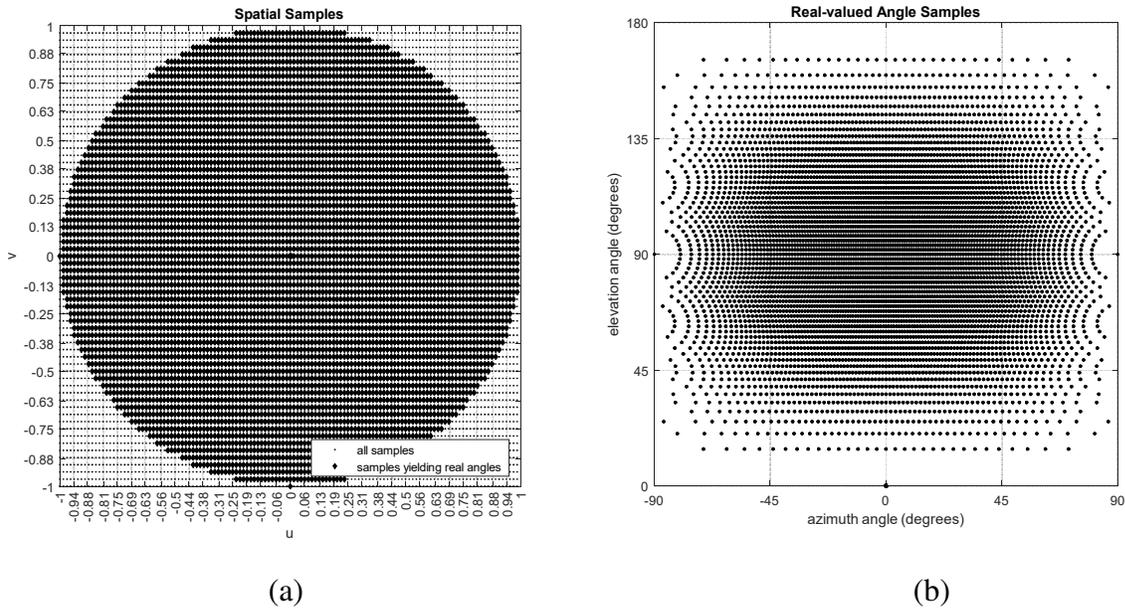

(a)           (b)

**Fig. 3**. (a) Uniformly sampled spatial variables, and (b) the corresponding real angles for which $(u^2 + v^2) \leq 1$ is satisfied.

The corresponding non uniform samples in the angular domain calculated below are illustrated in Fig. 3.

$$\phi_{m',n'} = \sin^{-1}\left(\frac{u}{\sin(\theta_{m'})}\right) = \sin^{-1}\left[\frac{\frac{2m'}{Mq_\phi} - 1}{\sqrt{1 - \left(\frac{2n'}{Nq_\theta} - 1\right)^2}}\right] \quad (19)$$

$$\theta_{n'} = \cos^{-1} v = \cos^{-1}\left(\frac{2n'}{Nq_\theta} - 1\right) \quad (20)$$

where the above equation is valid only for real angles for which $u^2 + v^2 \leq 1$ is satisfied, and where $q_\phi$ and $q_\theta$ are positive integer oversampling factors for $\phi$- and $\theta$-axes, respectively (Fig. 3). Now let us sort those angular samples into a single (row) vector, and evaluate the steering vector at each row yielding the angle beamforming matrix $\boldsymbol{B}$, the received signal pattern can be calculated

$$\boldsymbol{g} = \boldsymbol{B}\boldsymbol{r} \quad (21)$$

where $\boldsymbol{g}$ is the column vector for the received signal pattern, $g$ is calculated in (16), $\boldsymbol{r}$ is defined in (15), and where $\boldsymbol{B}$ is the angle beamforming matrix with a size $[(M-1)(N-1)q_\phi q_\theta)]$ by $P$. For the hardware implementation of a multi-resolution $\boldsymbol{B}$, both oversampling ratios, $q_\phi$ and $q_\theta$, can be set to be multiples of 2 and the largest of those can be used to calculate a vector look-up table for the list of orthogonal set of Fourier coefficients.

To obtain a 2D pattern with 512 azimuth and 256 elevation angle samples, $\boldsymbol{g}$ should have 131,072 row elements. For a virtual array of 192 elements $\boldsymbol{B}$ is expected to have 25,165,824 complex coefficients to be calculated or stored in the radar hardware for a general sparse array. However, for a USA those exponential coefficients become regular, and the number goes down to only 512 values. This brings a great advantage to USAs for their hardware implementations.

### IV. MULTI-OBJECTIVE DESIGN AND OPTIMIZATION OF UNIFORM SPARSE ARRAYS

For a uniform sparse array both the physical and virtual array elements are thinned and $r(p)$ is non zero only for its $P_s$ elements where $P_s < P$ [10]. Deleting the zero equations (21) simplifies to

$$\boldsymbol{g}_s = \boldsymbol{B}_s \boldsymbol{r}_s \quad (22)$$

where both columns of $\boldsymbol{B}_s$, and rows of $\boldsymbol{r}_s$ are reduced to $P_s$, and the *thinning/sparsity ratio* is defined as $t = P_s/P \leq 1$, and where a fully populated array this thinning ratio is 1.

#### A. Optimizing Parameters

In practice radar design engineers more than often need to satisfy multiple contradicting objectives for the same desired antenna array. The main optimization constraints are; 1) usable field of view (uFOV), 2) beamwidth (BW), 3) the total number of physical elements (*N*), and 4) the peak-to sidelobe ratio (PSLR). In addition to those here are some additional practical considerations.

*Physical size limitations*: The TX and the RX antenna elements in practice have finite physical dimensions, namely their width and heights which limit the minimum inter-element spacing values, and those spacing values limit the horizontal and vertical uFOV values, respectively. A densely packed ULA and URA are directly affected by this limitation. This is effective whenever any size dimension is larger than λ/2.

*Mutual coupling*: The TX and the RX groups should often be physically separated by defining group separation distance to decrease the inter-group mutual coupling [1]. Here we assume that the mutual coupling among the same type of elements are calibrated digitally.

*Antenna element sharing of different arrays*: For multi-functional radars some of the TX and the RX elements are often shared between different scan modes. The antenna array design and optimization for all scans need to be done simultaneously. As a practical approach, the physical elements for a simpler scan mode could be forced to be used on some other complicated antenna configuration. This provides an effective utilization of the array aperture by an initial enforcement of some of the element positions in the design and optimization of a sparse array accordingly.

*Hardware implementation constraints*: Antenna elements are fed by transmission lines or waveguide structures whose layouts are usually implemented on a separate neigboring hardware board. The transmitted and received signals' layout should also be fed from another layer. As a design choice a central region can be preferred to keep all the transmission lines approximately equal in length. Then, this central region needs to be defined as a forbidden zone for the array elements.

In this work, thinning of fully populated uniform MIMO antenna arrays to form effective sparse arrays are examined as for their capability of improving usable field of view (uFOV), beamwidth (BW), and peak-to-side lobe-ratio (PSLR) using fewer physical antenna elements, simultaneously. The detailed definitions for those parameters are given in the following sub-sections.

i. *Peak-to-side lobe ratio (PSLR):*

Further to this discussion for a given target range and velocity the peak-to-side lobe-ratio (PSLR) should be considered. Maximum skin-return for a target is given by

$$g_{peak} = |\boldsymbol{g}_s(\phi_t, \theta_t)| = \max_{\phi,\theta \text{ in FOV}} |\boldsymbol{g}_s(\phi, \theta)| \quad (23)$$

where $(\phi_t, \theta_t)$ is the direction of the brightest signal observation in the FOV, and the maximum side lobe level is given by

$$g_{msll} = |\boldsymbol{g}_s(\phi_s, \theta_s)| = \max_{\phi,\theta \text{ in FOV}} |\tilde{\boldsymbol{g}}_s(\phi, \theta)| \quad (24)$$

where $(\phi_s, \theta_s)$ is the direction of the largest side lobe in the uFOV, and where $\tilde{\boldsymbol{g}}_s$ is obtained by setting $\boldsymbol{g}_s$ to zero inside its main lobe region. This side lobe have the potential to create false targets even this side lobe is outside of the (operational) FOV when FOV ≤ uFOV.

The peak-to-side lobe ratio (PSLR) can be calculated by

$$PSLR(\phi_t, \theta_t; \phi_s, \theta_s) = 20 \, log_{10}\left(\frac{g_{peak}}{g_{msll}}\right). \tag{25}$$

Note that for uniform and isotropic elements, assuming mutual coupling to be shift-invariant, and (in the presence of multi-targets) ignoring the interference between the skin returns of neighboring targets the PSLR becomes independent of the target direction.

$$PSLR = 20 \, log_{10}\left(\frac{g_{peak}}{g_{msll}}\right). \tag{26}$$

ii. *Beamwidth for uniform arrays:*

The first-null-, and the half-power-beamwidth can be calculated as follows [22]

$$BW_{fn} = \sin^{-1}\left(\frac{1}{L_\lambda}\right) \tag{27}$$

$$BW_{hp} = 0.886\left(\frac{1}{L_\lambda}\right) \tag{28}$$

where $N$ is the number of uniform linear array (ULA) elements, $BW_{hp}$ and $BW_{fn}$ are the half power beam width (HPBW), and the first null beamwidth calculated in radians, and where $d_\lambda$ and $L_\lambda = (N-1)d_\lambda$ are inter-element spacing and aperture length in terms of wavelength, respectively. Eqns (27) and (28) are valid for the cross sections of uniform rectangular (URA) arrays for their horizontal and the elevation cross sections, separately.

iii. *Side lobes, Grating lobes and Usable FOV:*

ULA and URAs, by definition, have constant inter-element spacings, and at any target angle the inter-element phase differences are also constant which are ideally zero at the target angle and all elements contribute constructively. However, for field angles retreating from the main lobe moving along one of the axes, these phase differences with respect to the array center begin to rotate for the furthest elements the fastest. First null occurs when the two elements furthest to the array center are out-of phase with respect to this center by approximately $\pm\pi$ yielding the first minimum. Those far elements are $d(N-1)/2$ meters away from the array center with a phase difference of $\pm\pi d_\lambda(N-1)$ the first null is observed approximately at $\pm \sin^{-1}(1/d_\lambda(N-1))$ radians, respectively. This approximation is accurate within 0.6, 0.4, 0.2, and 0.1 degrees for $N = 15, 18, 25,$ and 35, respectively. Retreating further from the broadside angle while furthest elements' phase rotations turning back to in-phase, we observe the second to the last elements to be approximately out-of phase creating the second minimum. In this narrow angular region bounded by those two minima all except the last two elements are mostly in-phase causing the largest side lobe. Side lobes gradually decrease due to increasing incoherence between the elements until we reach the half way to the first grating lobe. For an odd numbered ULA all radiations cancel each other completely leaving the only one in the center to yield the minimum side lobe level of $1/N$. This phenomenae occurs independent of the number of elements or their spacings, and this first side lobe sets the PSLR approximately to –13.8 dB.

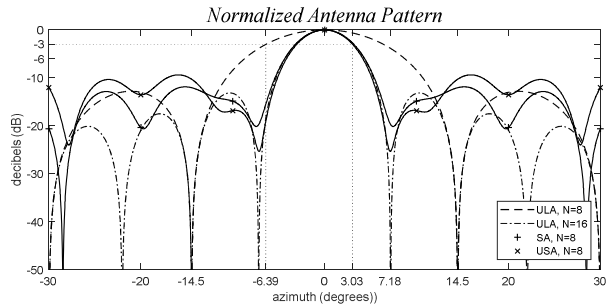

**Fig. 4**. Uniform linear arrays (ULA), $d_\lambda = 0.5$, (*dashed*) $N = 8$, (*dashed dot*) $N = 16$, and some examples to (*cross*) a sparse array (SA) with $d_\lambda = 0.5$, and $N = 8$ with nonuniform positions $d_\lambda = 0, 1.07, 2.52, 3.79, 4.48, 4.98, 5.86, 7.5$, and (*plus*) a uniform sparse array (USA) with $d_\lambda = 0.5$, and $N = 8$ using the positions of (dashed dot) as reference for which are thinned to 0, 3, 5, 7, 8, 10, 14, and 15.

Simple heuristics helping us to understand the interactions between ULA and URA elements also help us to understand why sparse arrays have much lower skin-return near (and even inside) the expected main lobe. Due to the irregularity of sparse element positions we expect more than just two elements to be out of phase away from the main lobe as illustrated in Fig. 4. In-phase interactions between elements become rapidly incoherent yielding a much smaller beamwidth. However, due to the conservation of radiated power the side lobes cannot be supressed but spread across the half-hemisphere, or can be optimized to be pushed outside of a given FOV.

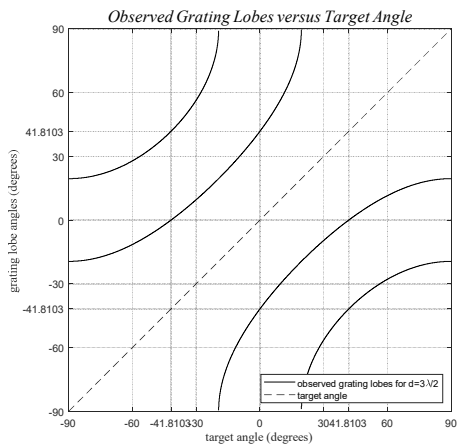
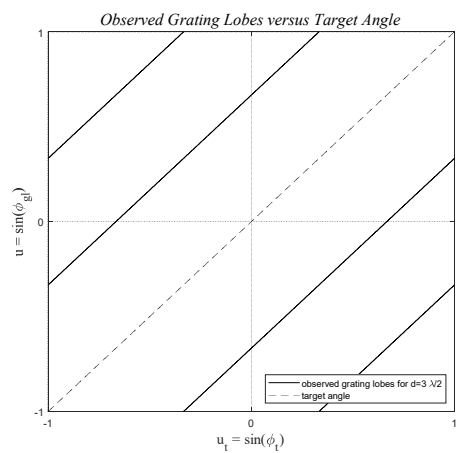

(a) (b)

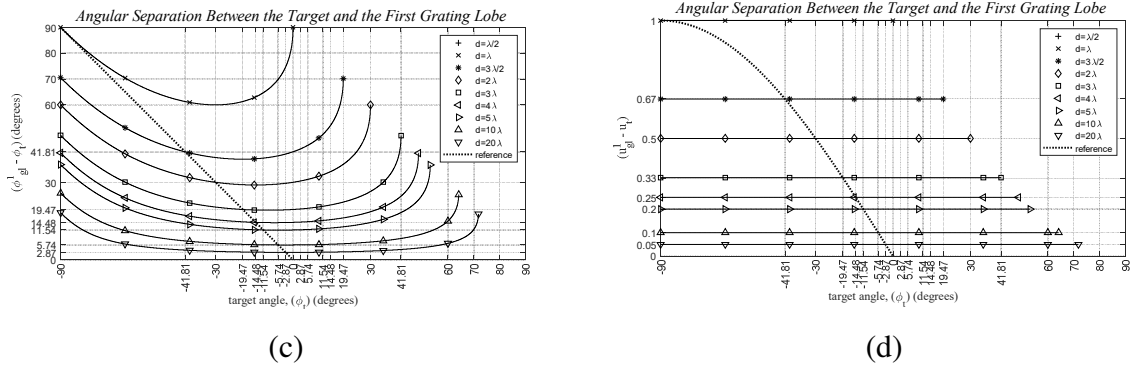

(c)                                              (d)

**Fig. 5**. Grating lobe angles as a function of the target angle. (*a,* and *b*) The first two grating lobes with respect to the target angle for a ULA with $d_\lambda = 3\lambda/2$, , (*c,* and *d*) the first grating lobe for various inter-element spacings where the dotted line shows the locus of points for the first false target image when the target is at broadside. Plots are shown in degrees and in the (*u, v*) plane, respectively.

Grating lobes occur at angles when *all* of the array elements are perfectly in-phase, and they are constructively contributing to the received signal pattern equivalent to the main lobe. Large side lobes, however, are *not* grating lobes, they require only the complex sum of all the contributions to be comparatively large, allowing some inter-element phase mismatching, and even some distructive interferences between the elements.

No grating lobe appears for $d_\lambda \leq \lambda/2$. Otherwise, for larger spacing values each target creates its own false target image(s) for which each image corresponds to a wrapped phase difference of some multiples of $\pm 2\pi$. False target images, or equivalently the grating lobes of a target, occur at angles with respect to its source target angle. The number of observable grating lobes increases as the inter-element spacing increases as shown in Fig. 5.

Usable FOV (uFOV) can be defined as the maximum grating lobe-free angular extend around broadside where outside of the uFOV a flipped (with respect to the origin) replica of the interior pattern which carries no additional information.

TABLE I

Inter-element spacings and the usable FOV values[1]

| Spacings (λ) | 0.5 | 0.51 | 0.53 | 0.58 | 0.65 | 0.78 | 1 | 2 | 20 |
|---|---|---|---|---|---|---|---|---|---|
| uFOV (±deg) | 90 | 80 | 70 | 60 | 50 | 40 | 30 | 14.48 | 1.43 |

For a fully populated ULA a target located at $\phi_t$ creates grating lobes at angles

$$\phi_{gl} = \sin^{-1} \gamma_n, \quad \text{for } n: -1 < \gamma_n \leq 1, \text{ and for } n = 0, \pm 1, \pm 2 \ldots \quad (29)$$

---

[1]One sided angular extend.

where $\gamma_n = (n/d_\lambda + \sin \phi_t)$ and $\phi_{gl} \neq \phi_t$ which yields the usable FOV (uFOV)

$$\phi_{uFOV} = \pm \mathrm{asin}(1/2d_\lambda), \text{ for } n: -1 < 1/2d_\lambda \leq 1, \text{ and for } n = \pm 1, \pm 2 \ldots \quad (30)$$

For $d_\lambda = \lambda/2$, and $\phi_t = \pi/2$, the first grating lobe occurs for $n = -2$ at $\phi_{gl} = -\pi/2$ allowing the usable angular extend with $\phi_{ufov} = \{-\pi/2, \pi/2\}$ (Fig. 1). Similarly for $d_\lambda = \lambda$, $\phi_t = \pi/6$, the closest grating lobe occurs for $n = -2$ at $\phi_{gl} = -\pi/6$, then the usable angular region becomes $\phi_{ufov} = \{-\pi/6, \pi/6\}$ as suggested by (30). Here note that in practice, the physical size of elements limit the minimum possible inter-element spacing values both on the horizontal and the vertical. For an element width of two wavelengths azimuthal uFOV is approximately limited to 14.48 degrees as shown in Table I. All targets outside the uFOV are expected to create false images stacked inside the uFOV, this becomes a problem when the antenna gains for the individual elements do not attenuate outside the uFOV sufficiently.

Above discussion shows that ULA and URAs often become unpractical even under these two constraints. Additional degrees of freedom are necessary. Sparse arrays can be designed to provide optimum solutions both for avoiding the grating lobes by their irregular element positions. They also provide very high angular resolution by allowing much smaller inter-element spacing values.

In this section, a practical design procedure for grating lobe-free sparse array with high angular resolution is proposed.

TABLE II

**Algorithm**: Sparse array optimization

***desired parameters***:
- array dimension (1D/2D/3D), and let us assume 1D as an example.
- available number of TX and RX elements, $N_{tx}$, $N_{rx}$, respectively,
- uFOV, and BW both for azimuth and elevation,

***constraints***:
- physical dimensions of antenna elements, $w_{tx}, h_{tx}, w_{rx}, h_{rx}$
- forbidden zones for mutual coupling, $y_{mc}, z_{mc}$
- enforced element positions

***initialize k = 0***:
- calculate required virtual aperture length,  $2Y_{\lambda,max}$ using (27) and (28)
- calculate required minimum inter-element spacings,  $\Delta d_{\lambda,min}$ using (29) and (30)
- determine reference full-grid positions,  $y_n$
- set positions to the enforced list of positions
- (optional HIA): calculate uniformly distributed set of $\Delta d_\lambda$  $\Delta d_\lambda^n$

***iterate k until*** ($k < K_{max}$) ***or*** ($PSLR_{best} > PSLR_{desired}$)
- random shuffling and random perturbations of $\Delta d_{\lambda,y}^n$
- add non-overlapping TX and RX positions satisfying constraints until all positions are calculated
- calculate the received signal pattern , and calculate the PSLR,
- update array positions and $PSLR_{best}$ if $PSLR_{new} < PSLR_{best}$

***end***

***repeat:*** outer loop is used if desired and hyper-parameters are also to optimized in a given range of values.

## B. Design and Optimization of a Uniform Sparse Arrays

Examining Section (IV.A) it is safe to say that for uniform arrays a desired uFOV BW can be improved only by increasing the antenna length ($L_\lambda$), or the aperture size ($A_\lambda$) and so with a larger number of elements ($N$). Thus, for given FOV and $N$, BW improvement is not possible with uniform linear and rectangular arrays. This is due to the fact that required FOV and BW values are sufficient to determine the size of the fully populated aperture and which could well be over the practical limits. Additional requirements for a specific implementation becomes even more stringent once the physical size of elements and the mutual coupling constraints are also to be considered. Therefore there is a need for an efficient way to design and construct TX/RX arrays utilizing thinning of arrays, and allowing larger apertures with much less elements, hopefully without receding the first two constraints considerably.

Let us assume that initial constraints are only the maximum number of elements and available physical antenna aperture, and all other parameters are to be optimized without any loss of generality. Radar scan type under consideration will determine the dimension of the array. Hardware constraints limit the available number of array elements. Further, operational environment might be requiring a minimum BW value. Operating frequency and the prefered implementations determine the element size requirements. Initial simulations for mutual couplings can provide insight about some horizontal and vertical separation distances, $L_{mc,y}$, and $L_{mc,z}$ for the two axes to keep those coupling under some threshold. Physical and virtual element sharing of a previously designed array would enforce some positions as an initial value for the optimization (See Table II.)

At this initial step, the desired BW and uFOV values determine the minimum aperture length/size and minimum inter-element spacing, respectively which yield the corresponding grid space for the reference fully populated virtual array using (27) through (30) (Table II). Reader can refer to the literature on various approaches in constructing of such virtual arrays. Available number of TX and RX elements determines the targeted thinning ratio as discussed in (22). Enforcement of initial element positions should be done on the same grid space.

As an iterative process, random or some type of preferred search algorithm is performed to determine successful arrays for their PSLR where each candidate is to satisfy all the constraints. Iteration is terminated if desired PSLR, or maximum number of iterations is reached. Further stopping criteria is met if last three PSLR values are withing significance (0.5 decibels).

The desired and the constructing parameters can also be optimized using an outer loop. In general, this optimization can well be constructed for a different set of constraining parameters. There is a vast amount of work offering improved convergence including, evolutionary algorithms and gradient descent type analytical approaches [23] which will be studied in the future. Here we propose a heuristic approach to illustrate fast convergence to a satisfactory 'locally optimum' array.

*iv. A heuristic initialization approach (HIA) for fast convergence:*

Firstly, low PSLR requires the grating lobes to be moved outside of operational FOV. This sets the grid sizes using (30), and Table I. Secondly, element positions need to be optimized for smaller PSLR. For a large side lobe overall in-phase contributions are expected to be large, and this generally occurs when the frequency of a certain inter-element spacing value is much higher than others. Thus, large side lobes can be spread across the half-hemisphere by setting the spacing values to be irregular (Table II). However, the total radiated power being conserved the overall sidelobe levels are expected to rise. PSLR reduction becomes possible when the Empirical Cumulative Distribution Function (ECDF) is forced to a smooth. One can initialize the spacing values by forcing their ECDF to be linear (uniform distribution) as proposed below.

A linear USA with $N$ elements has $(N-1)$ inter-element spacing values, $d_n$, which can be set to be empirically uniform [23].

$$d_n = d_{min} + (n-1)\Delta d \quad \text{for} \quad n = 1 \ldots (N-1) \tag{31}$$

where $d_{min}$ is the minimum inter-element spacing, for $N \geq 3$

$$\Delta d = \frac{X_{max} - X_{min} - (N-1)d_{min}}{\sum_{k=1}^{N-2} k} \tag{32}$$

and where $(N-1)$ spacings are placed between $N$ elements which are located between $X_{max}$ and $X_{min}$.

Let us first set the reference grid space to $\lambda/2$ for a grating lobe-free pattern. Ignoring the element size constraint the minimum spacing can be set to the reference grid size, $d_{min} = \lambda/2$. For $L_\lambda = 11$, and $N = 5$ we calculate $\Delta d = 3\lambda/2$, and spacings are linearly increasing as given by $2d_n/\lambda = \{1, 4, 7, 10\}$ yielding element positions $2x_n/\lambda = \{0, 1, 5, 12, 22\}$ for $X_{min} = 0$. This yields a linear ECDF, $F_D(d_n) = (2 + 2d_n/\lambda)/12$. All positions are in the reference grid space.

Secondly, using the same reference grid space, for an element size of $2\lambda$ to avoid any overlapping we need $d_{min} = 2\lambda$. Then, we calculate $\Delta d = \lambda/2$, and spacings are linearly increased as $2d_n/\lambda = \{4, 5, 6, 7\}$ yielding positions $2x_n/\lambda = \{0, 4, 9, 15, 22\}$. This also yields a linear ECDF, $F_D(d_n) = (-3 + 2d_n/\lambda)/4$. No grating lobes are expected since the ECDF is smooth, and the positions are located on a uniform grid size of $\lambda/2$. Please note that in both examples a random shuffling of the spacing ordering will change the element positions but not the ECDF. Each alternative configuration will yield to a different signal-return pattern. This requires search iterations for the optimum spacing ordering for the best PSLR.

Similar to the second example, let us now set $L_\lambda = 10$. We can recalculate $\Delta d = \lambda/3$, and spacings are given by $3d_n/\lambda = \{6, 7, 8, 9\}$ with element positions, $3x_n/\lambda = \{0, 6, 13, 21, 30\}$ which do not coincide with the reference grid points. For a general sparse array there is no constraint on the element positions as proposed in Table II. However, a 'uniform' sparse array (USA), by definition which is proposed in this paper, requires all of its element positions to be in some uniform grid space. For this third example, we can change the reference grid size to $\lambda/3$, or move the elements to some neighboring grid points without causing any overlapping. Both will provide the first group of initial array candidates for the optimization search. Element positions can further be perturbed further to move around the grids finding a better 'local optimum' PSLR value as described in Table II. The ECDF for the optimum spacings is expected to be a smooth function with minimal repetitive spacing values to spread the grating lobes into side lobes. Numerical examples of the ECDF are given Fig. ECDF in Section V.

   v.   *Sparse arrays with minimized mutual coupling:*

Spare array design procedure described above allows TX and RX elements to be located anywhere inside a common aperture, preferably both expanding over the aperture for smaller BW. However, this could be a problem if the mutual coupling among the TX and RX groups is high, then the two element groups should be separated to use different sections of the aperture. We can study this case by modifying the beamforming equation (6) by taking the mutual coupling contributions into account

$$g(u,v) = \int_S C(x_\lambda, y_\lambda; u, v)\, r(x_\lambda, y_\lambda) \exp[j2\pi(x_\lambda u + y_\lambda v)]\, dx_\lambda\, dy_\lambda, \qquad (33)$$

and physical aperture sparsing can be done as in (22)

$$\boldsymbol{G_s} = \boldsymbol{B_s C_s\, r_s} \qquad (34)$$

where $\boldsymbol{C_s}$ is the square mutual coupling matrix of size $P_s$. The above equation can be simplified by ignoring the mutual couplings between TX and RX elements with separation distances larger than some $y_{mc}$, and $z_{mc}$ for the horizontal and vertical, respectively. Under these assumptions, sparse array optimization can be forced to have an additional constraint by defining a 'forbidden zone' to push TX and RX elements away into separate regions which would simplify (34) to (22), approximately.

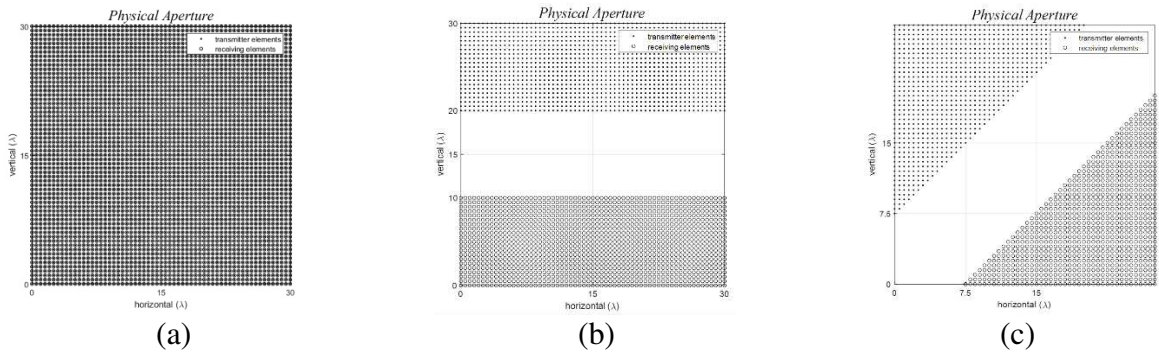

|         (a)         |         (b)         |         (c)         |

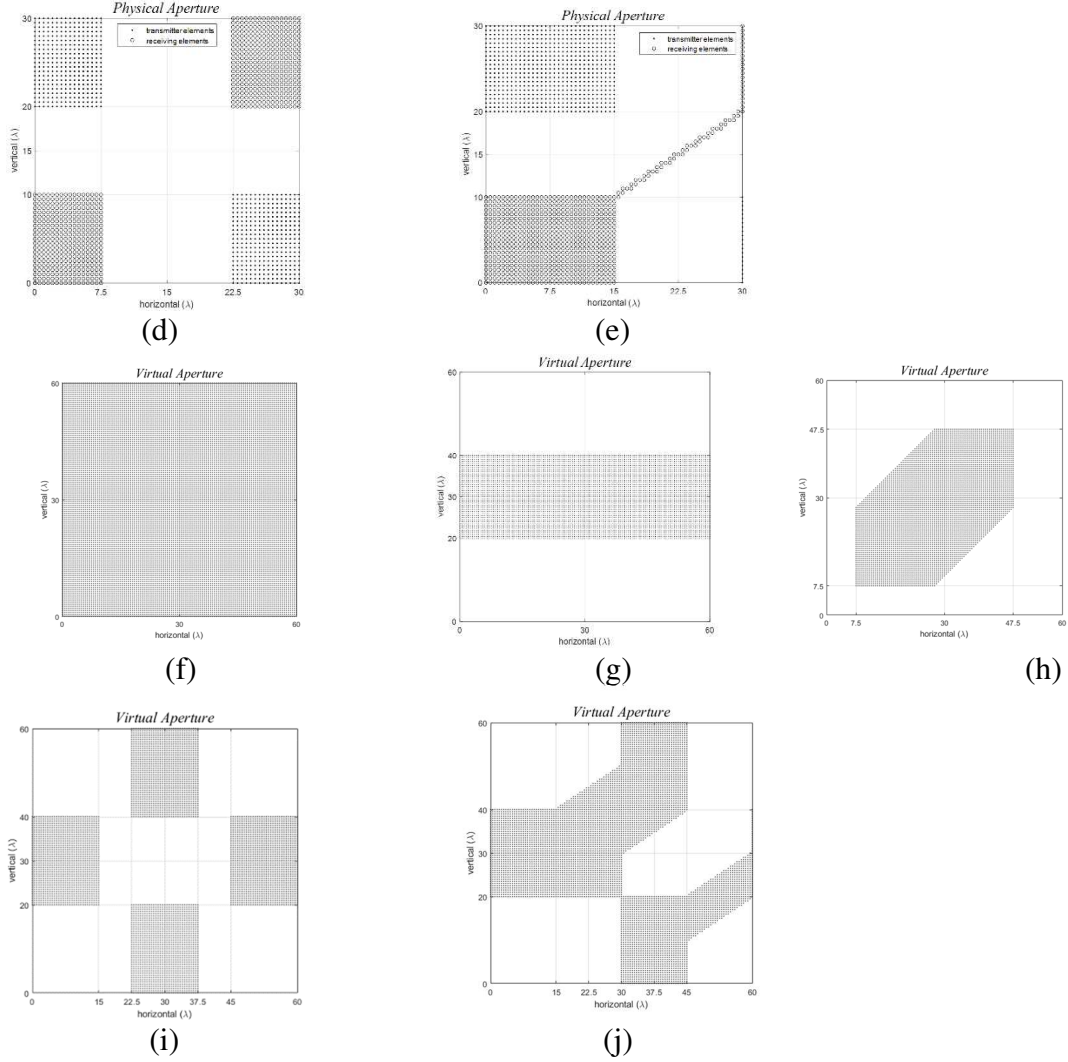

**Fig. 6.** Some example array structures with mutual coupling forbidden zones, (*a – e*) physical apertures, and (*f – j*) their virtual apertures. Physical and virtual apertures for; (*a-f*) fully-populated URA with no forbidden zones (co-located TX and RXs), (*b-g*) two-vertical, (*c-h*) two-diagonal, (*d-i*) four-corners and (*e-j*) improved four corner structures, respectively. Forbidden distances, $y_{mc}$ and $z_{mc}$ are selected to be $15\lambda$ and $10\lambda$, respectively.

Alternative approaches to define forbidden zones for minimizing the mutual coupling between the TX and RX elements, and the formation of virtual apertures are illustrated in Fig. 6 for a physical aperture of $30\lambda \times 30\lambda$ and a minimum element spacing of $\lambda/2$. Mutual coupling zones are illustrated for a rectangular of size $15\lambda \times 10\lambda$ along the azimuth and elevation. Consideration of the physical element sizes to avoid overlapping further causes some positions to be unusable. This will create aperture loss and beam width spreading which need careful consideration.

Let us define the performance metrics for virtual aperture and BW efficiency

$$\alpha_{ap} = A_{vrx}/(\beta_v A_{phy}) \qquad (35)$$

$$\alpha_{bw} = BW_{observed,hp}/BW_{theory,hp} \tag{36}$$

where $A_{phy}$ and $A_{vrx}$ are the physical and virtual aperture sizes, the virtual aperture gain, $\beta_v$, is 2 and 4 for the 1D and 2D cases, respectively, the aperture loss factor, $\alpha_{ap} \leq 1$ and the BW spreading factor, $\alpha_{bw} \leq 1$ which are calculated for the examples of Fig. 6 given in Tables III and IV.

TABLE III

Aperture Loss Factors

| $(y_{mp}, z_{mp})/\lambda$ | $(0,0)$ | $(3,2)$ | $(15,10)$ | $(20,15)$ |
|---|---|---|---|---|
| $(a, f)$ Shared aperture | 1 | 1 | 1 | 1 |
| $(b, g)$ Vertical | 0.50 | 0.47 | 0.39 | 0.17 |
| $(c, h)$ Diagonal | 0.75 | 0.68 | 0.34 | 0.26 |
| $(d, i)$ Four corners | 0.75 | 0.71 | 0.35 | 0.06 |
| $(e, j)$ Improved four corners | 0.75 | 0.71 | **0.43** | **0.16** |

TABLE IV

HPBW Spreading Factors[2]

| $(y_{mp}, z_{mp})/\lambda$ | $(0,0)$ | $(3,2)$ | $(15,10)$ | $(20,15)$ |
|---|---|---|---|---|
| $(a, f)$ Shared aperture | (1, 1) | (1, 1) | (1, 1) | (1,1) |
| $(b, g)$ Vertical | (1, 2) | (1, 2.13) | (1, 3.03) | (1, 4) |
| $(c, h)$ Diagonal | (1, 2) | (1, 2.17) | (1, 3.03) | (1, 4.76) |
| $(d, i)$ Four corners | (1, 1) | (1, 1) | (1, 1) | (1, 1) |
| $(e, j)$ Improved four corners | (1, 1) | (1, 1) | **(1, 1)** | **(1, 1)** |

Both the aperture loss and the BW spreading factors are unity for standard ULA and URAs when TX and RX elements can be co-located on a shared aperture and their physical sizes are ignored as shown in Table III and IV (*a, f*). The available physical apertures should be separated to TX and RX sub-apertures to minimize their mutual couplings. Defining sub-apertures is illustrated in (*b - j*) for two-vertical, two-diagonal, and four corner regions which all provide a pool of positions for the algorithm given in Table II. The efficiency of the better four corners case is further improved in (*e, j*).

---

[2] Aperture lengths along the horizontal and vertical axes are used to estimate the beamwidth as suggested in (27), and (28). HPBW spreading factors, $\alpha_{bw}$, are given both for azimuth and elevation, respectively, except it is given for the two diagonals for the case (*c, h*).

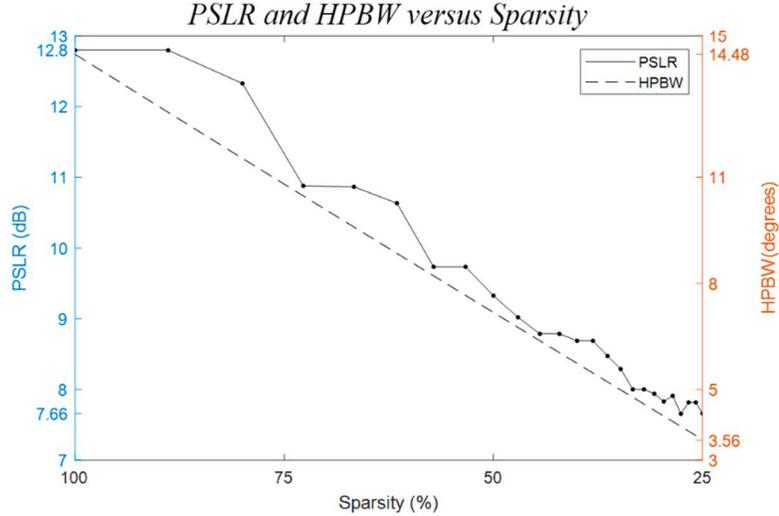

**Fig. 7.** PSLR and HPBW as a function of sparsity for a ULA with an aperture length of $32\lambda$.

### C. *Disadvantages of Sparse Arrays:*

Optimum sparse array design of Section IV.*B.* illustrated in Fig. 3 suggests uniformly distributed inter-element spacings which in effect attenuates and spreads the grating lobes into side lobes, and thus, sparse arrays can offer solutions with no grating lobes. However, this grating lobe power is spread across the uFOV increasing the side lobe levels and lowering the PSLR. This becomes unavoidable especially for wide uFOV and very small thinning ratios, as expected. In general, sparse arrays often require effective sparse array processing to reduce PSLR especially for the multi-target case. Such practical approaches are possible but kept outside the scope of this work, and will be studied in our next paper.

## V. RESULTS AND CONCLUSION

### A. *Fullly populated uniform arrays:*

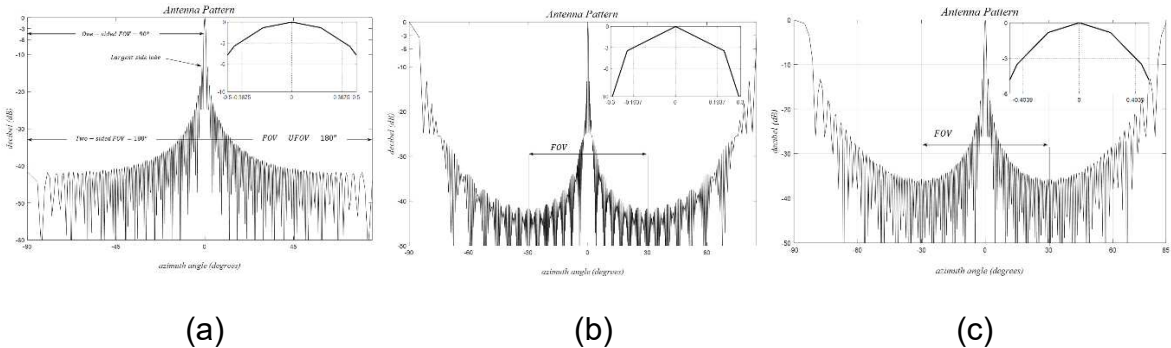

(a)  (b)  (c)

**Fig. 8**. Received signal patterns for uniform linear MIMO arrays with different parameters; (a) $N_{vrx} = 121, d = \lambda/2$, (b) $N_{vrx} = 61, d = \lambda/2$, (c) $N_{vrx} = 61, d = \lambda$.

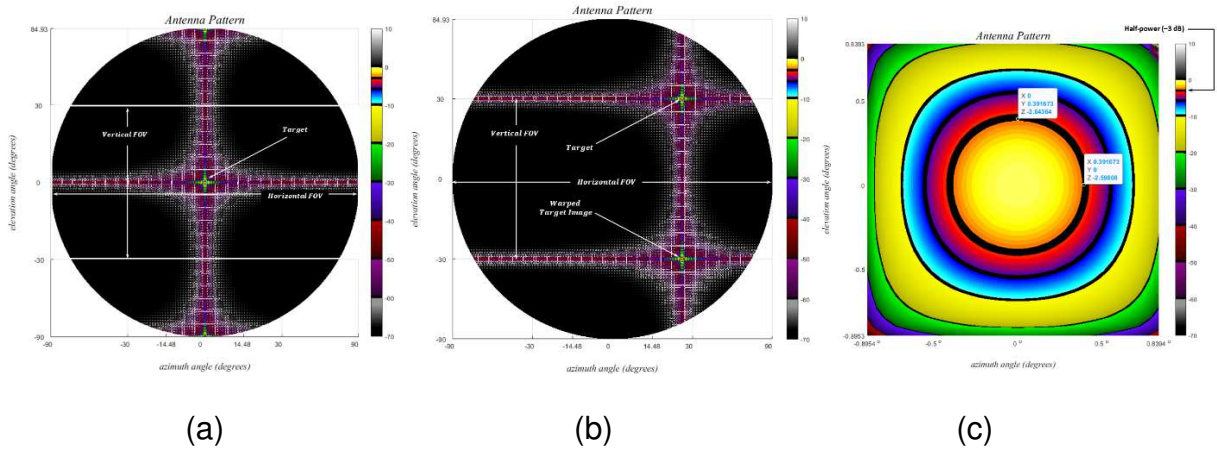

(a)　　　　　　　　　　(b)　　　　　　　　　　(c)

**Fig. 9**. Received signal patterns for a uniform rectangular MIMO array; $(M_{vrx}, N_{vrx}) = (121, 61)$, $d = \lambda/2$, (a) $(\phi_t, \theta_t) = (0,0)$, (b) $(\phi_t, \theta_t) = (0, 30^0)$, (c) close up view of the beam width region.

Grating lobes, uFOV and BW are examined in Fig. 8. Received signal patterns show the number of VRXs, inter-element spacings, aperture lengths, FOV, and BW to be (a) 121, $\lambda/2$, 60$\lambda$, 180º, 0.3875, (b) 121, $\lambda$, 120$\lambda$, 90º, 0.1937, and (c) 61, $\lambda$, 60$\lambda$, 90º, 0.4039, respectively. BW and $L_\lambda$ are observed to be inversely proportional also suggested by (27) and (28), and for $d_\lambda \geq 1/2$, $d_\lambda$ Grating lobes, uFOV angles are also inversely proportional suggested by (29) and (30). Note that the PSLR is approximately constant for cases. It is obvious that to improve FOV and BW simultaneously the number of elements must be increased where degrees of fredom is two. The uFOV shifts with the target angle without any distortion in the $(u, v)$ plane as shown in Fig. 9 in agreement with Fig. 2 (b) and (d).

## B.  Uniform Sparse Arrays with Large Antenna Elements:

Here are some numerical results to illustrate creating USAs with no grating lobes which are optimized with or without mutual coupling forbidden zones. For the optimization of Version A's 12 RXs Version B's 8 RX positions are enforced as the initial condition. Physical aperture size is limited to $35\lambda \times 35\lambda$ and elements sizes are $2\lambda \times 5\lambda$ both for Ankara and Coruh arrays and their versions A and B.

Ankara arrays are the two examples of USA with no forbidden zones. Version A and B have 16 and 8 RXs, respectively. Version A is enforced to use B's 8 RXs and 12 TXs, and its remaining 8 RXs are optimized separately.

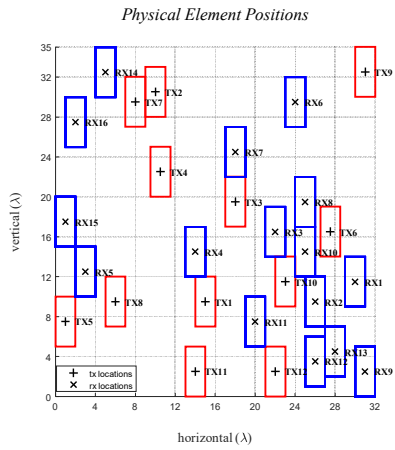
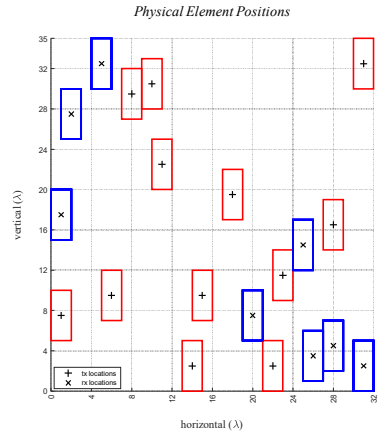

(a)            (b)

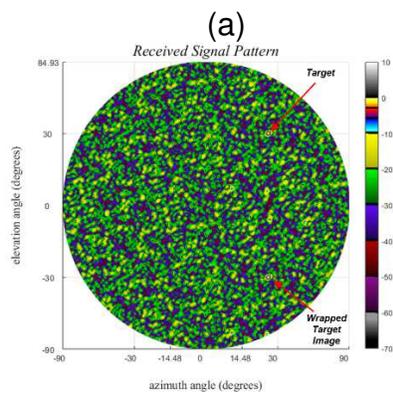
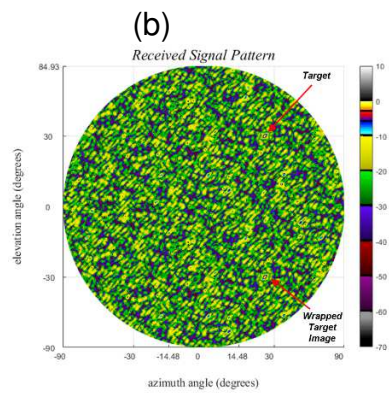

(c)            (d)

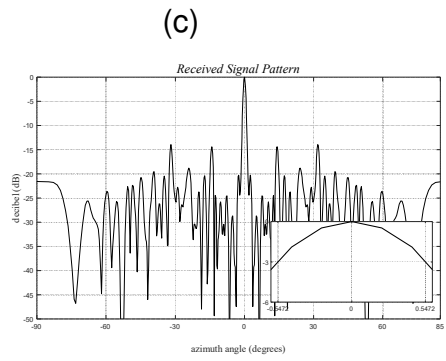
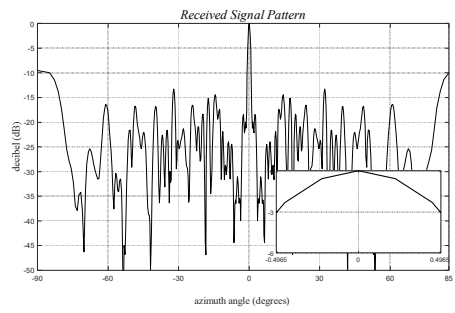

(e)            (f)

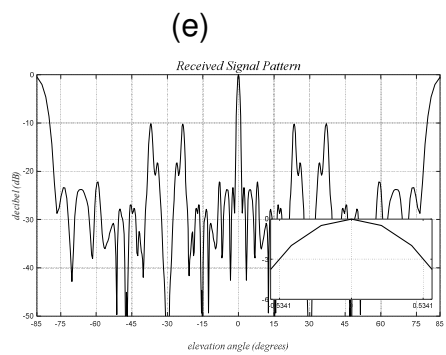
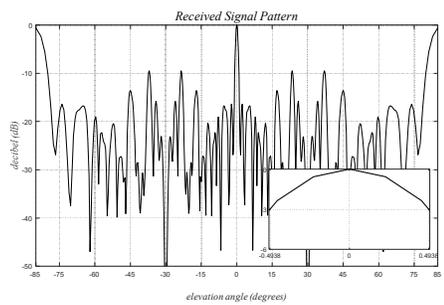

(g)            (h)

**Fig. 10.** Uniform sparse arrays, $d_x = \lambda/2$, $d_y = \lambda$, $w_{rx} = w_{rx} = 2\lambda$, $h_{rx} = h_{rx} = 5\lambda$, no forbidden zones are defined. (*left*) Ankara A, (*right*) Ankara B, (*a*, and *b*) Physical element positions, (*c – h*) 2D and 1D received signal patterns in the $(u, v)$ planes with tick values converted to degrees. Single target angles $(\phi_t, \theta_t)$ are $(30^0, 30^0)$, and $(0^0, 0^0)$ for 2D and 1D plots, respectively.

Ankara USAs are designed on a reference grid of size $\lambda/2 \times \lambda$ using physical element sizes $2\lambda \times 5\lambda$ and are grating lobe-free only along the horizontal. The PSLR values are 11.23 dB, 8.64 dB, and one-sided azimuth and elevation HPBW values are $[0.6^0, 0.53^0]$, $[0.5^0, 0.5^0]$ for A, B arrays, respectively.

TABLE V

Calculated Mutual Coupling Between TX and RX elements

|      | RX-1 | RX-2 | RX-3 | RX-4 | RX-5 | RX-6 | RX-7 | RX-8 | RX-9 | RX-10 | RX-11 | RX-12 | RX-13 | RX-14 | RX-15 | RX-16 |
|------|------|------|------|------|------|------|------|------|------|-------|-------|-------|-------|-------|-------|-------|
| TX-1 | -88.651 | -85.267 | -54.271 | -47.927 | -54.705 | -49.511 | -41.019 | -70.726 | -77.11 | -86.105 | -74.747 | -52.62 | -56.303 | **-39.994** | -66.794 | -71.652 |
| TX-2 | -41.472 | -50.072 | -67.71 | -60.685 | -63.788 | -70.807 | -97.675 | -91.437 | -50.987 | -56.544 | -71.05 | -91.372 | -92.104 | -71.523 | -81.446 | -80.96 |
| TX-3 | -60.438 | -63.204 | **-36.883** | -42.819 | -53.835 | -59.261 | -84.826 | -93.292 | -78.314 | -67.282 | -63.33 | -43.031 | -66.323 | -52.488 | -65.601 | -64.677 |
| TX-4 | -58.255 | -57.184 | -57.22 | -84.873 | -104.18 | -85.229 | -67.048 | -72.424 | -67.98 | -42.327 | -54.027 | -66.866 | -59.862 | -61.781 | -68.997 | -71.877 |
| TX-5 | -101.1 | -79.955 | -70.271 | -71.816 | -76.042 | -66.099 | -55.319 | -81.208 | -101.59 | -72.681 | -71.798 | -54.888 | -54.139 | -59.64 | -79.994 | -88.144 |
| TX-6 | -78.928 | -77.229 | -40.512 | **-35.269** | **-34.785** | -46.212 | -53.284 | -80.857 | -90.673 | -53.201 | -70.32 | -49.973 | -66.873 | -70.457 | -73.024 | -81.748 |
| TX-7 | **-39.739** | -52.174 | -89.134 | -60.637 | -64.793 | -72.025 | -81.717 | -84.477 | -41.083 | -79.13 | -75.541 | -77.902 | -91.573 | -68.989 | -77.016 | -78.784 |
| TX-8 | -94.831 | -73.671 | -64.245 | -67.349 | -66.73 | -58.885 | -49.945 | -76.394 | -96.977 | -66.385 | -50.997 | -46.043 | -40.223 | -50.887 | -76.765 | -91.192 |
| TX-9 | -85.991 | -59.709 | -81.098 | -85.7 | -94.738 | -99.657 | -101.52 | -116.23 | -77.478 | -55.635 | -80.649 | -72.532 | -91.383 | -101.39 | -109.27 | -110.92 |
| TX-10 | -70.158 | -87.249 | -63.339 | -53.128 | -42.325 | **-36.389** | -42.047 | -68.371 | -81.219 | -82.842 | -77.985 | -54.899 | -55.164 | **-36.728** | -47.274 | -49.975 |
| TX-11 | -101.93 | -101.93 | -71.013 | -71.62 | -56.768 | -57.715 | -43.408 | -43.541 | -99.866 | -92.448 | -61.98 | -84.672 | -57.634 | -52.896 | -51.475 | -48.518 |
| TX-12 | -93.968 | -105.71 | -80.756 | -72.096 | -85.817 | -52.013 | -47.719 | **-31.543** | -85.286 | -80.022 | -75.091 | -55.215 | -83.873 | -60.007 | -42.539 | **-39.683** |

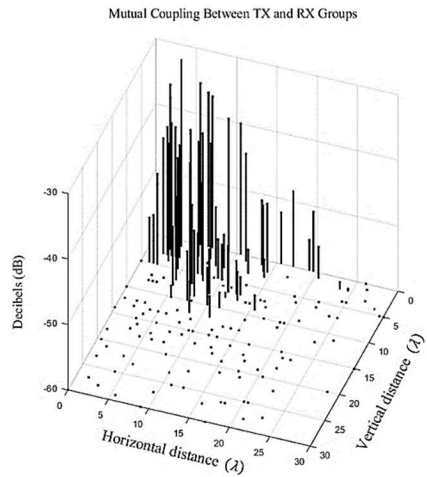

**Fig. 11.** Measured mutual coupling ratios between the TX and the RX elements for Ankara sparse array as a function of the distance between them.

   *i.*   *Simulated Mutual Coupling for Ankara A array:*

The mutual coupling between TXs and RXs within the same type of elements can mostly be compansated based on laboratory measurements, and accurate calibration of signals are possible. However, coupling among those two groups strongly depends on the environment and the target angles. The simulated results for Ankara array shown in Table V and Fig. 11 show that horizontal and vertical distances of 15 $\lambda$ and 10 $\lambda$ can be defined as the separation distances to minimize the mutual coupling.

### i. *USA with forbidden zones: Coruh A & B arrays*

In this section the improved four-corners approach proposed in Fig. 11 (*e* and *j*) is illustrated in the design of Coruh arrays. Forbidden distances, $y_{mc} = 10\lambda$, and $z_{mc} = 15\lambda$ are enforced to minimize the coupling effects. Version A and B elements, and physical element sizes are identical to Ankara USAs. They provide better array aperture and BW spreading efficiency as defined in Section (*IV.B.v.*)

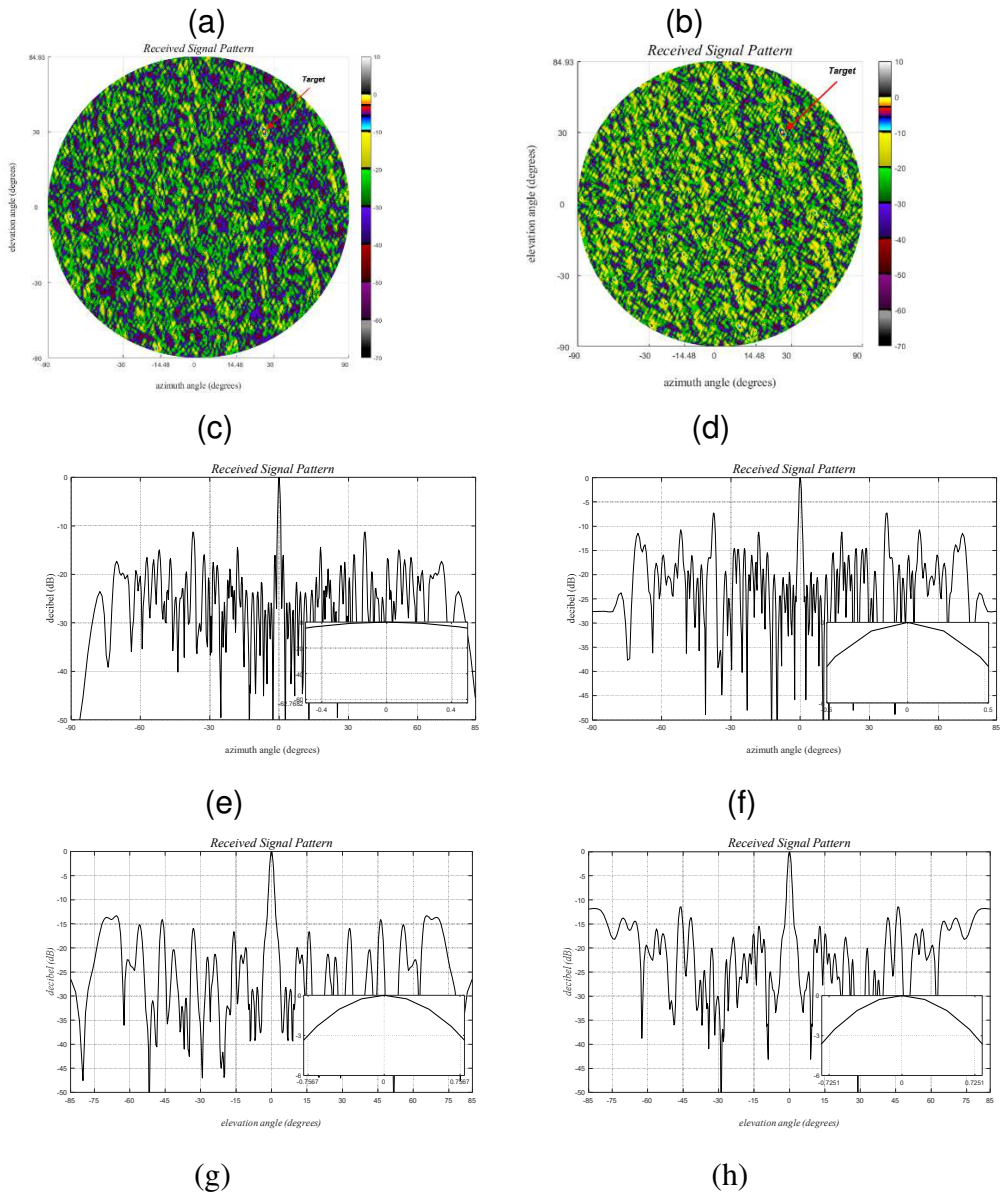

**Fig. 12.** Mutual coupling constrained uniform sparse arrays, (*left*) Coruh A, (*right*) Coruh B, $d = \lambda/2$, $w_{rx} = w_{rx} = 2\lambda$, $h_{rx} = h_{rx} = 5\lambda$, the forbidden distances, $y_{mc} = 10\lambda$, and $z_{mc} = 15\lambda$. (*a*,

and *b*) Physical element positions for a candidate array, (*c – h*) optimized 2D and 1D received signal patterns in the $(u, v)$ planes with tick values converted to degrees. Single target angles are $(\phi_t, \theta_t) = (0, 30^0)$.

Coruh USAs are designed on a reference grid of size $\lambda/2 \times \lambda/2$ using physical element sizes $2\lambda \times 5\lambda$ and are grating lobe-free both along the horizontal and vertical as shown for a typical candidate array in Fig. 12 (*a – d*). The optimized PSLR values are 11.10 dB, 9.14 dB, and one-sided azimuth and elevation HPBW values are $[0.4^0, 0.76^0]$, $[0.5^0, 0.73^0]$ for A, B arrays, respectively.

TABLE VII

Performance Parameter Values for Example USAs

|  | Ankara | | Coruh | |
|---|---|---|---|---|
|  | A | B | A | B |
| PSLR (dB) | 11.23 | 8.64 | 11.10 | 9.14 |
| BW-azimuth (deg) | 0.55 | 0.53 | 0.4 | 0.5 |
| BW-elevation (deg) | 0.49 | 0.5 | 0.76 | 0.73 |
| # of elements (tx, rx) | 12, 16 | 12, 8 | 12, 16 | 12, 8 |
| # of vrx's (generated, unique) | 192, 192 | 96, 96 | 192, 190 | 96, 96 |
| Thinning ratio (%) | 2.6 | 1.3 | 1.09 | 0.68 |
| uFOV (deg) | 180 | 60 | 180 | 180 |
| Reference URA size | $121 \times 61$ | | $156 \times 112$ | $130 \times 109$ |
| # reference URA elements | 7,381 | | 17,472 | 14,170 |
| Reference grid size ($\lambda$) | 0.5, 1.0 | | 0.5, 0.5 | |
| Physical aperture size ($\lambda$) | $32 \times 35$ | | $41.5 \times 34$ | |

The novel USA design approach proposed in Section IV provides with effective array solutions for low number of elements with better FOV and BW. Both Ankara and Coruh arrays allow utilization of large element sizes, and providing the received signal patterns to be grating lobe-free in any desired axis. It is shown that PSLR > 11 dB, and $BW < 0.5^o$ is possible with only for $N_{tx} = 12$, $N_{rx} = 16$, and $N_{vrx} = 192$. Other measured performance values are given in Table IV.

### i. The Empirical Distribution Functions for the inter-element spacings

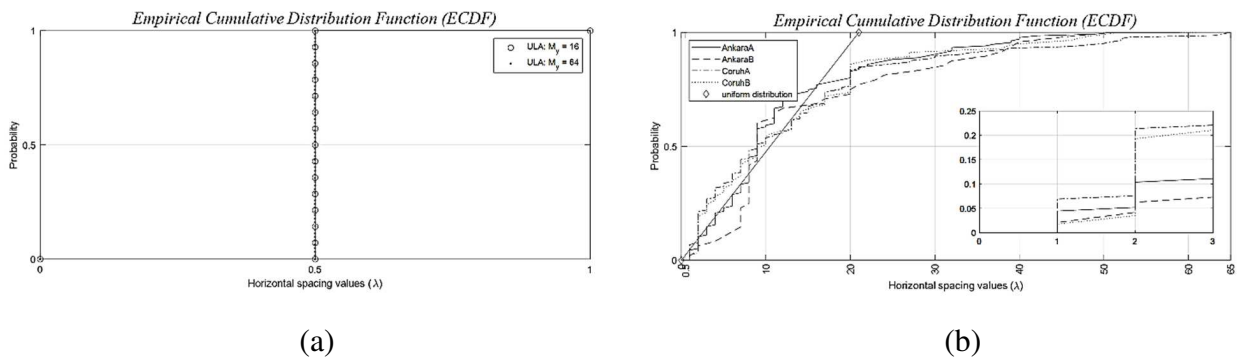

(a) (b)

**Fig. 13.** The empirical cumulative distribution functions (ECDF) for the horizontal inter-element spacings of, (*a*) ULAs with number of elements of 16 and 64, and (*b*) the USAs Ankara and Coruhwith versions A and B.

The inter-element spacings for ULAs are constant as shown in Fig. 13 (a) for $N = 16$ and 64 which need to be $\lambda/2$ or smaller in order to provide a grating lobe-free operation. However, when the physical size of elements are larger than this value the received signal pattern for ULAs will be the sum of all image replicas for each real-valued grating lobe angles at $\sin^{-1}(n/2w)$ shifted along the azimuth. Similarly for URAs, additional replica images will be added for each real-valued grating lobes at $\sin^{-1}(n/2h)$ shifted along the elevation as discussed in (29). Further, avoiding large side lobes is only possible if the sparse array has irregular spacing values. This happens when the ECDF is a smooth function as shown in Fig. 13 (b).

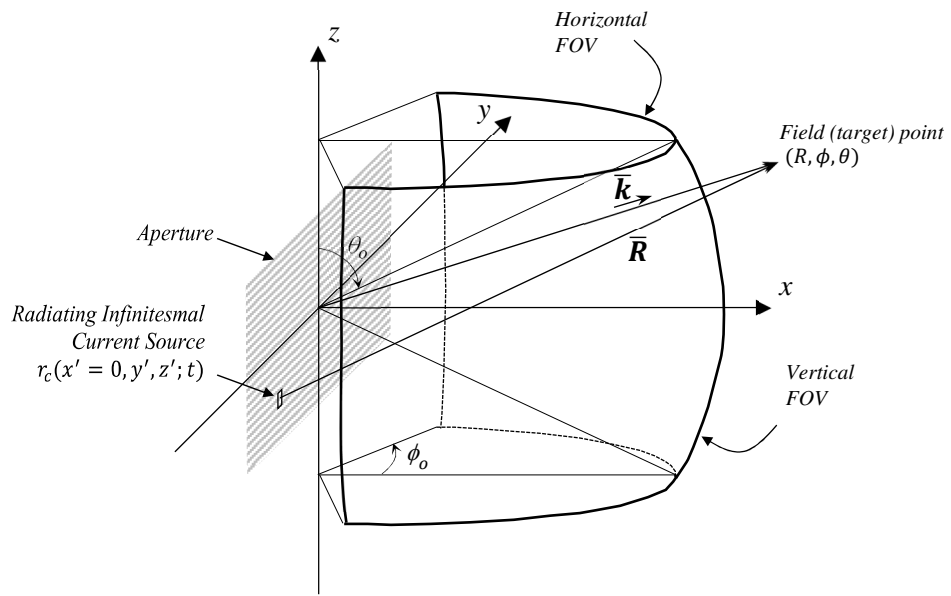

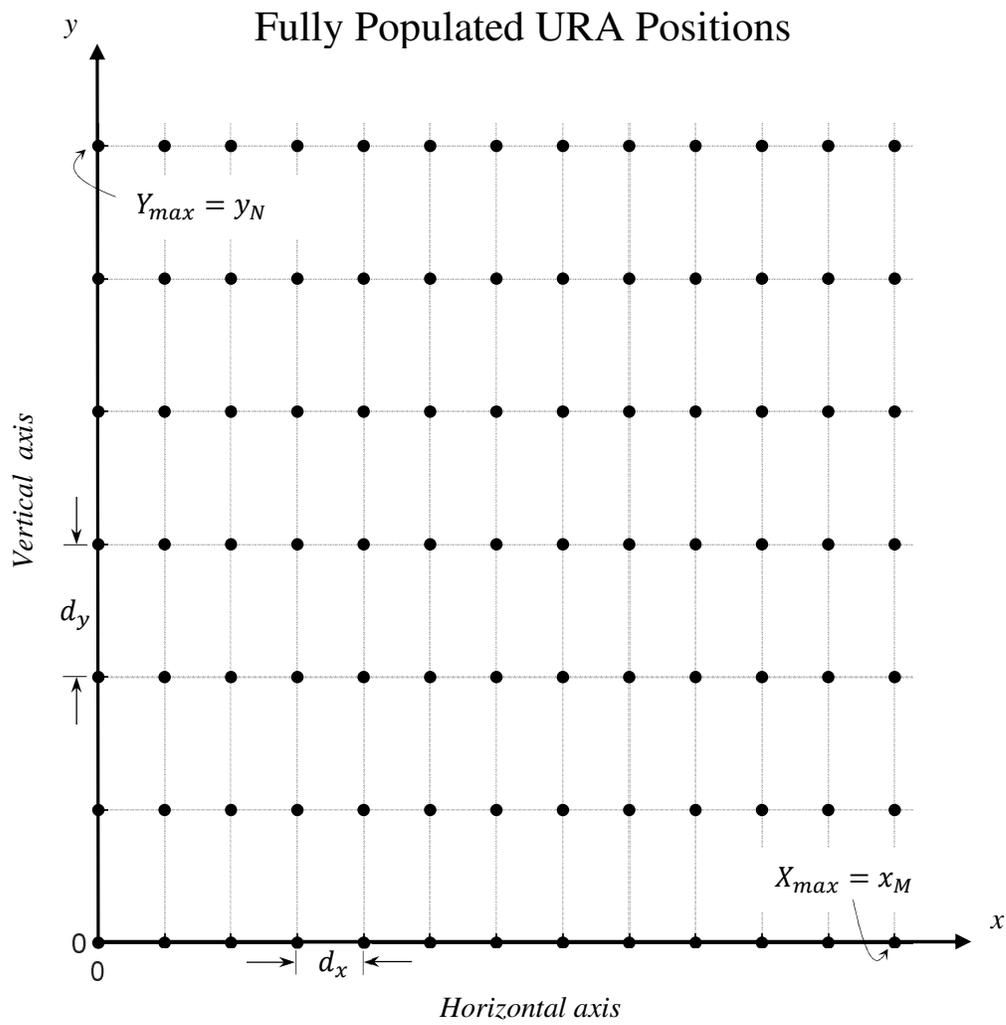

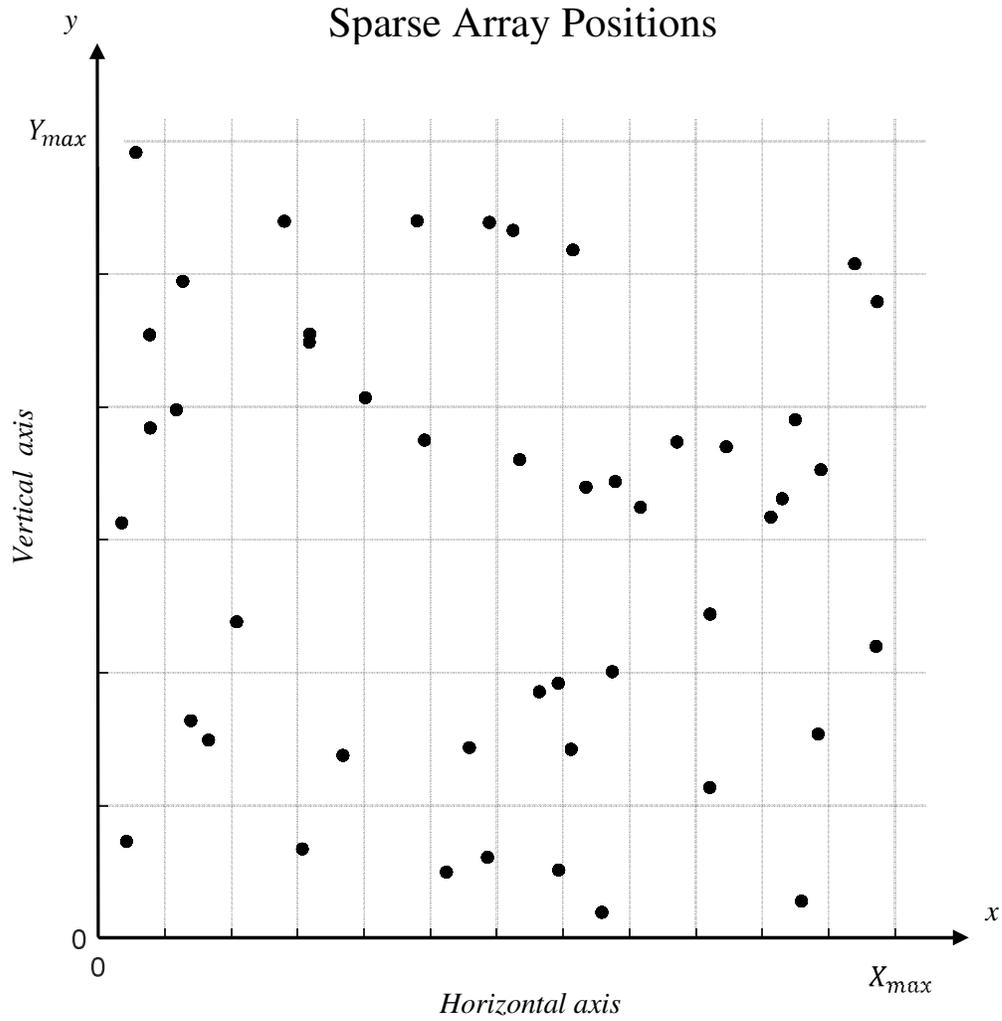

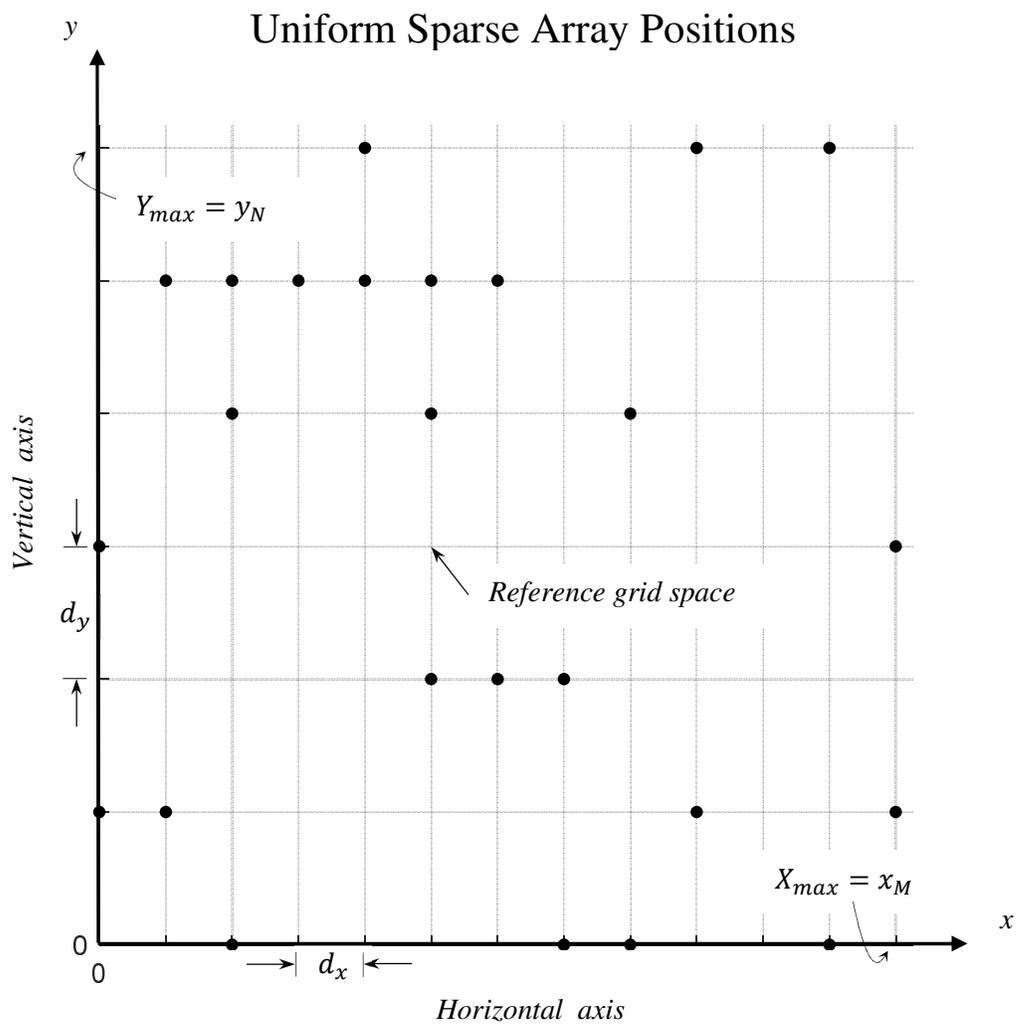

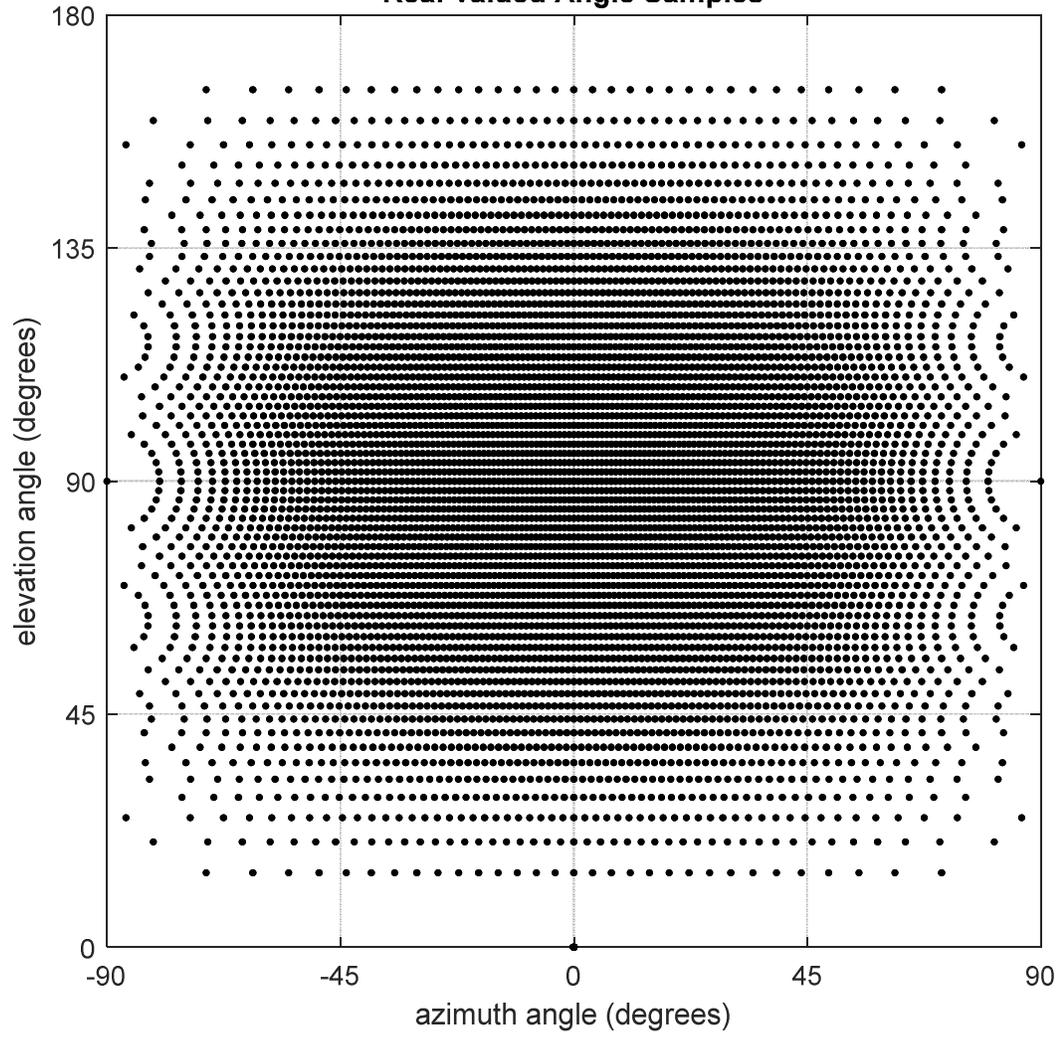

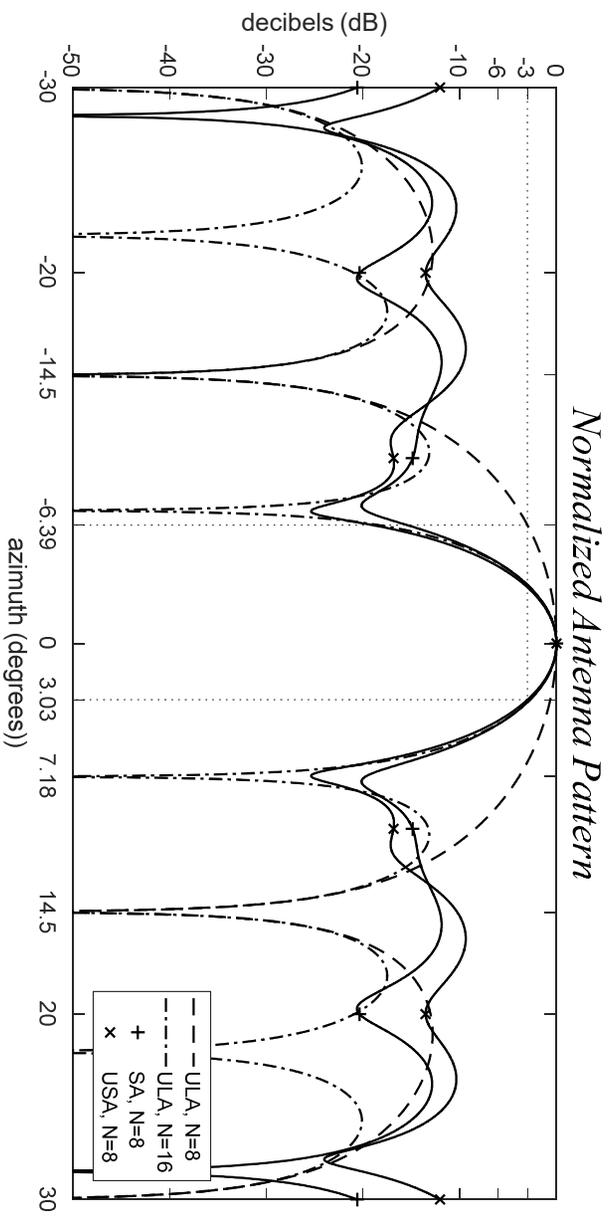

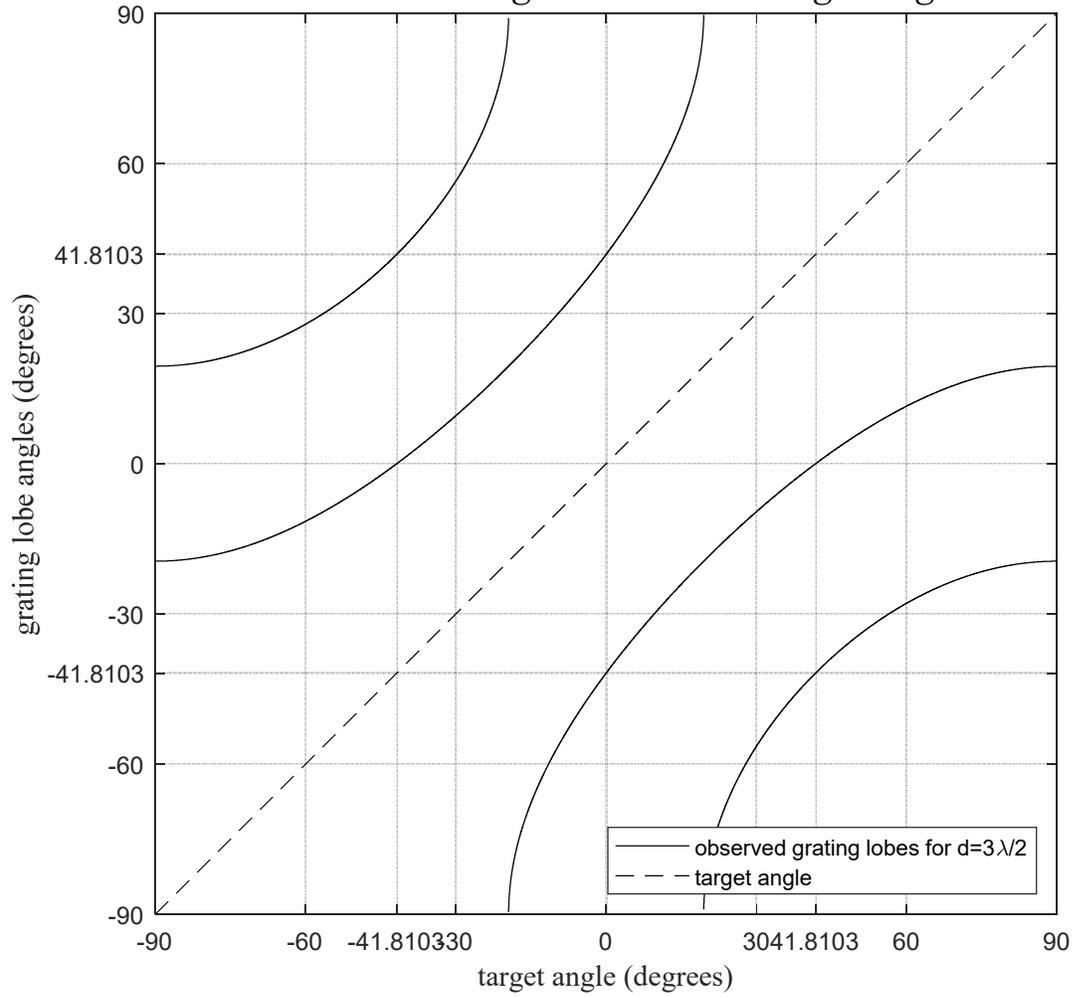

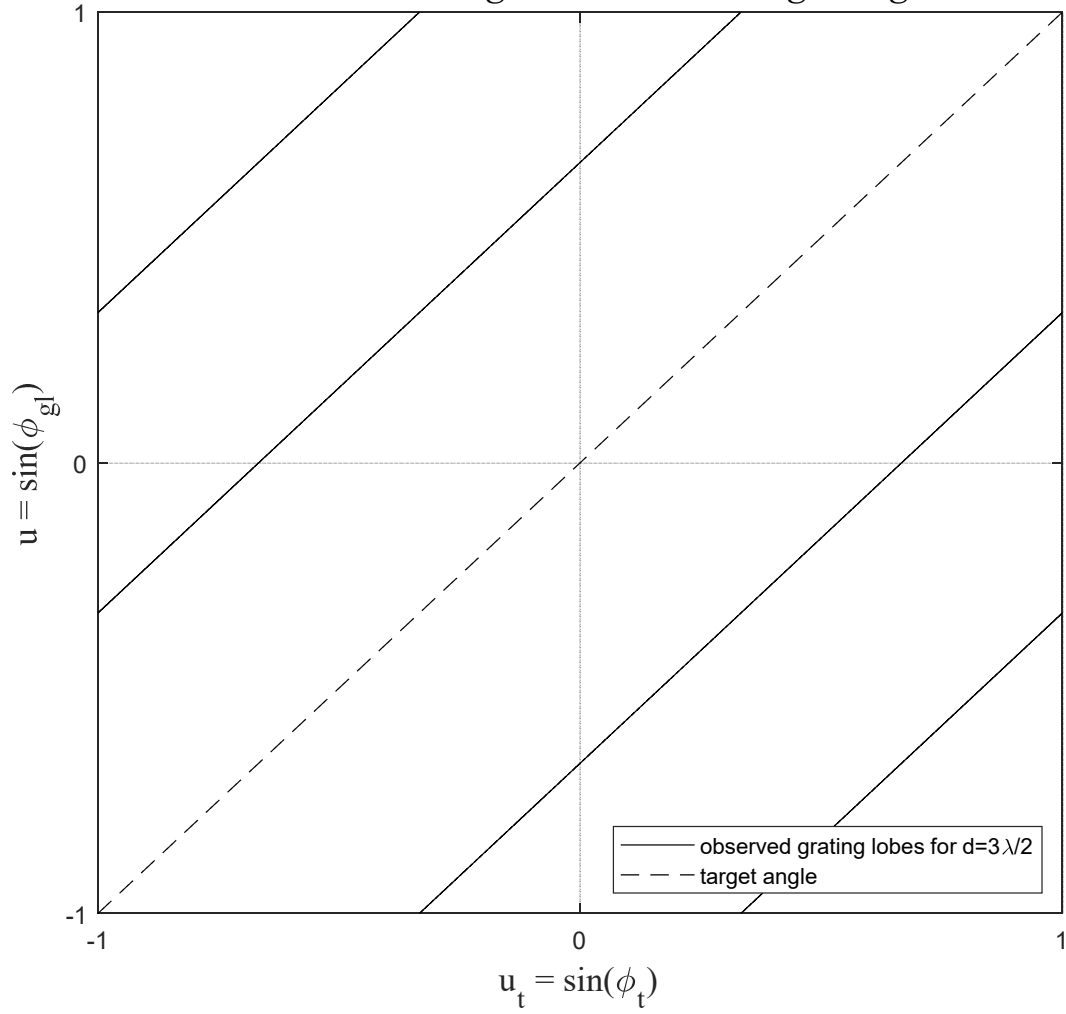

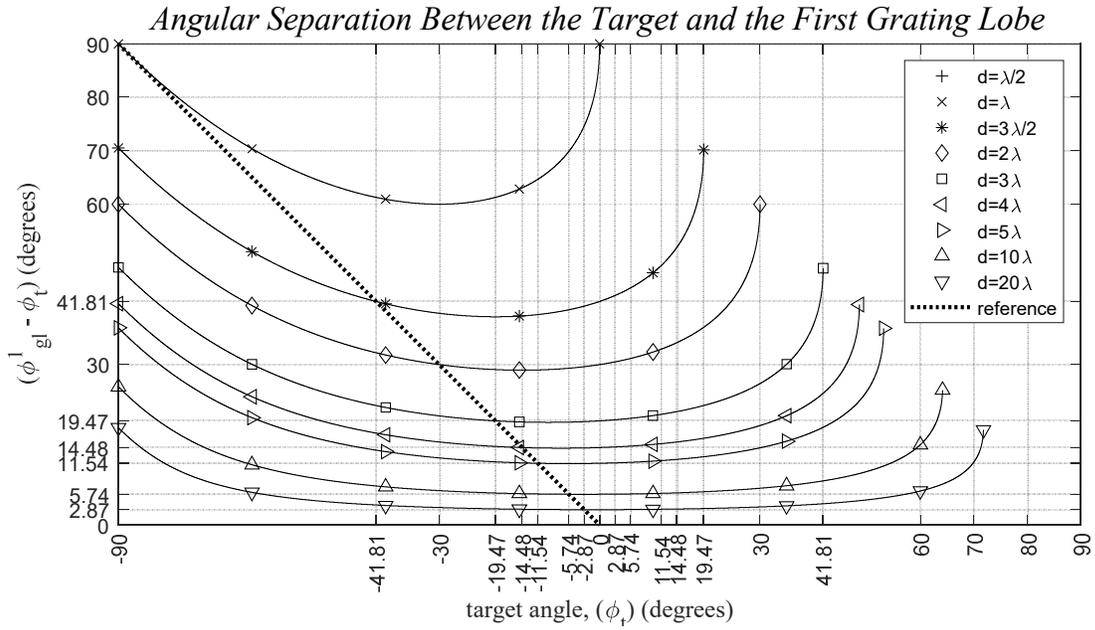

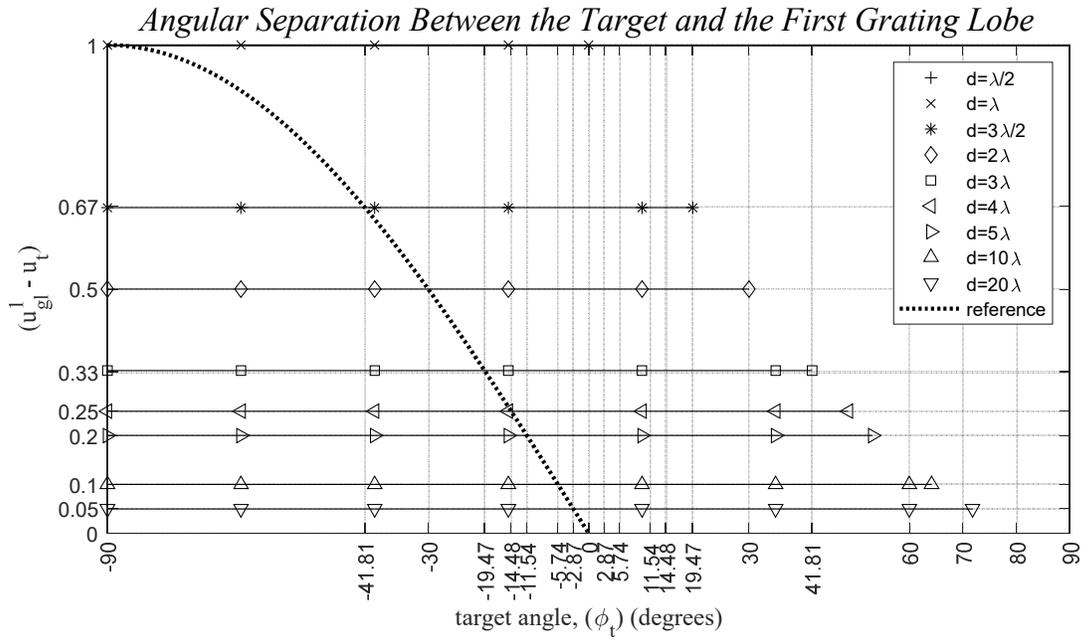

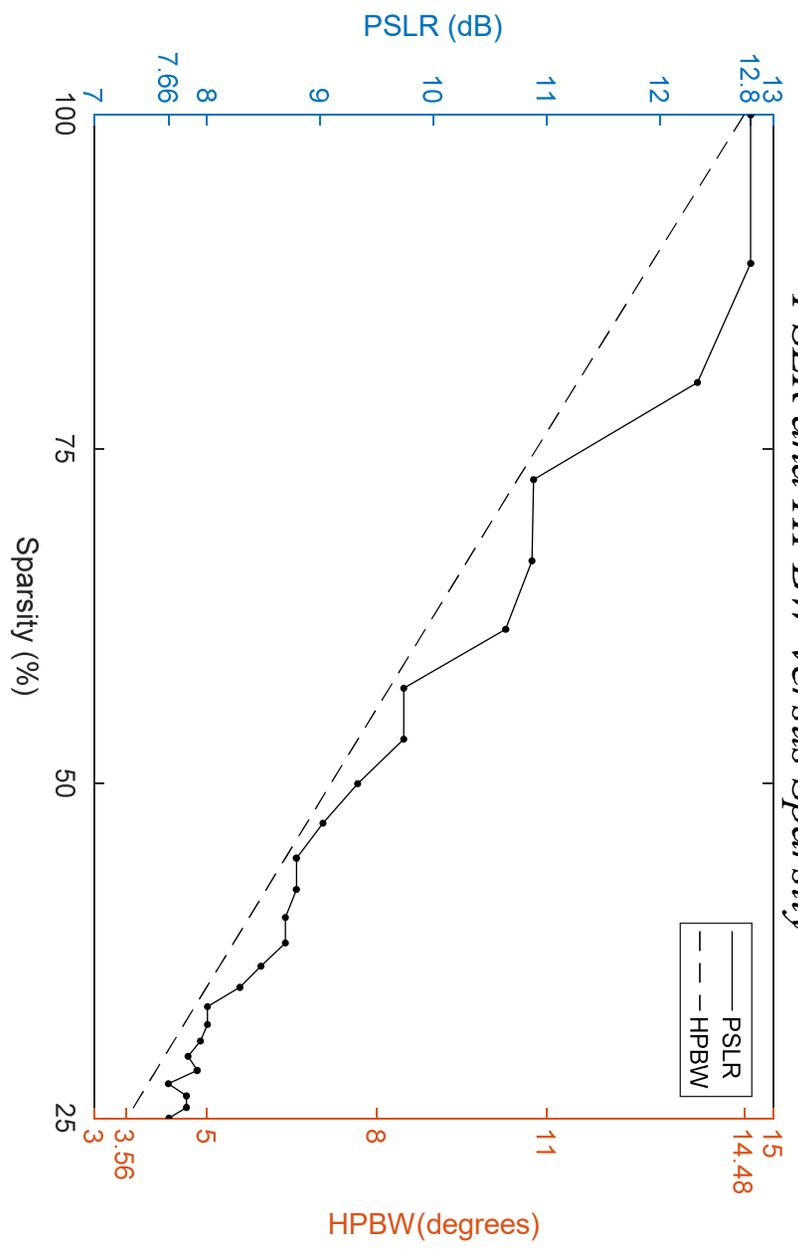

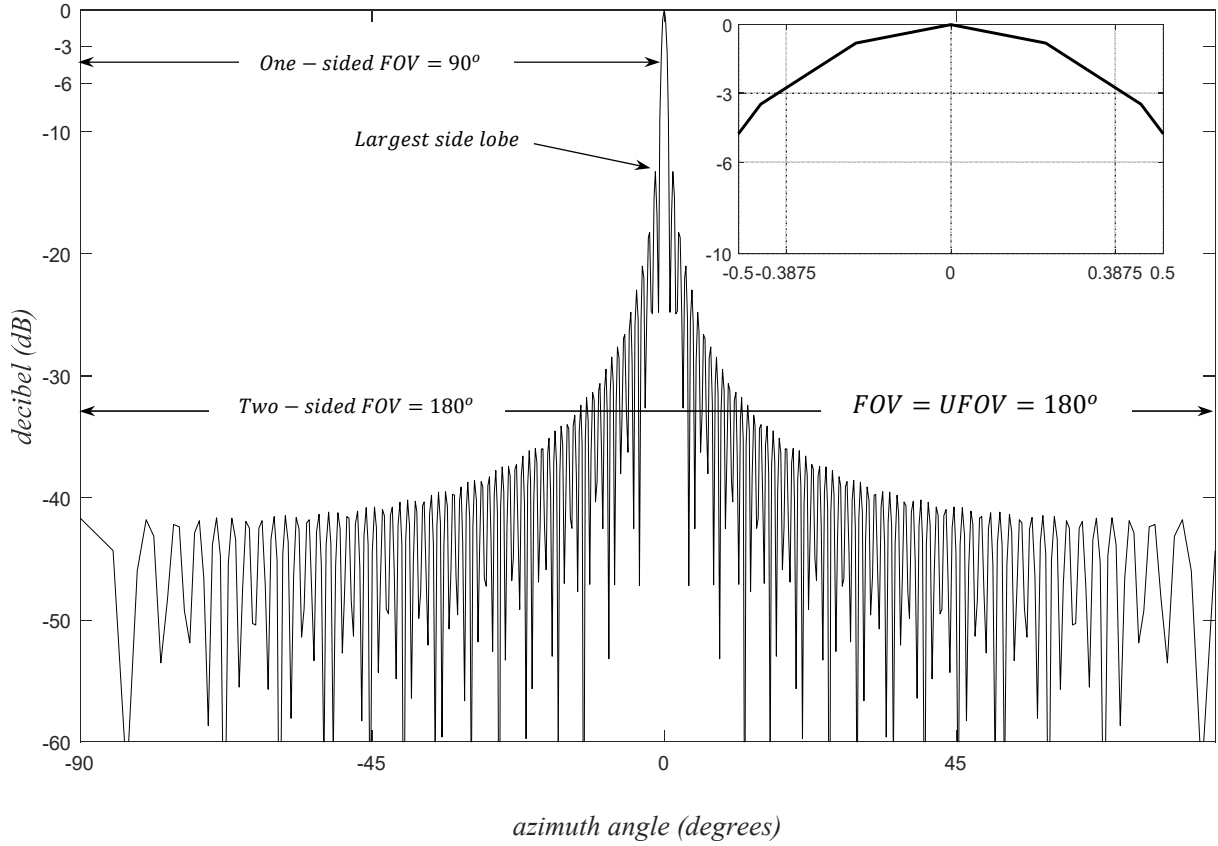

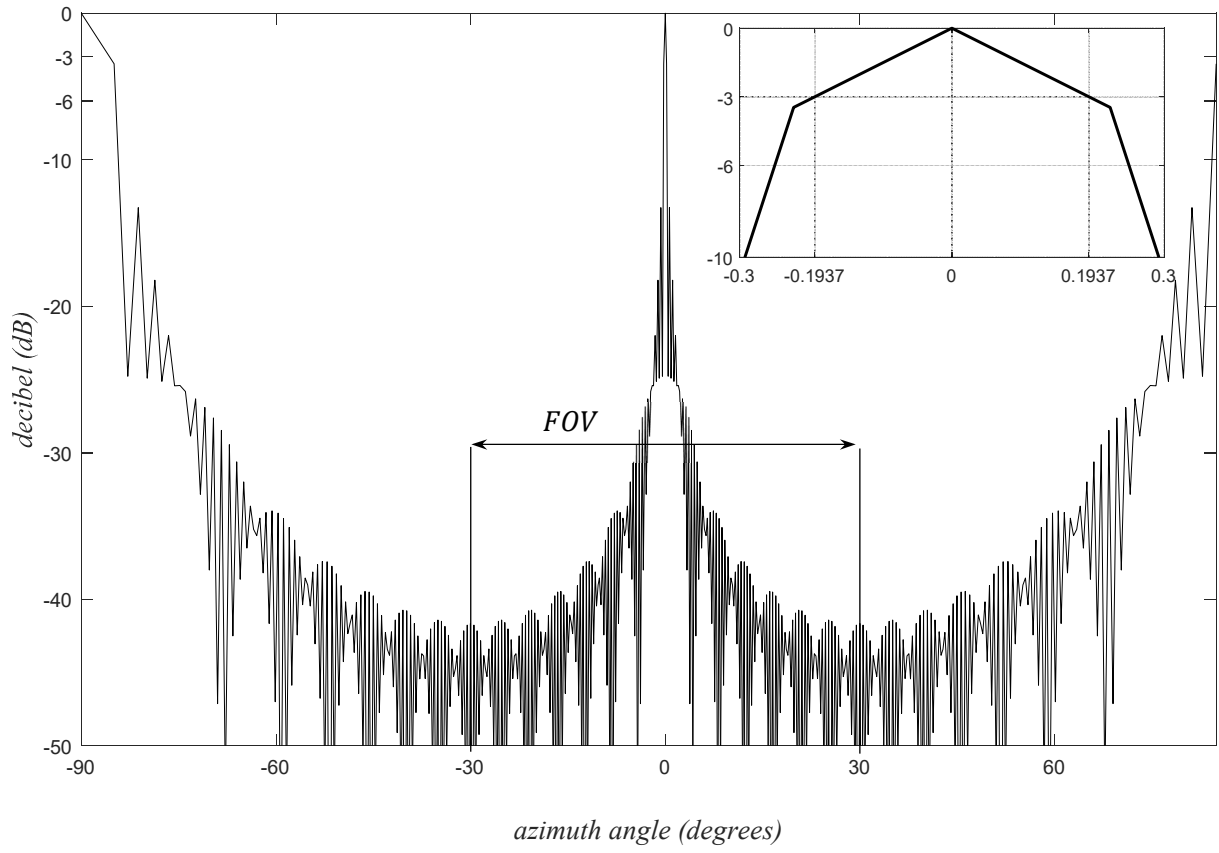

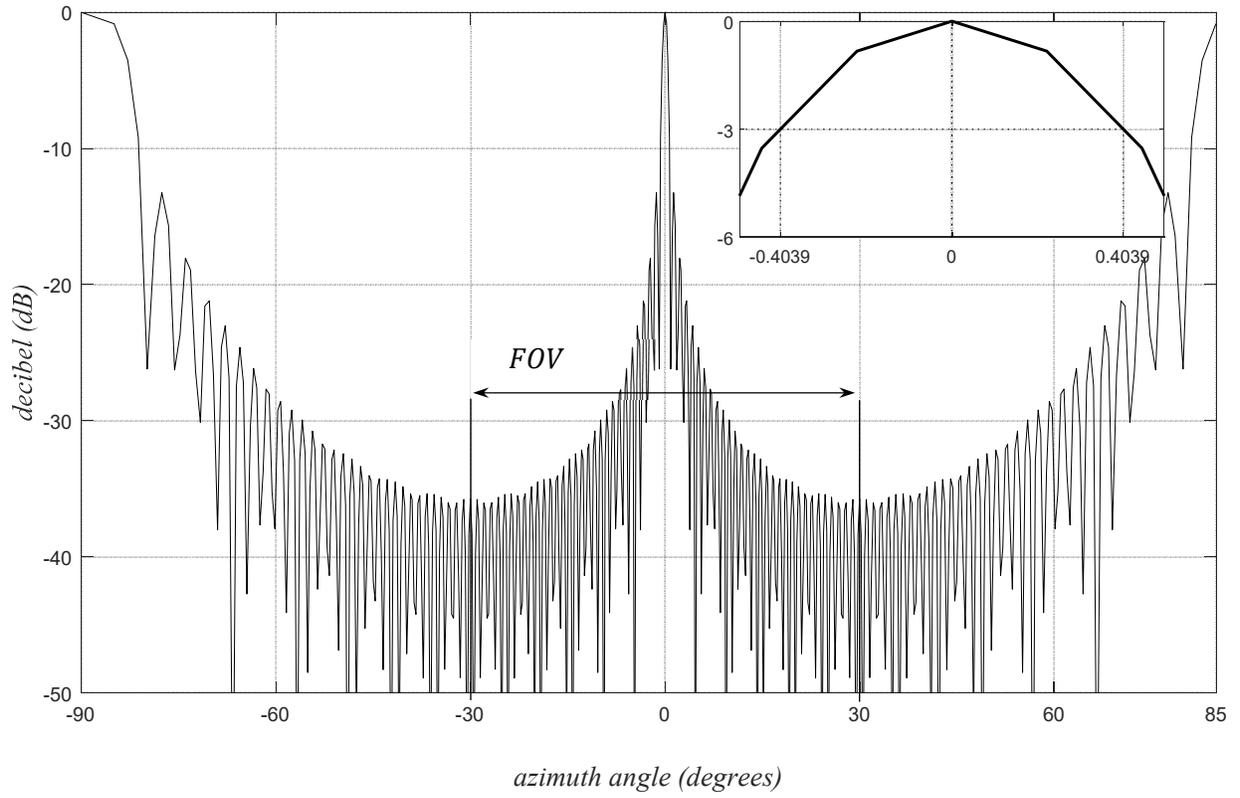

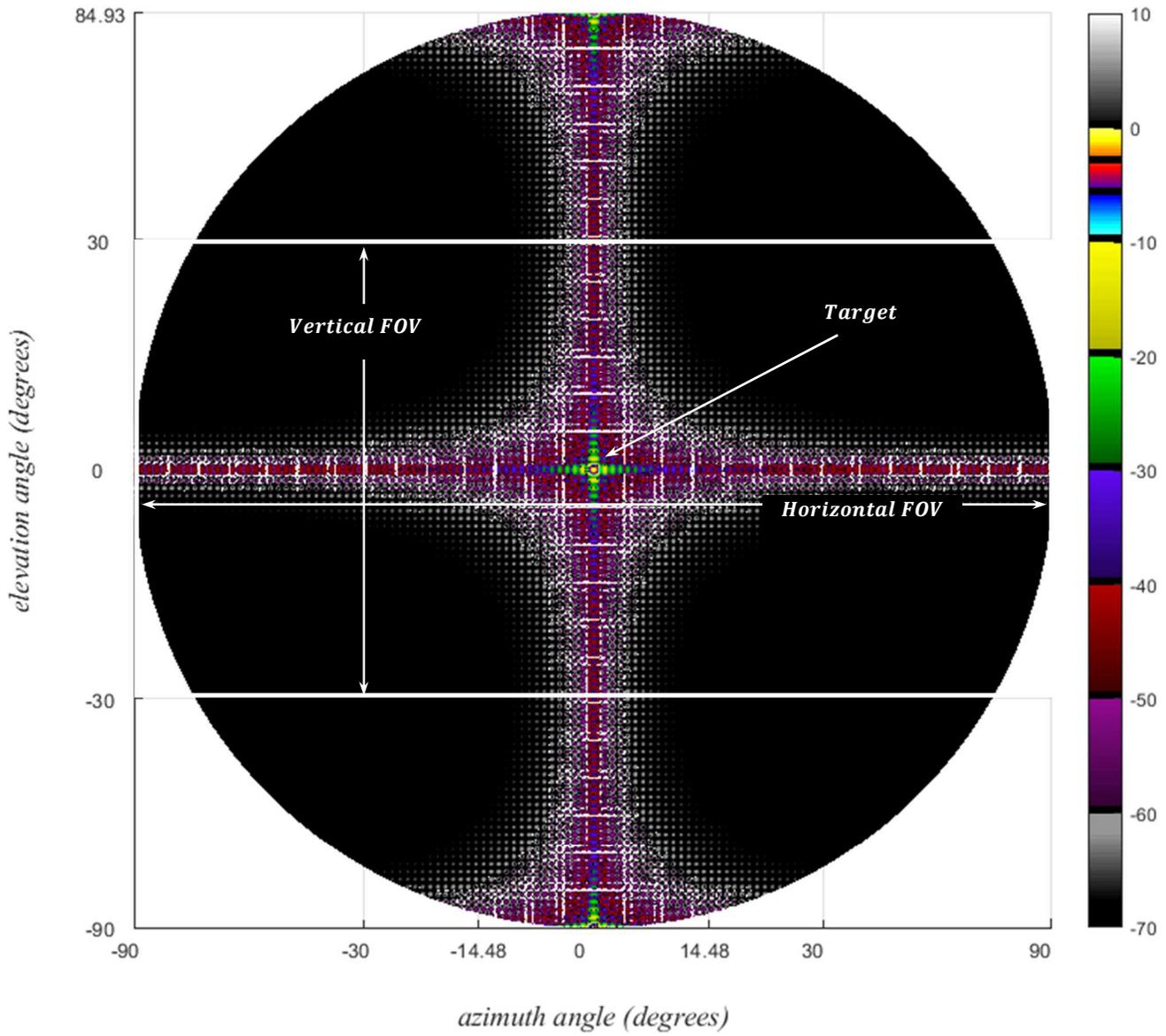

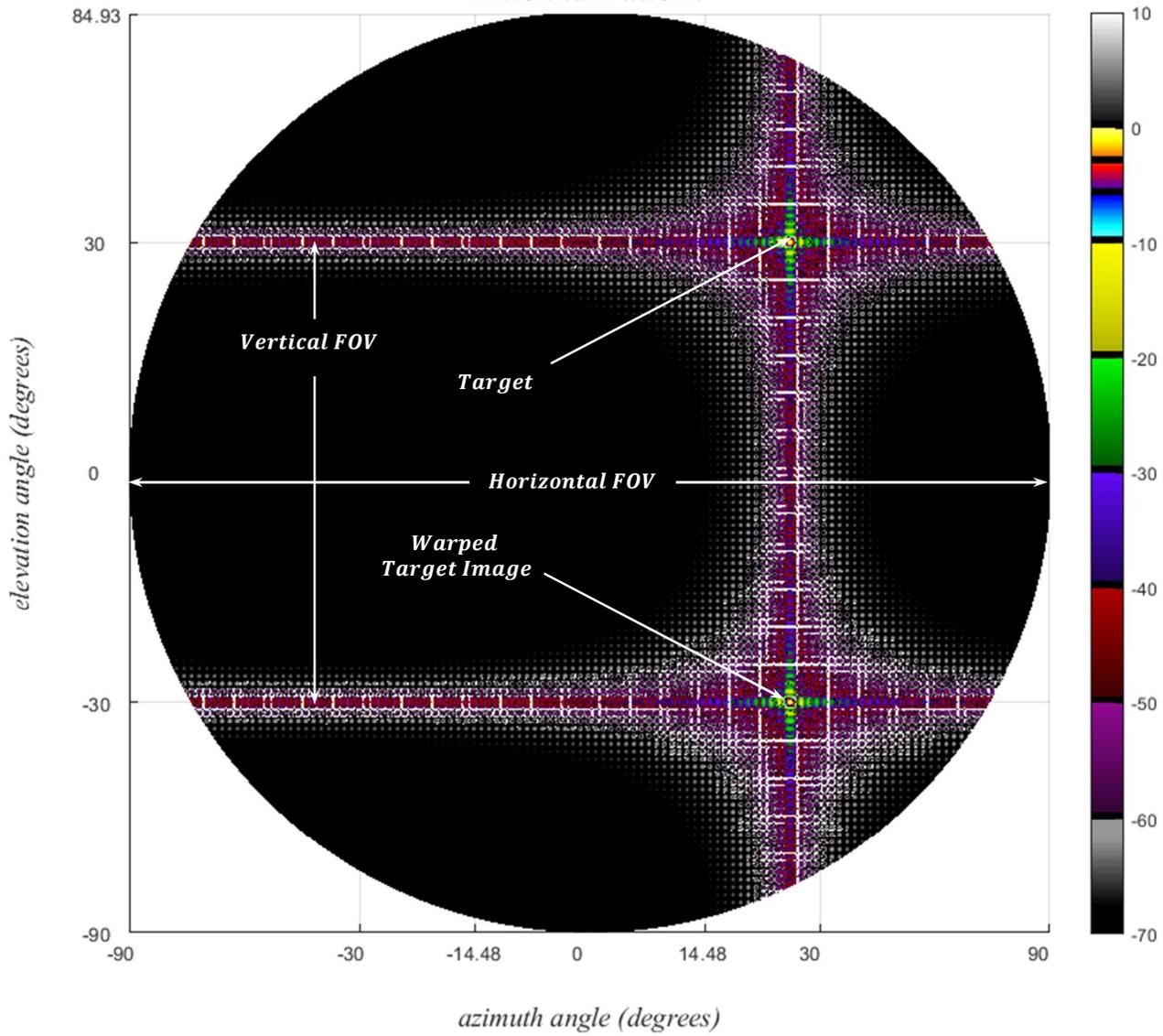

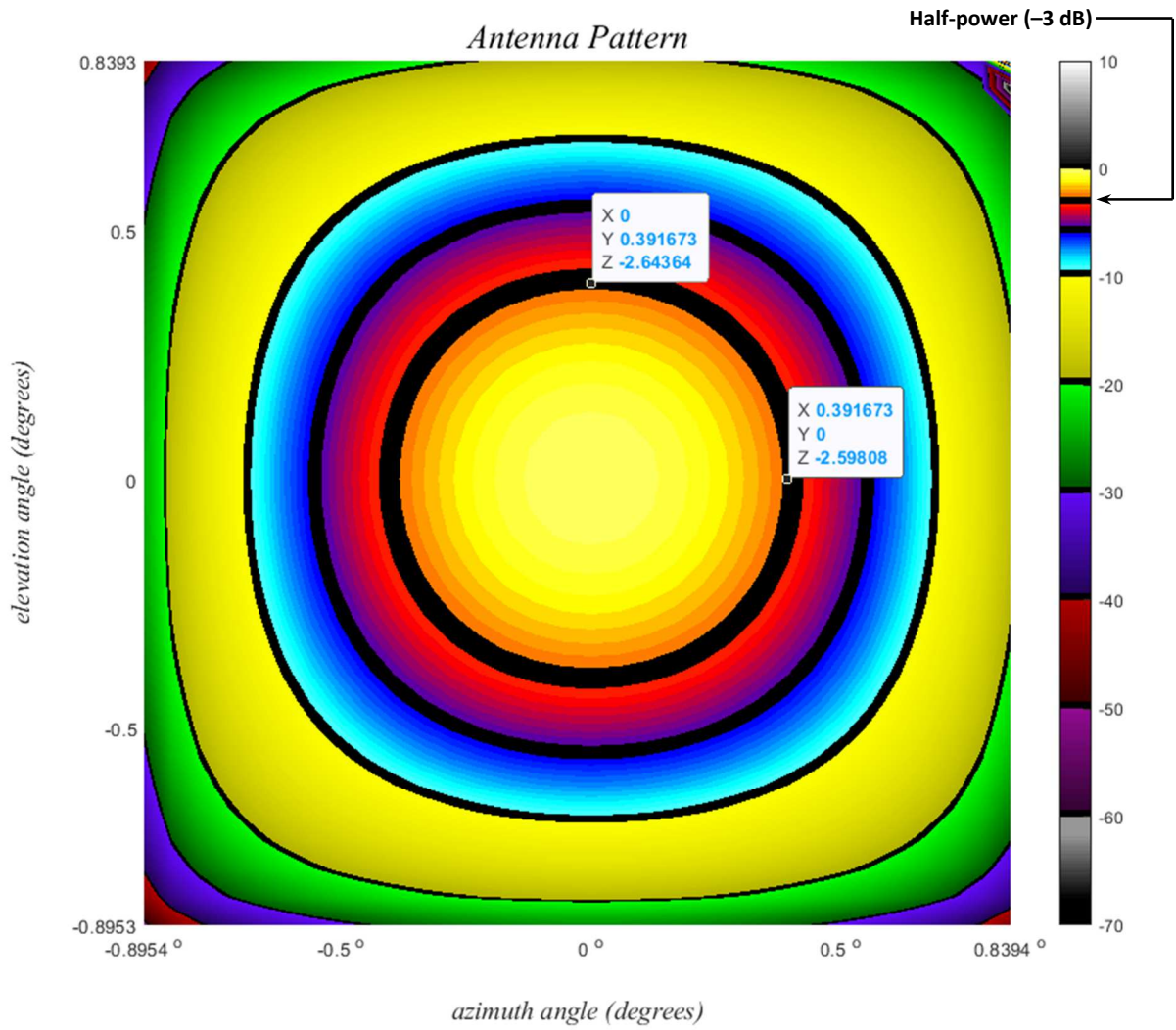

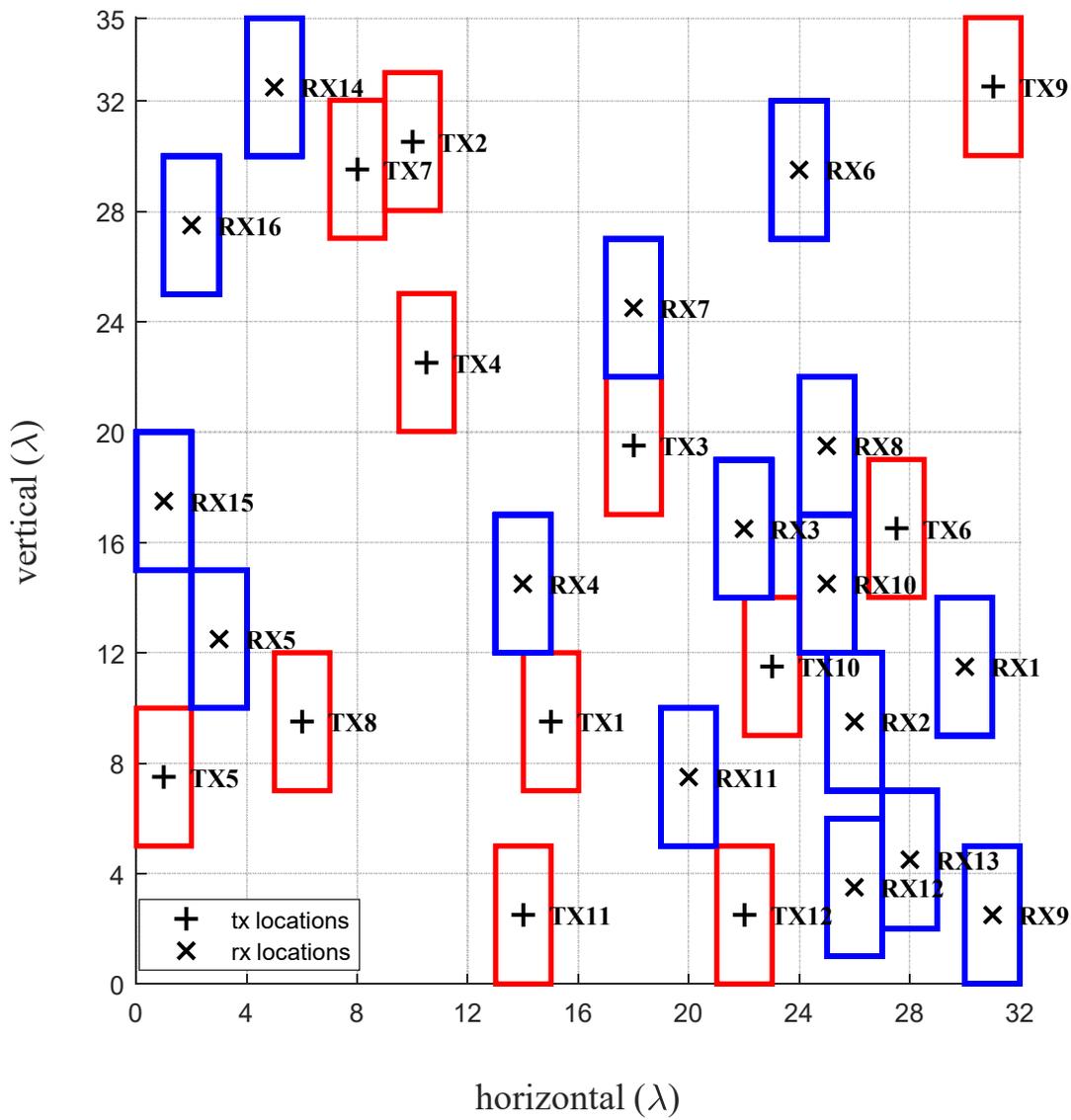

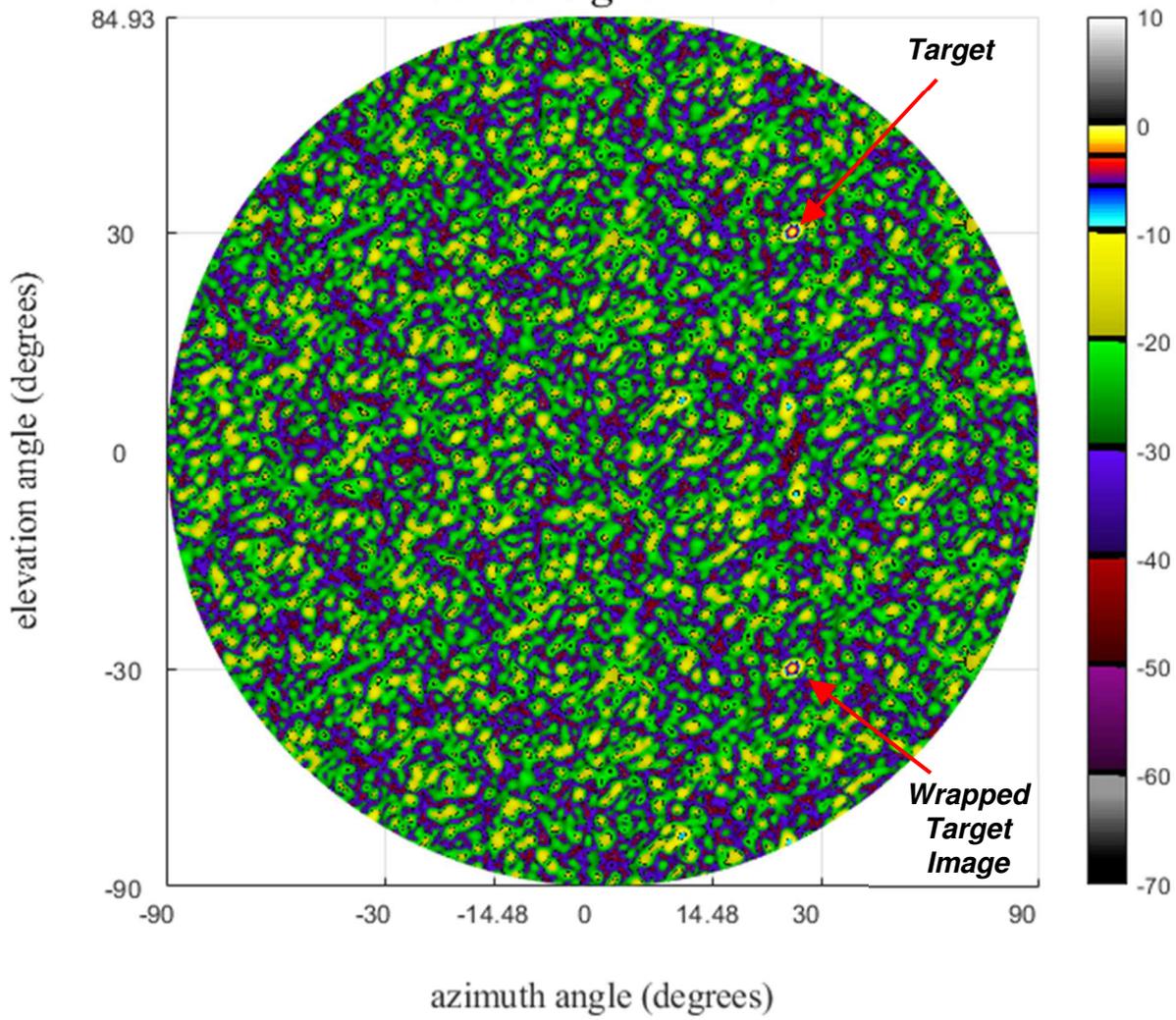

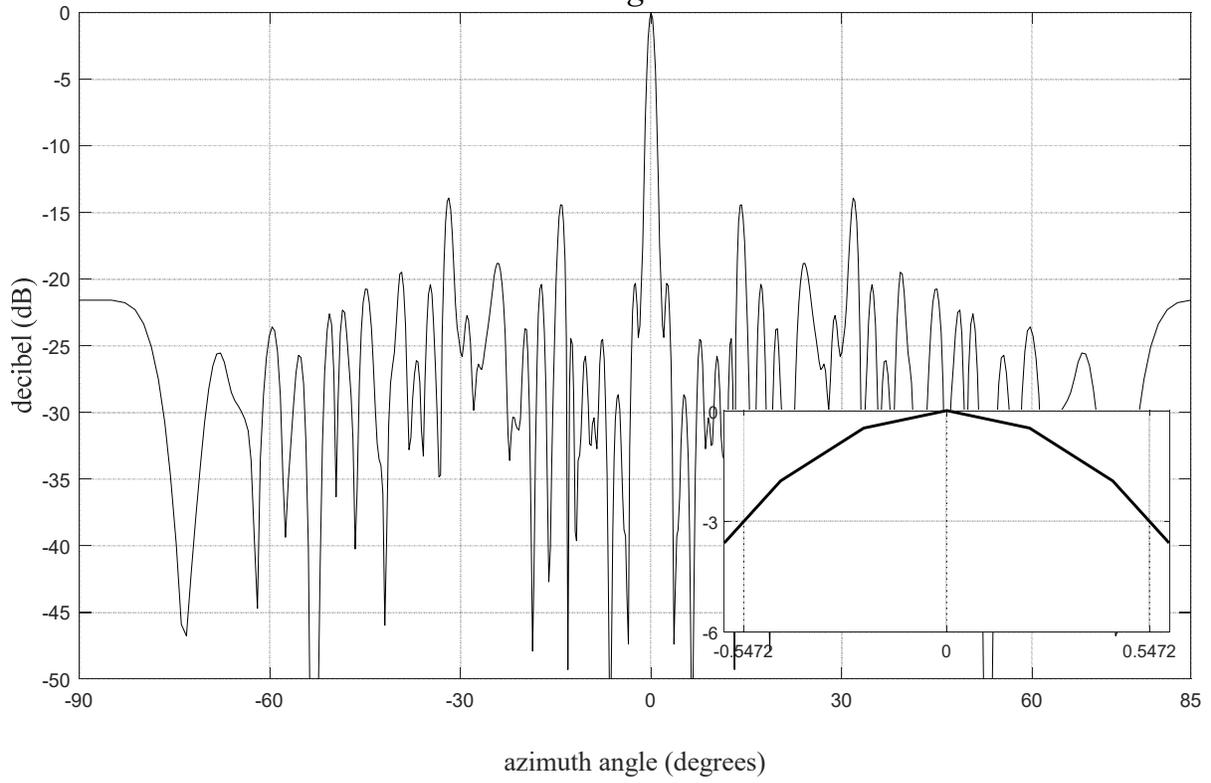

Received Signal Pattern

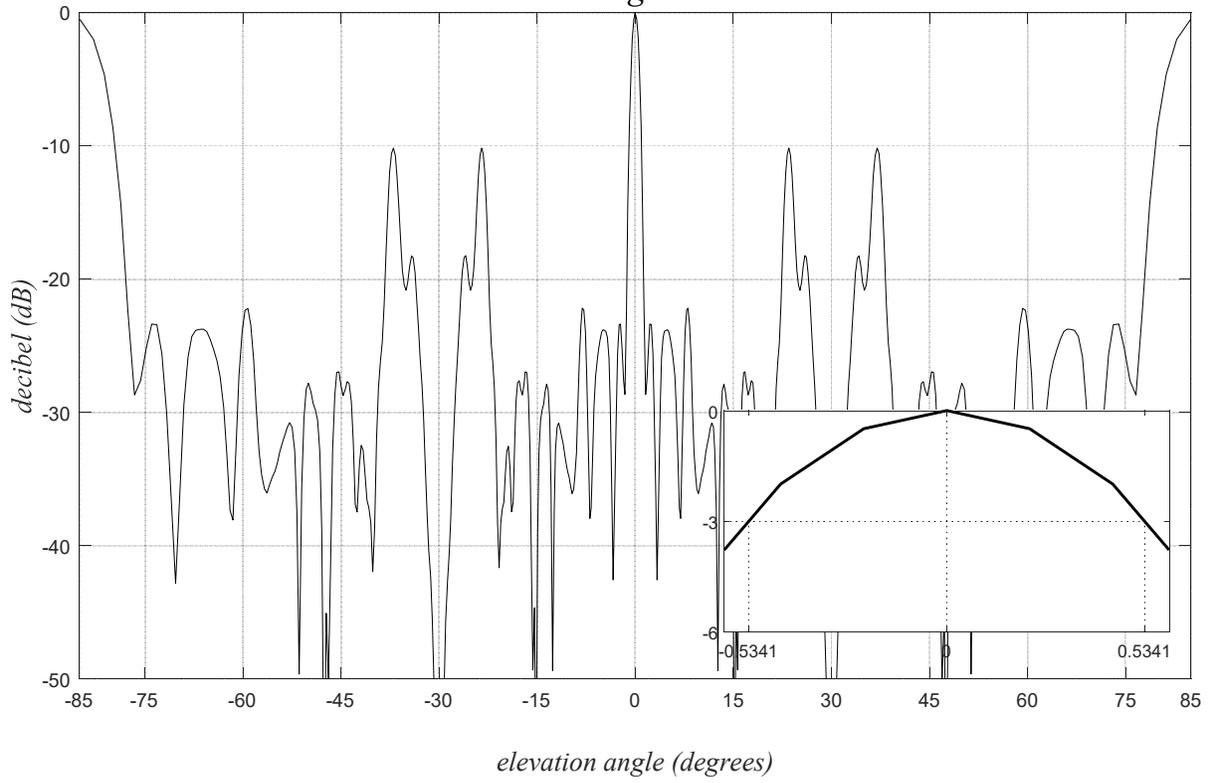

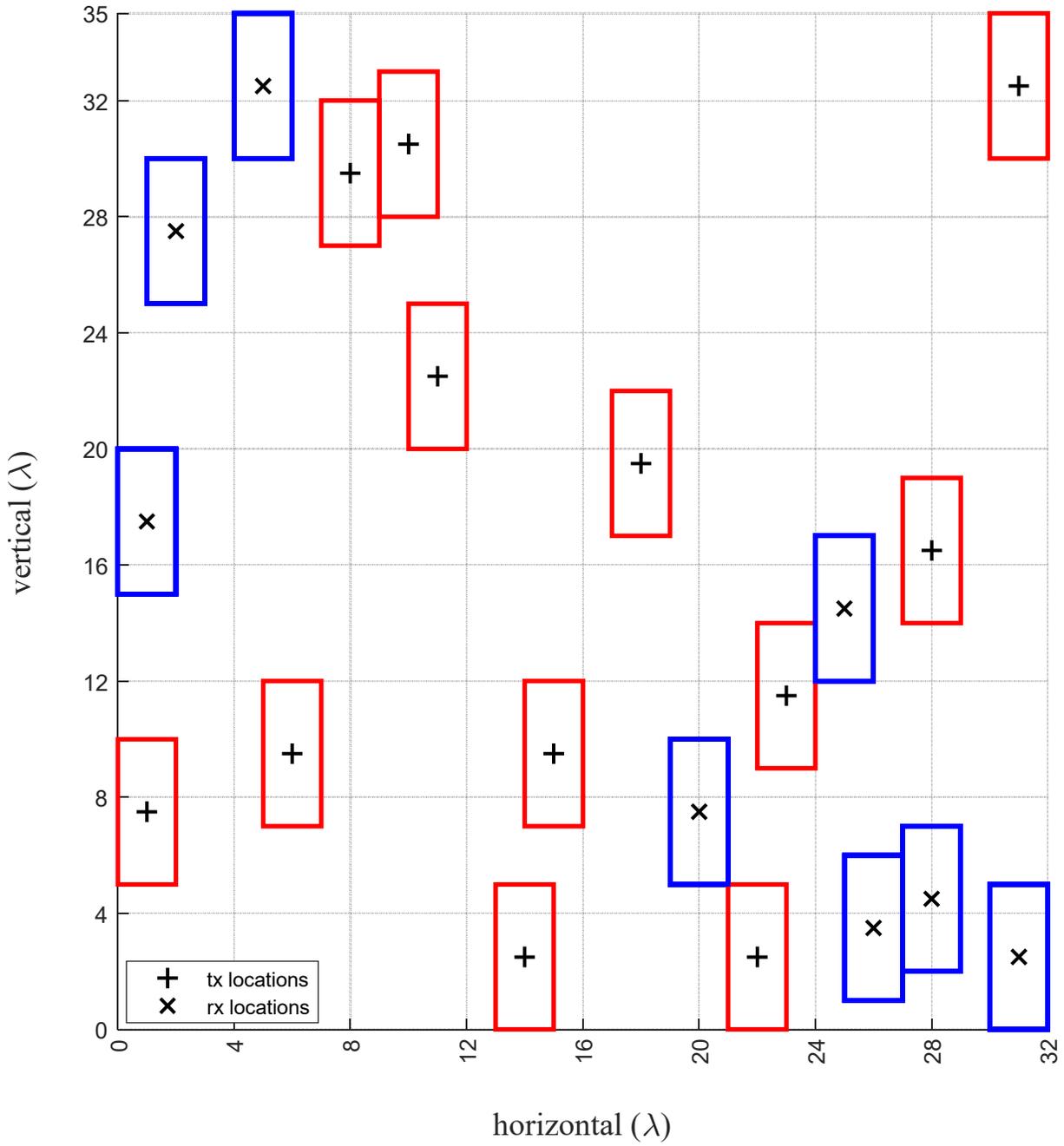

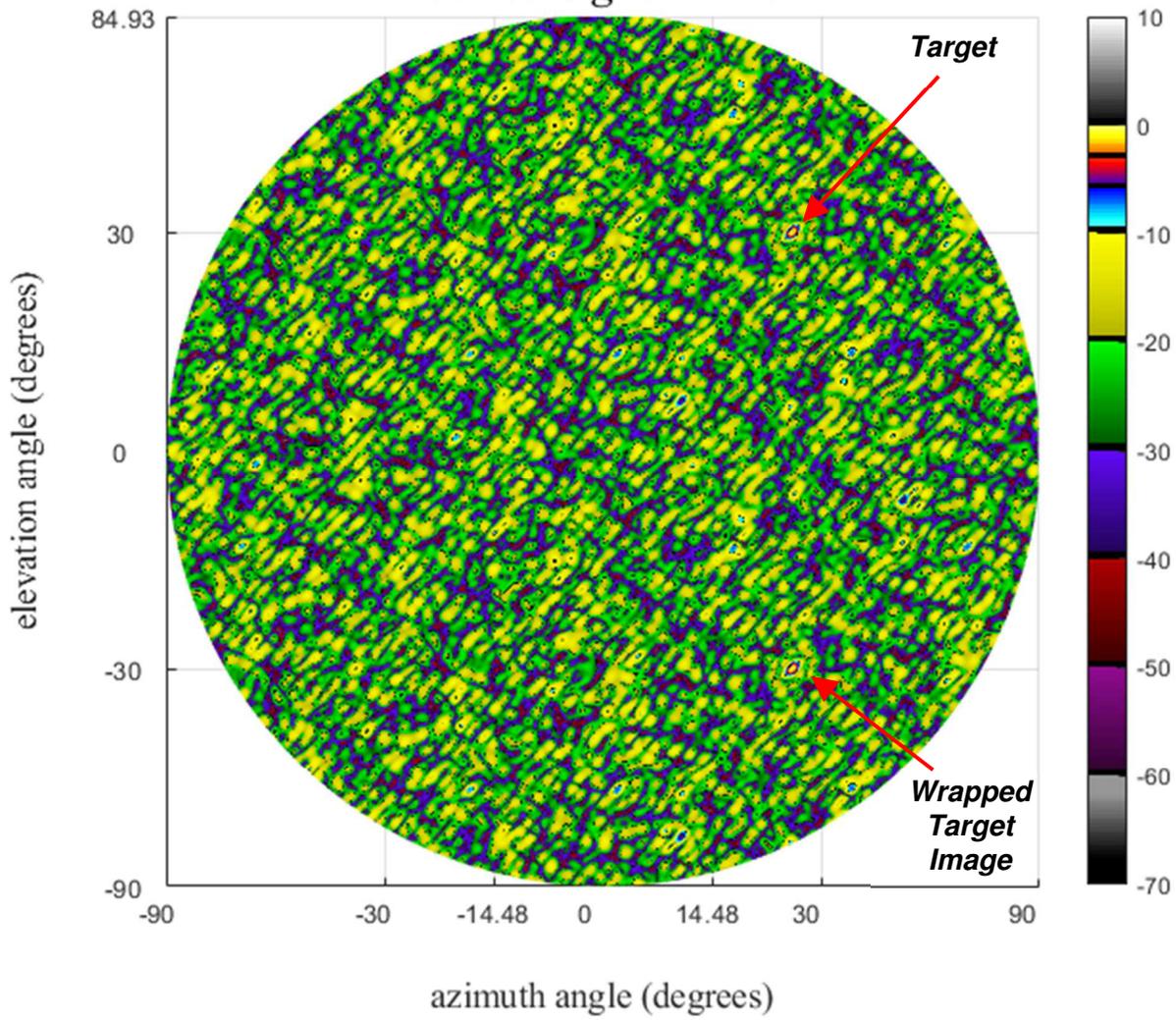

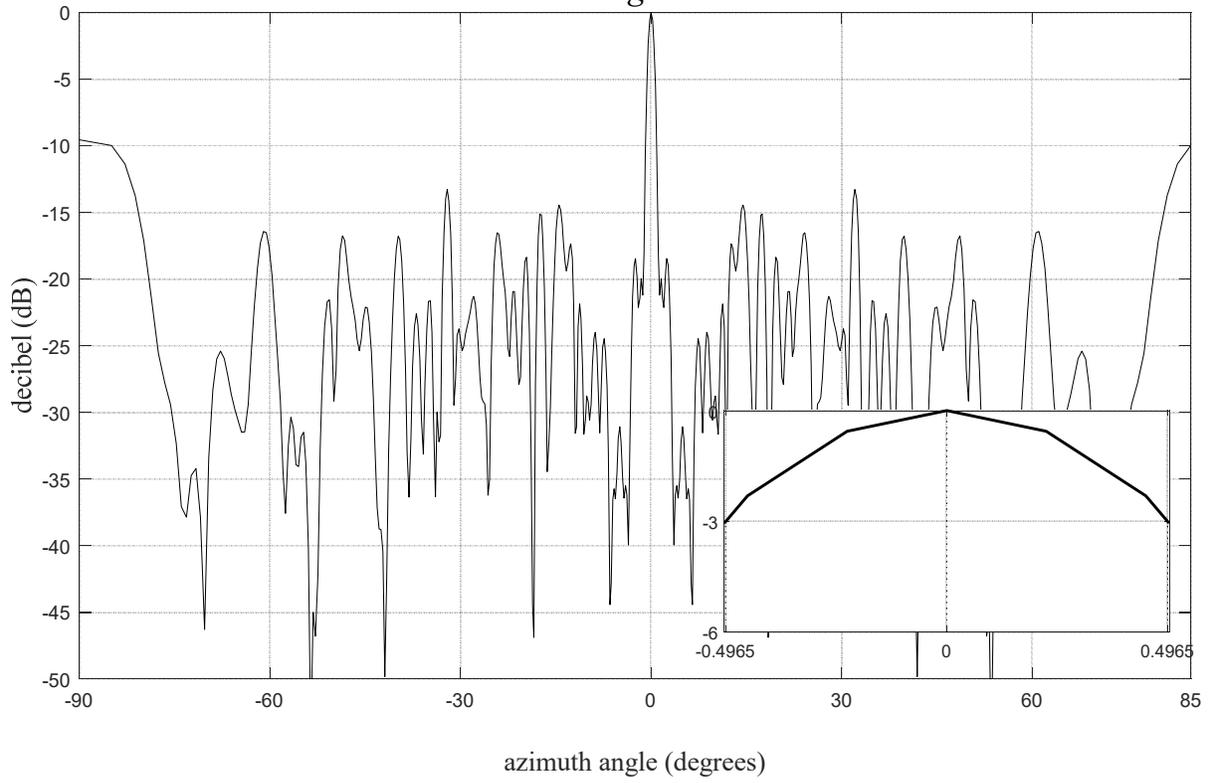

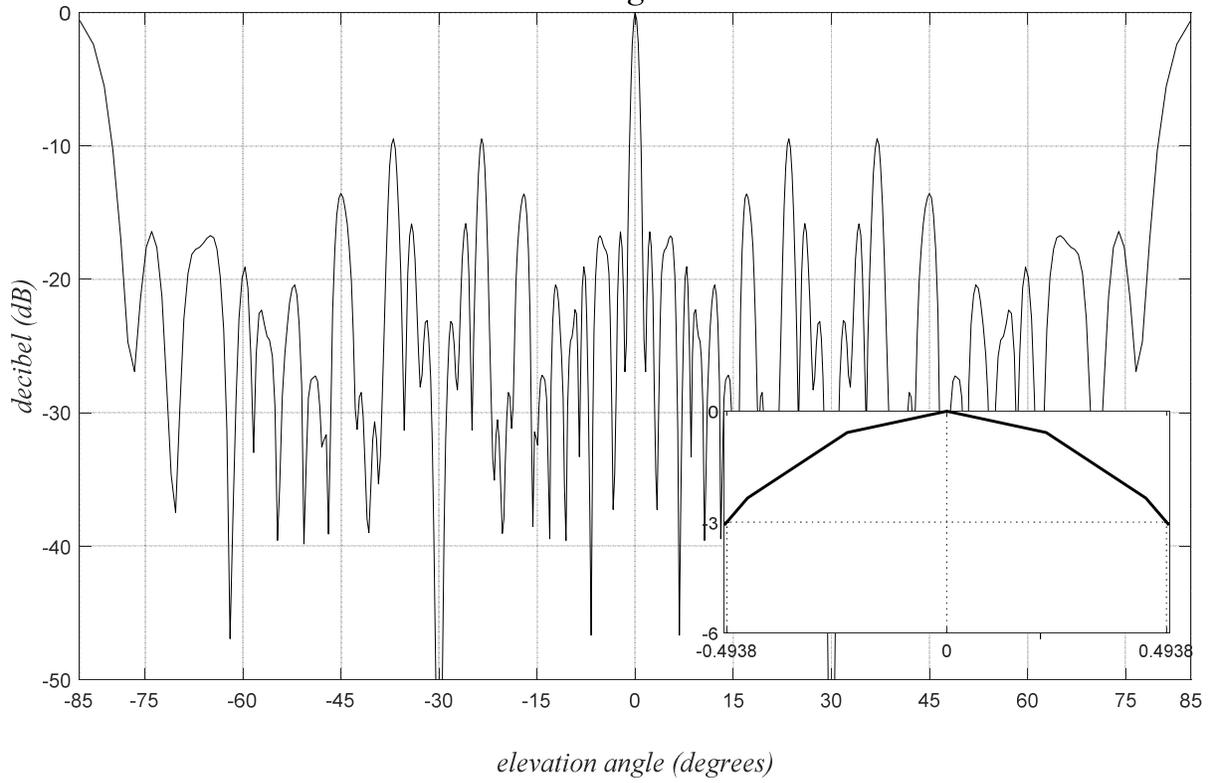

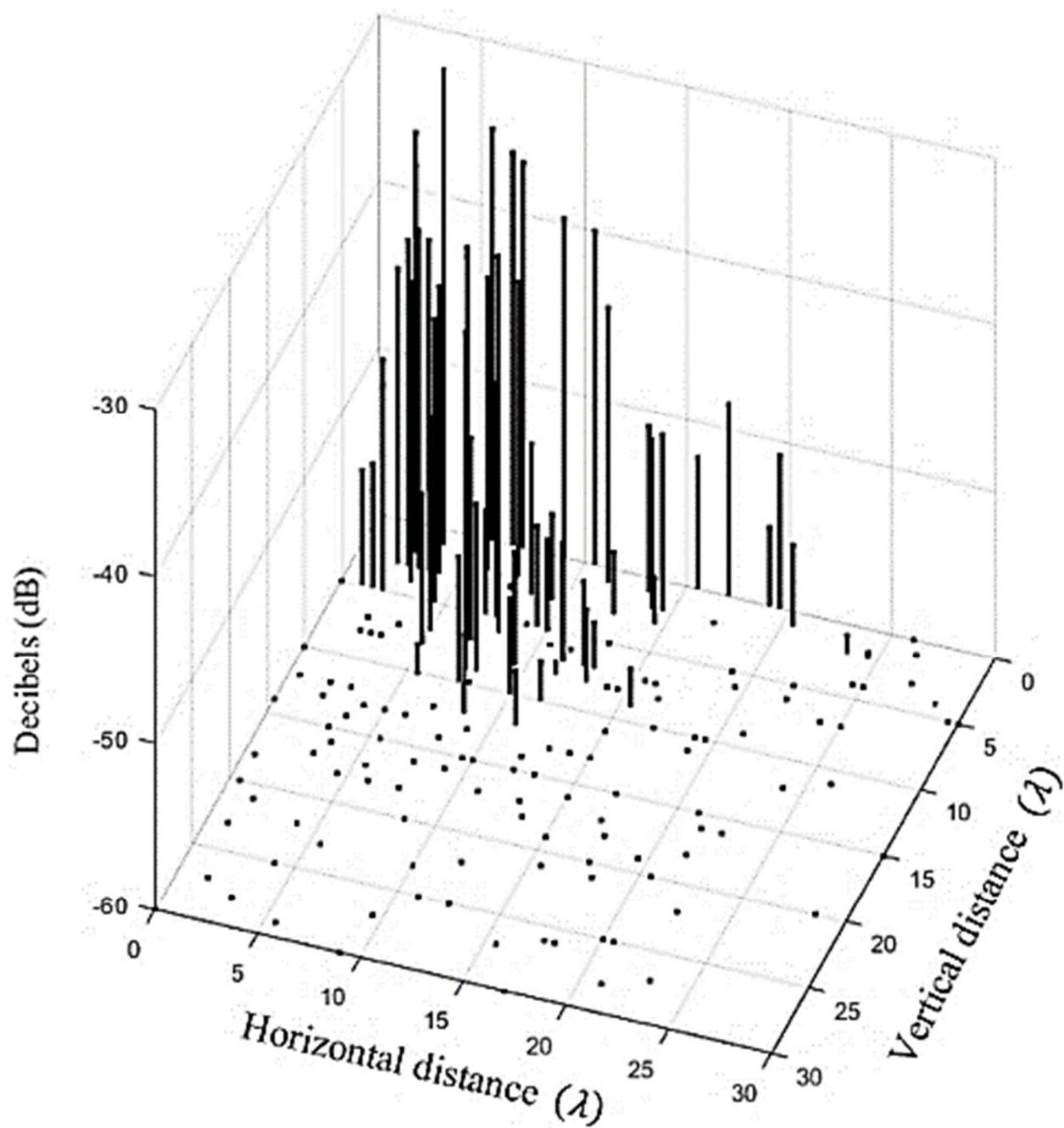

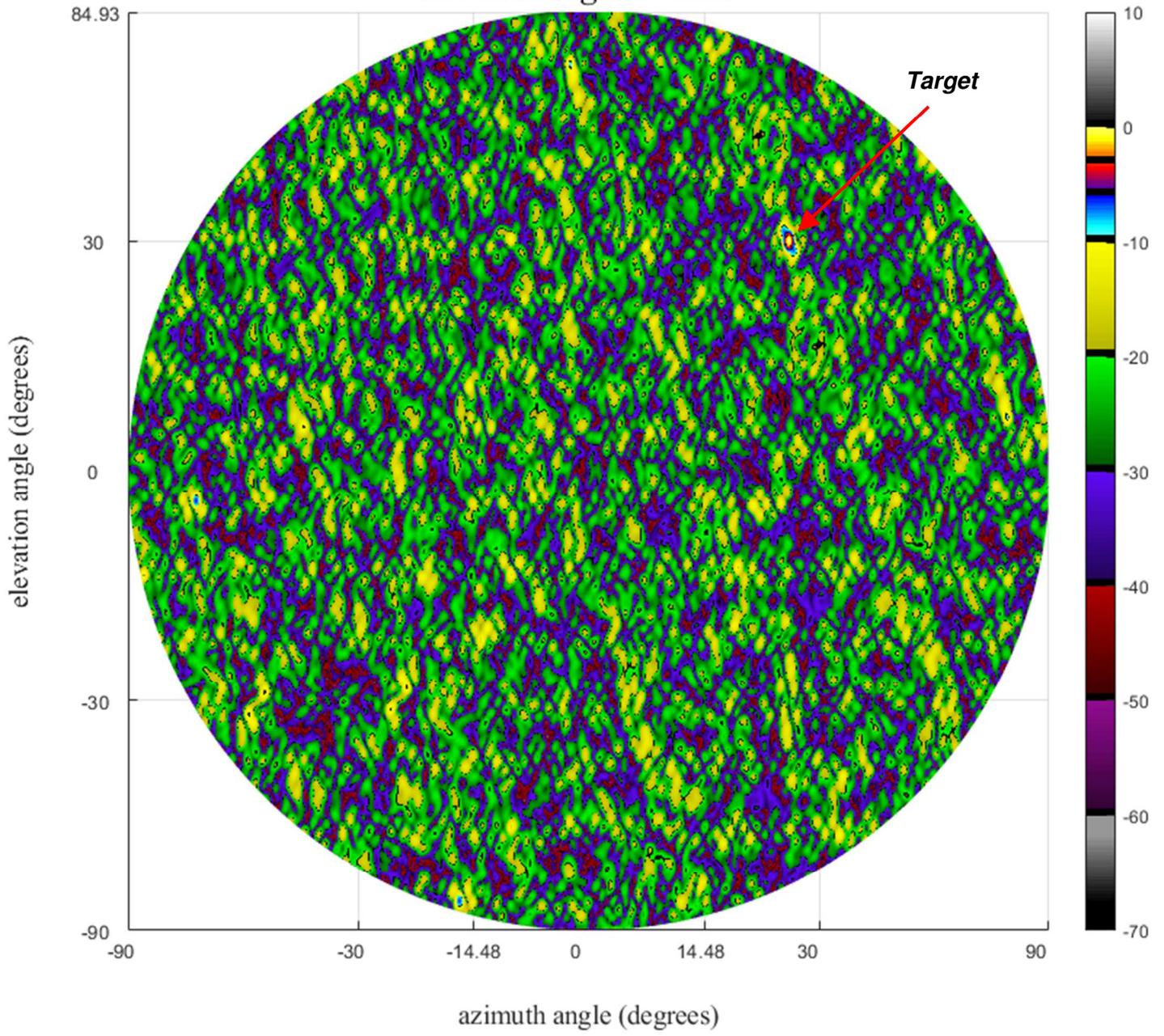

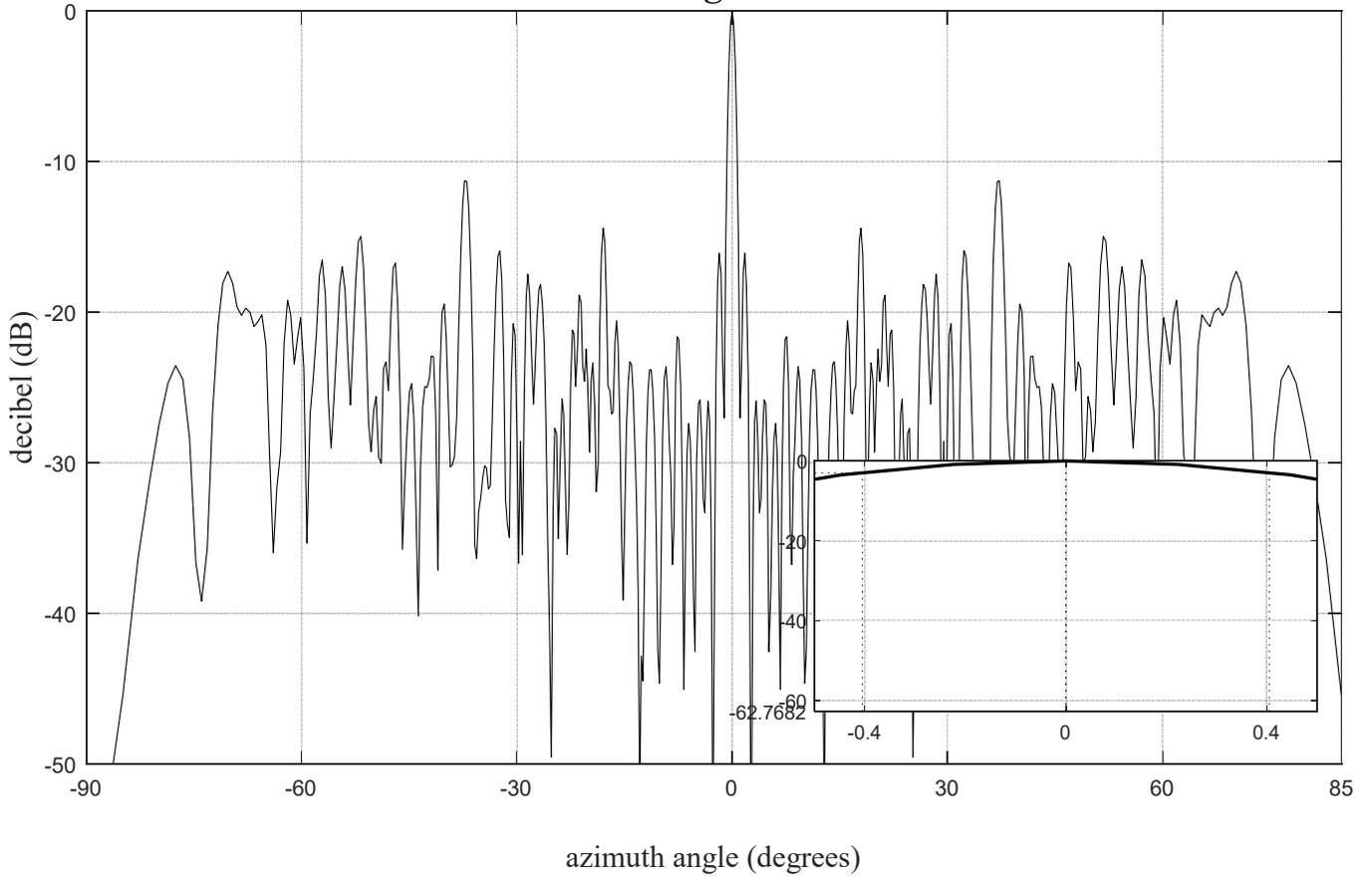

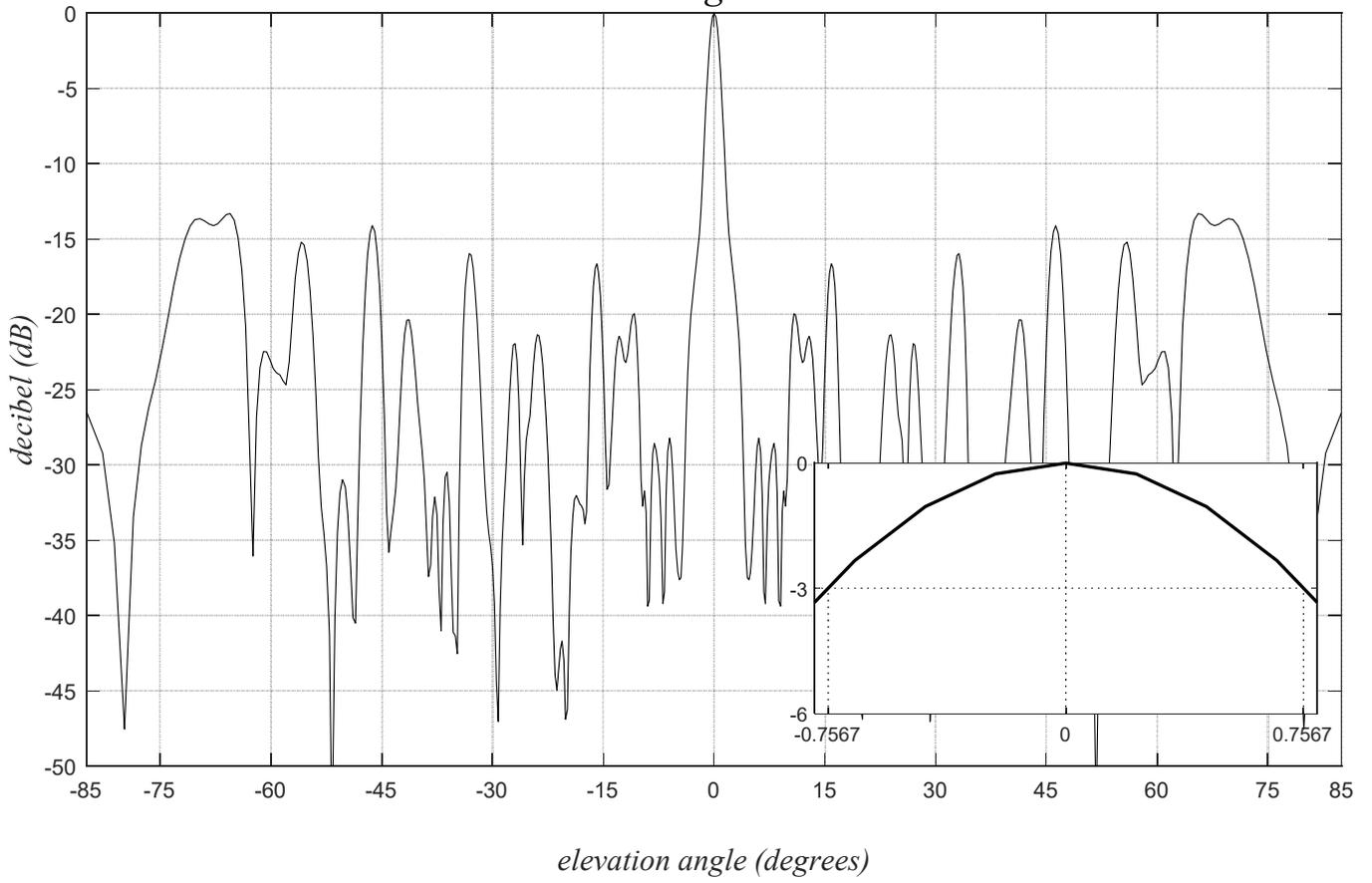

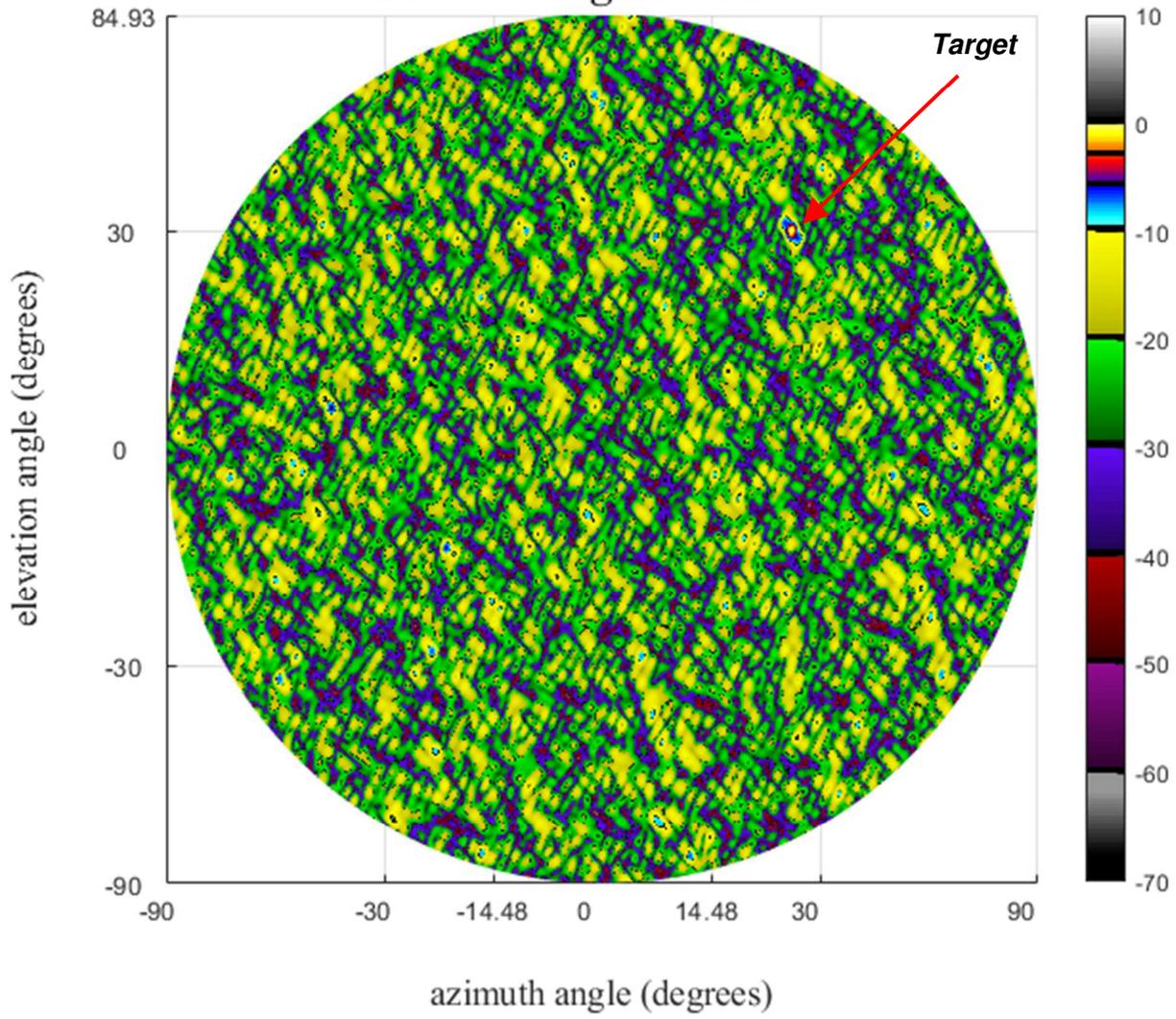

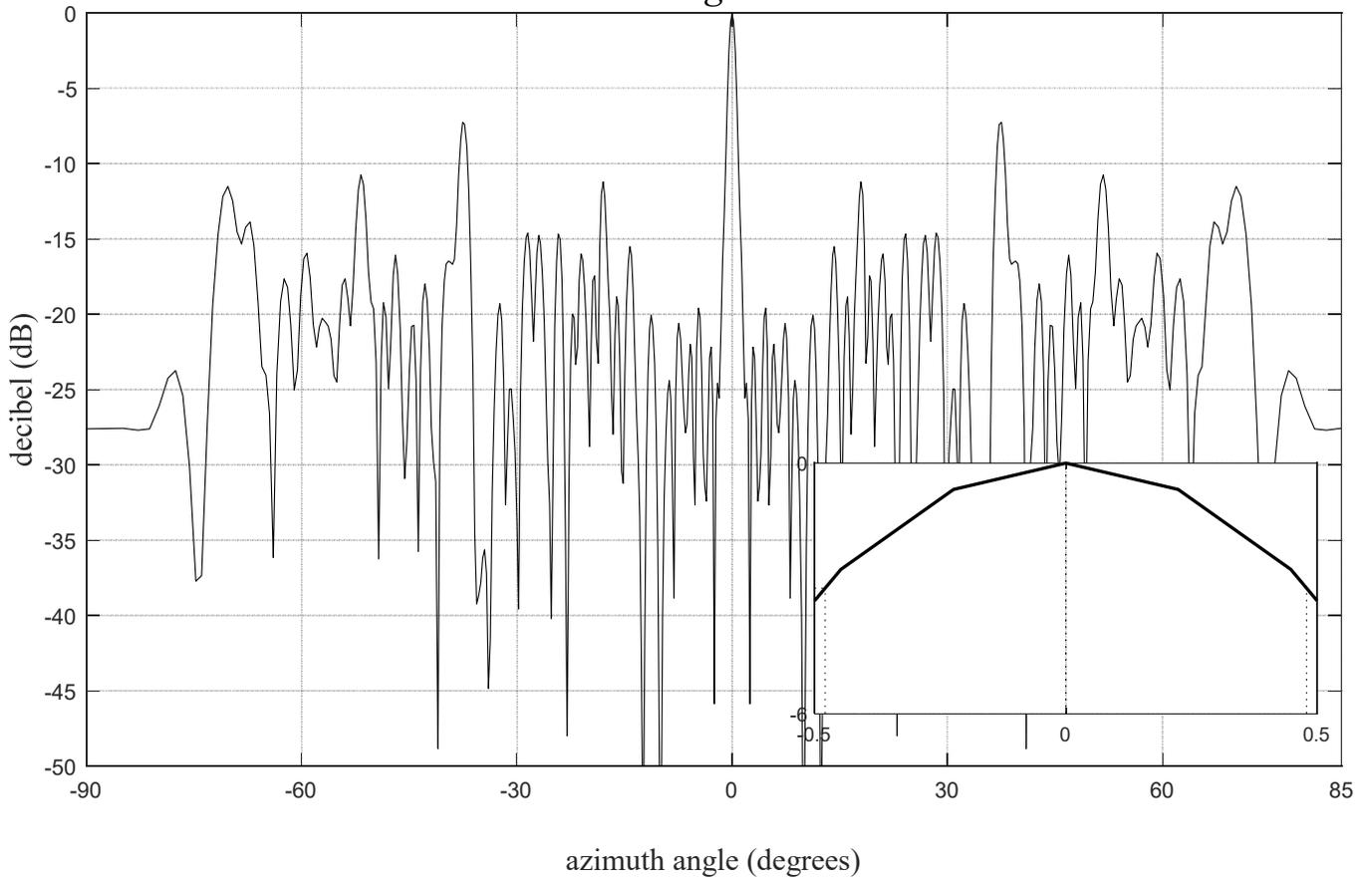

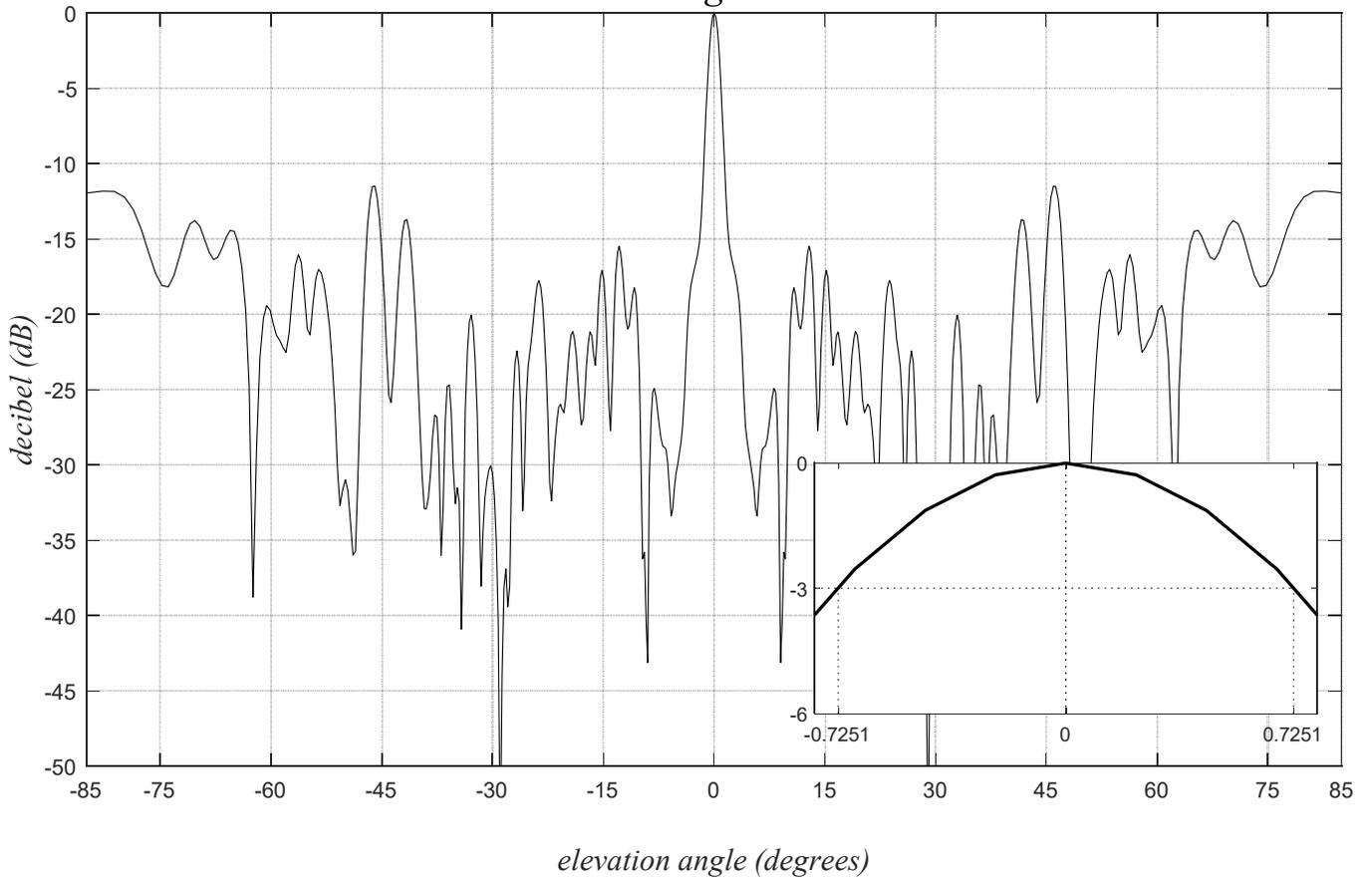

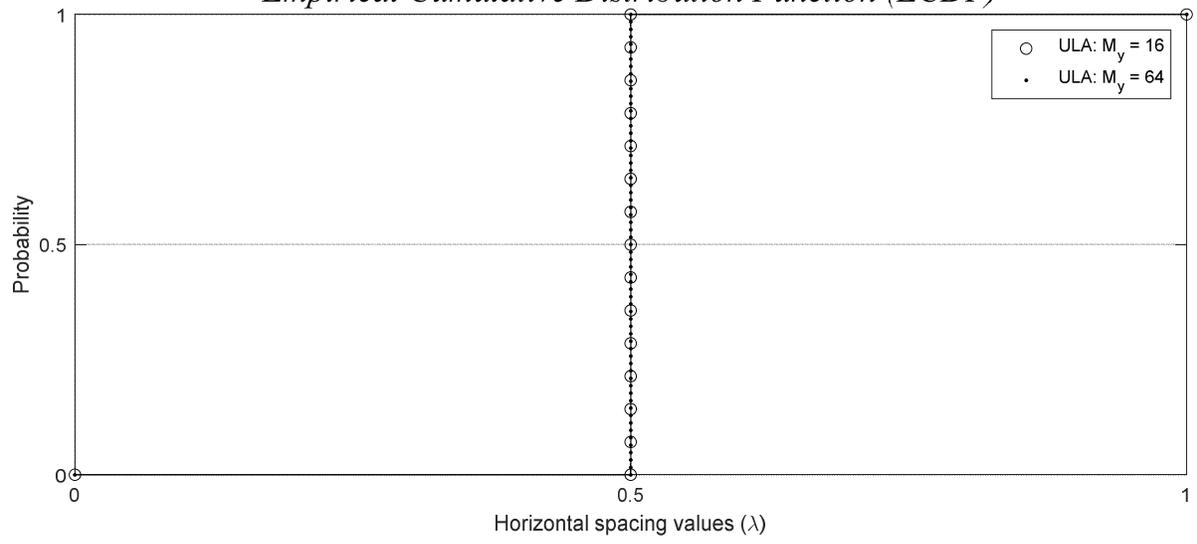

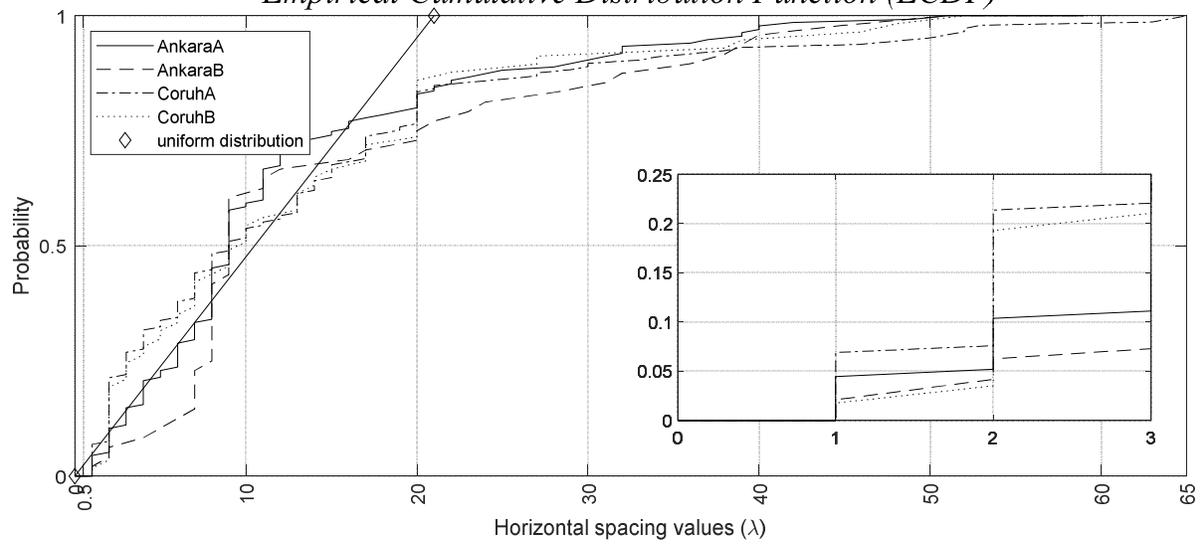

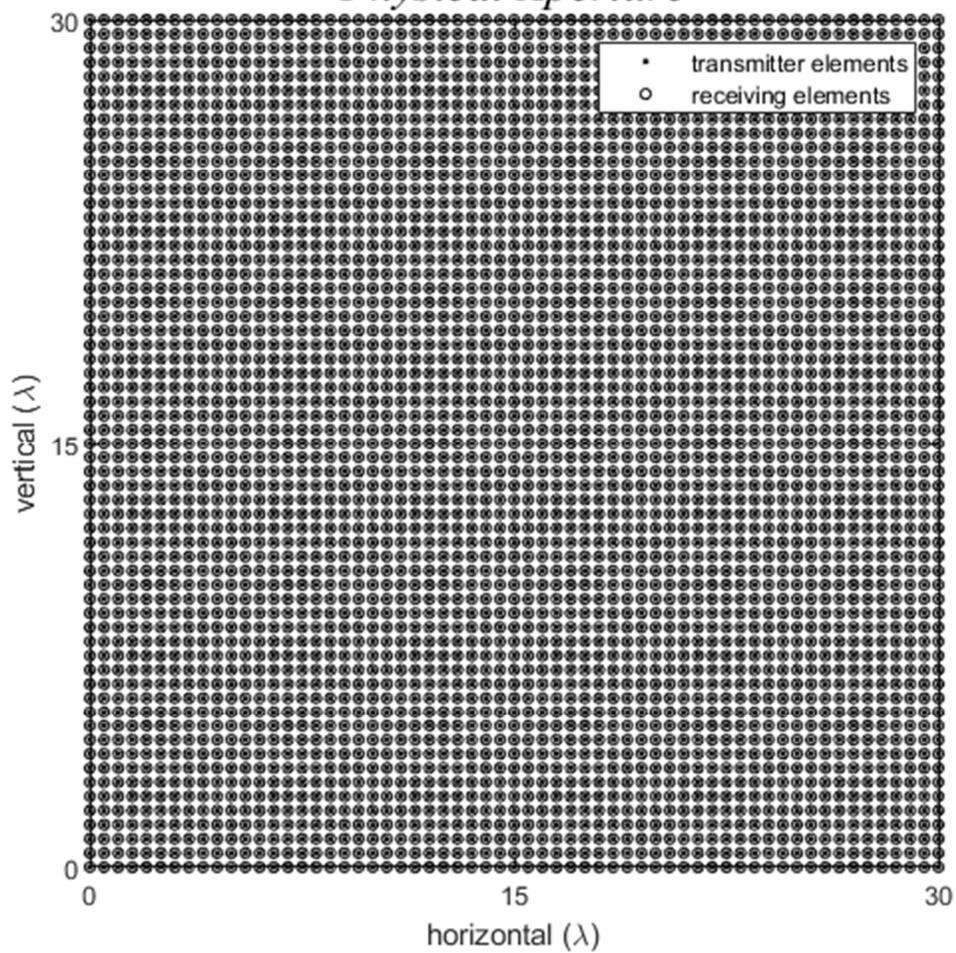

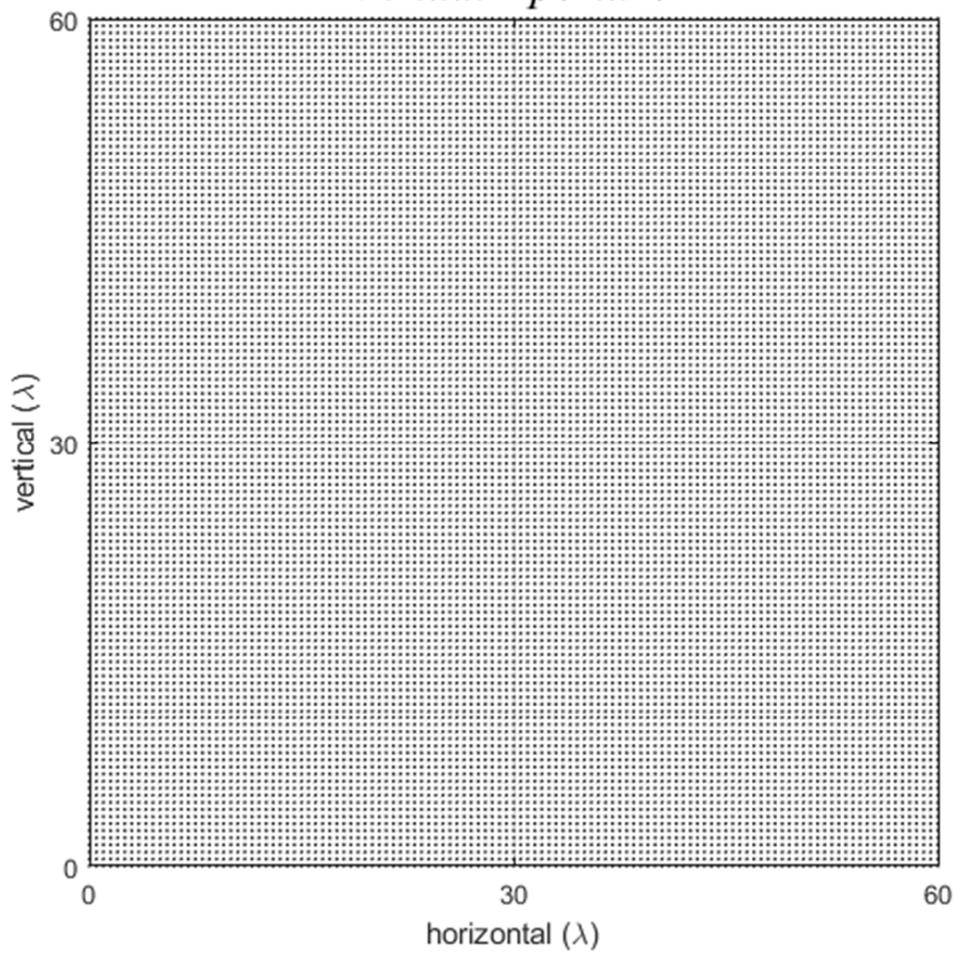

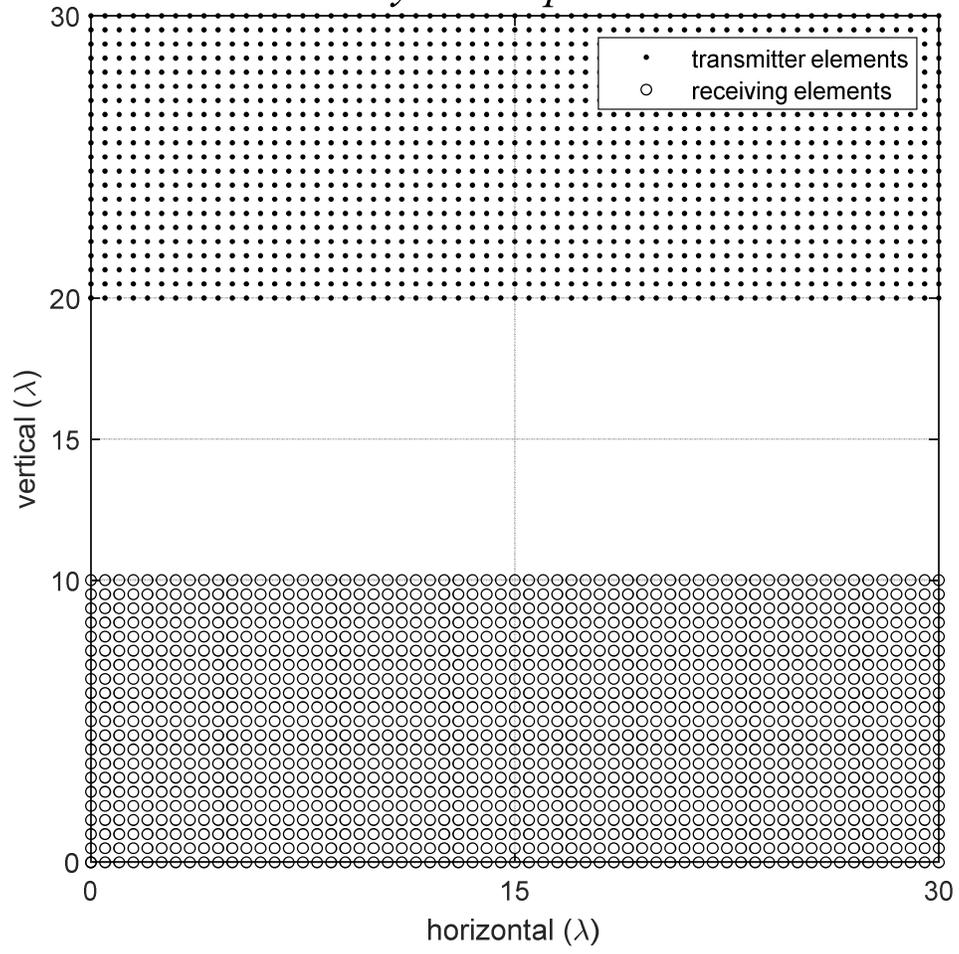

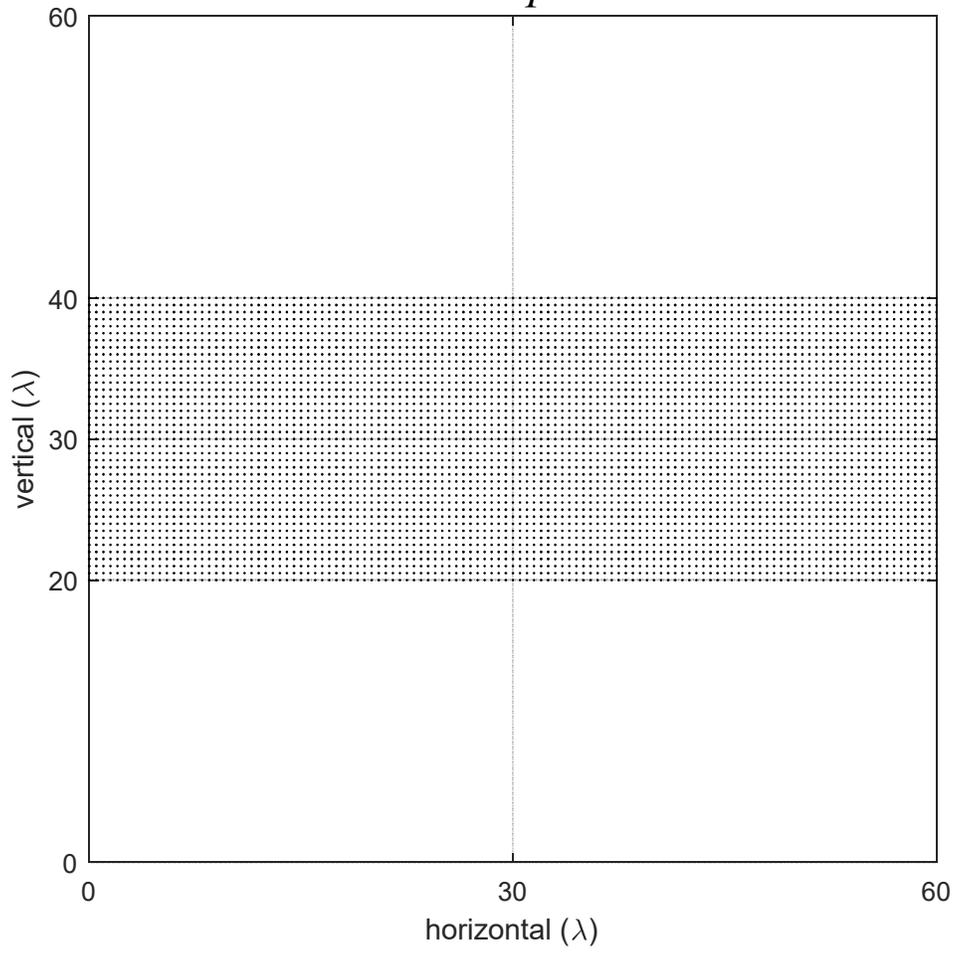

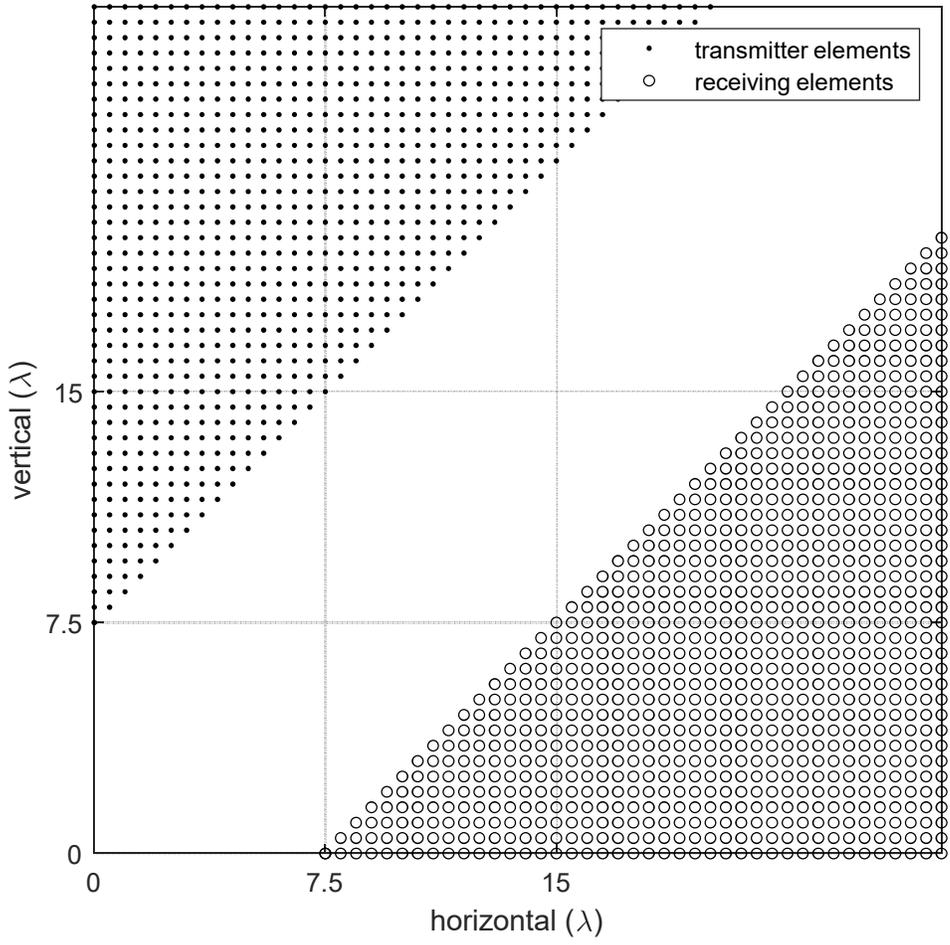

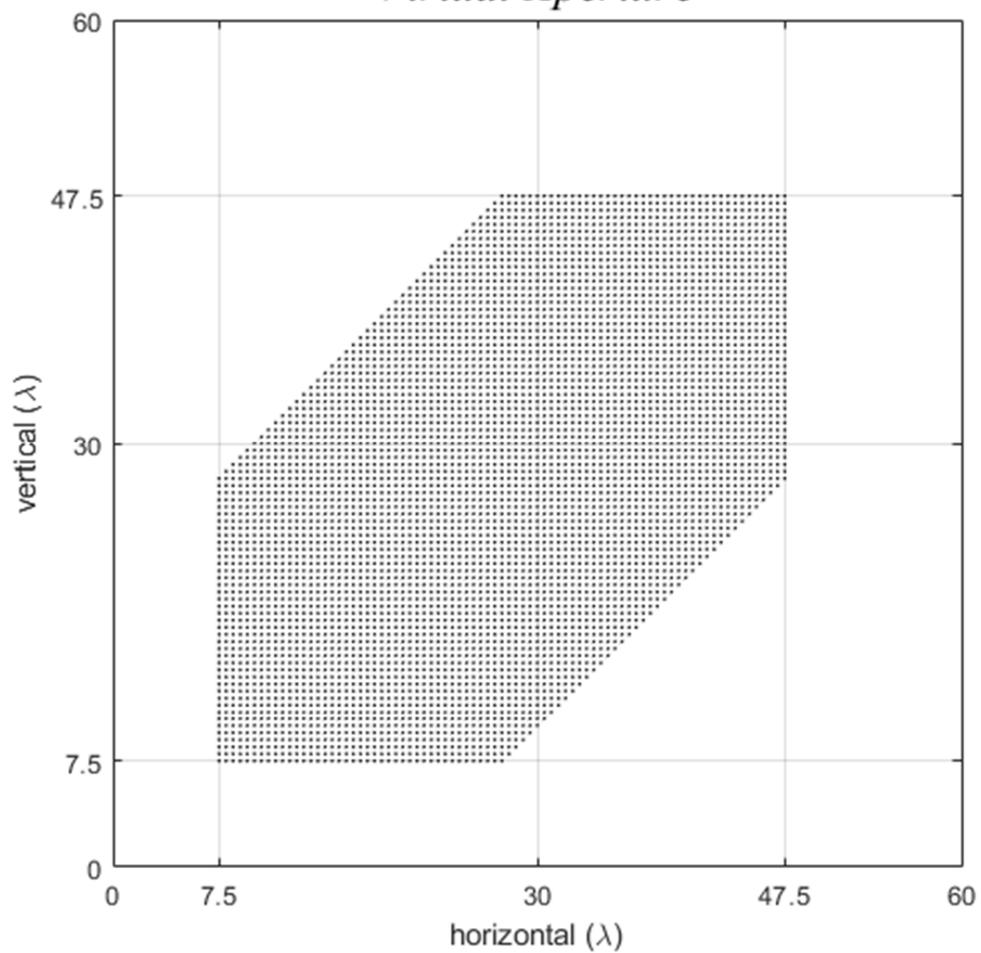

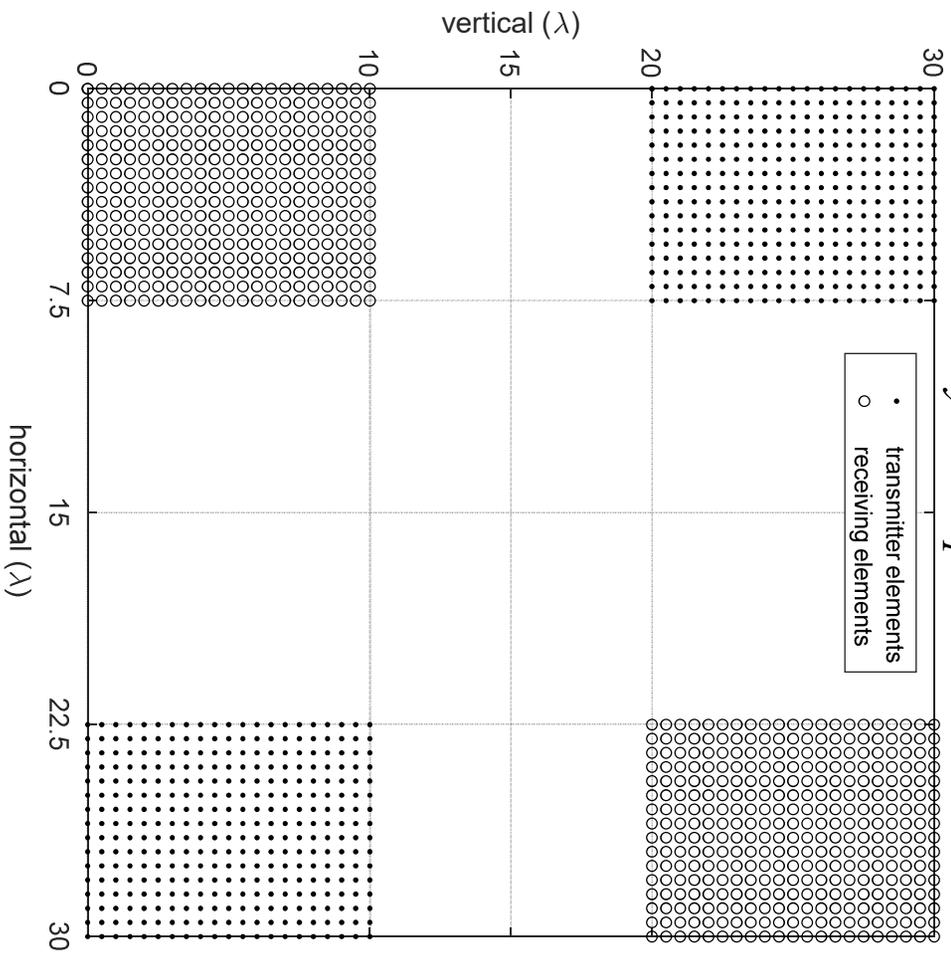

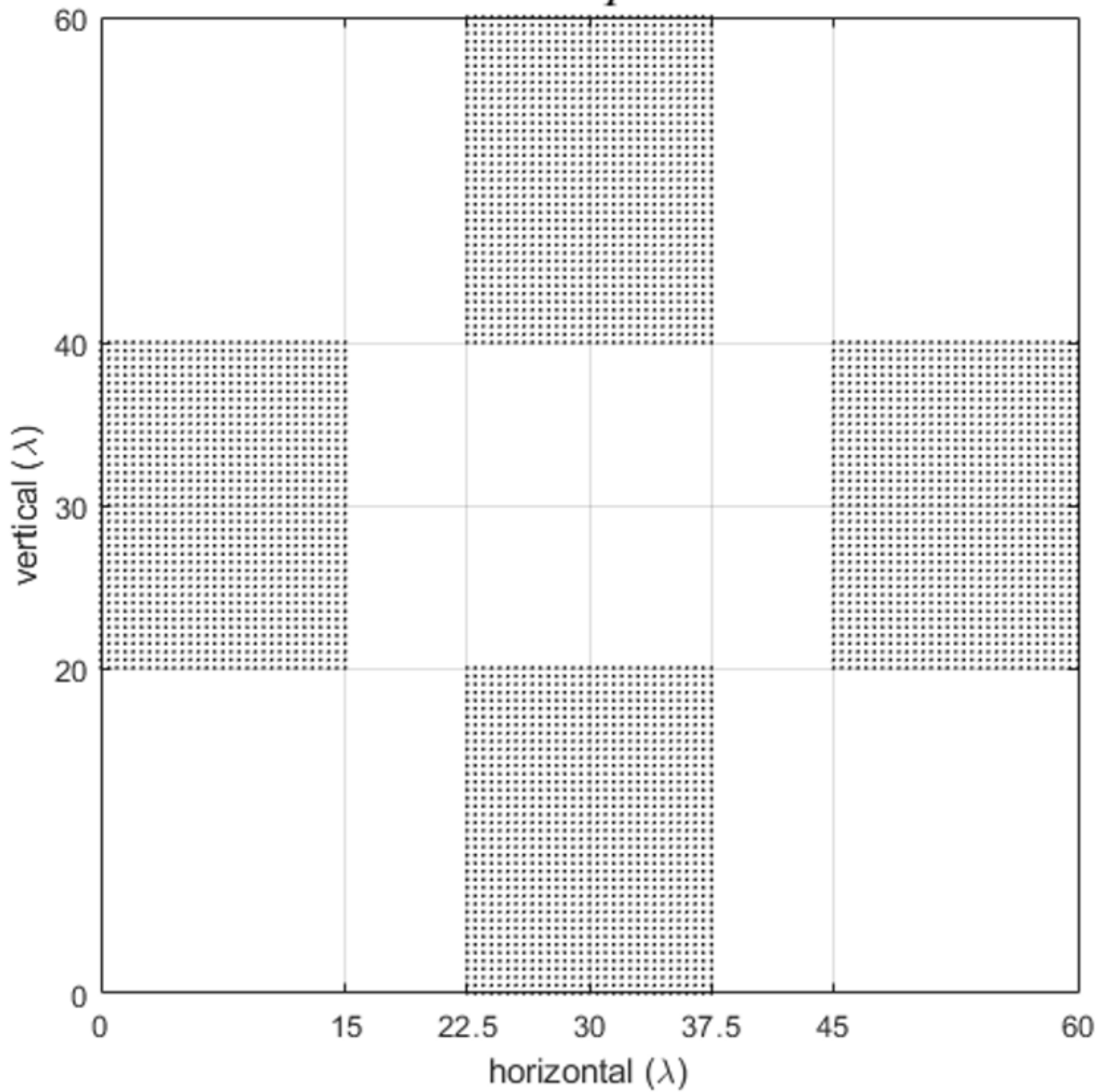

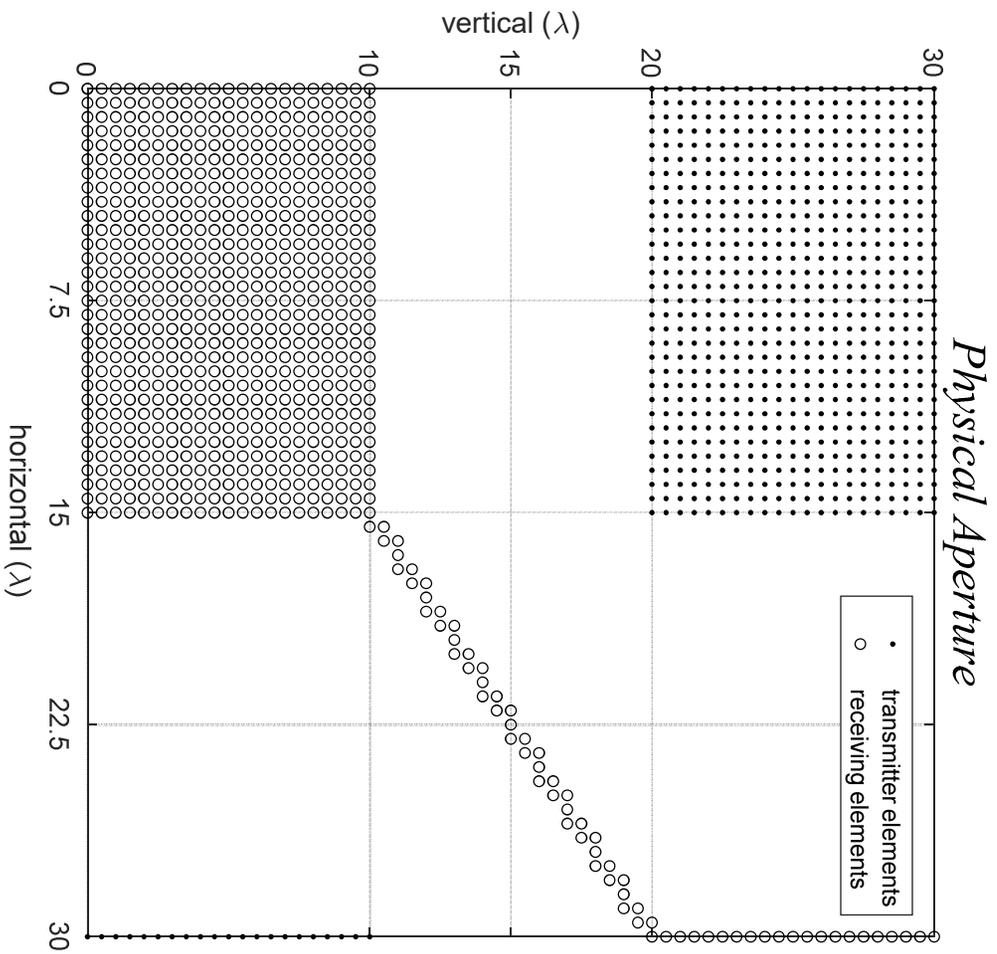

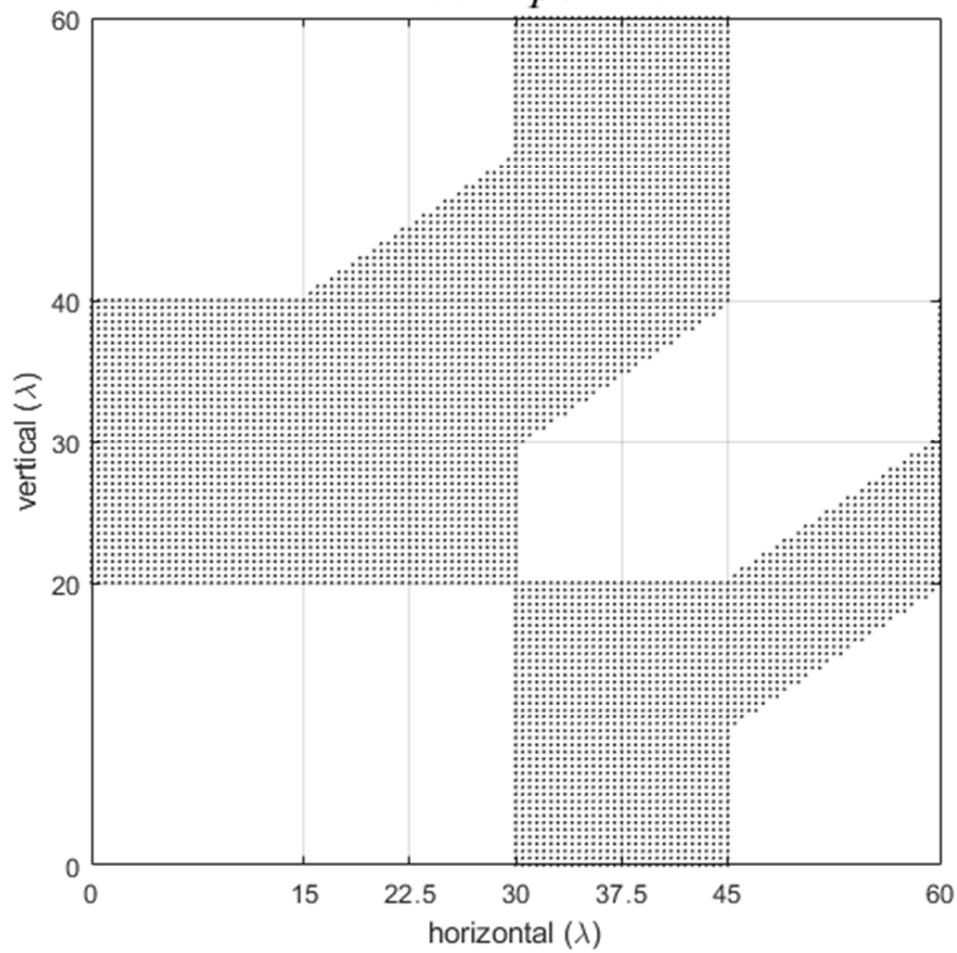